\documentclass[traditabstract]{aa}  
\usepackage{txfonts}
\usepackage{epstopdf}
\usepackage[switch]{lineno}
\usepackage{graphicx,longtable,lscape,natbib,amssymb,amssymb,amsmath}
\bibpunct{(}{)}{;}{a}{}{,} 
\newcommand{\ltsima} {$\; \buildrel < \over \sim \;$}  
\newcommand{\gtsima} {$\; \buildrel > \over \sim \;$}  
\newcommand{\lta} {\lower.5ex\hbox{\ltsima}}  
\newcommand{\gta} {\lower.5ex\hbox{\gtsima}}  
\newcommand{\ha} {H$\alpha$}

\newcommand{\ergs}{\>{\rm erg}\,{\rm s}^{-1}}
\newcommand{\ergshz}{\>{\rm erg}\,{\rm s}^{-1}\,{\rm Hz}^{-1}}

\newcommand{\kms}{$\rm{\,km \,s}^{-1}$}

\newcommand{\forb}[2]{\mbox{$[{\rm #1\, #2}]$}}
\newcommand{\oiii}{\forb{O}{III}}

\newcommand{\nii}{\forb{N}{II}\,}

\usepackage{color}

\begin{document}

\title{The MURALES survey. III.} \subtitle{Completing the MUSE
  observations of 37 3C low-z radio galaxies.}

\author{Barbara Balmaverde\inst{1} 
                \and Alessandro Capetti\inst{1}
                \and Alessandro Marconi\inst{2,3}
                \and Giacomo Venturi\inst{3,13}
                \and M. Chiaberge\inst{4,5}
                \and R.D. Baldi\inst{6} 
                \and S. Baum\inst{8}
                \and R. Gilli\inst{7}
                \and P. Grandi\inst{7}
                \and Eileen T. Meyer\inst{11}
                \and G. Miley\inst{9}
                \and C. O$'$Dea\inst{8}
                \and W. Sparks\inst{12}
                \and E. Torresi\inst{7} 
                \and G. Tremblay\inst{10}}
\institute {INAF - Osservatorio Astrofisico di Torino, Via Osservatorio 20, I-10025 Pino Torinese, Italy
\and Dipartimento di Fisica e Astronomia, Universit\`a di Firenze, via G. Sansone 1, 50019 Sesto Fiorentino (Firenze), Italy
 \and INAF - Osservatorio Astrofisico di Arcetri, Largo Enrico Fermi 5, I-50125 Firenze,Italy
 \and Space Telescope Science Institute, 3700 San Martin Dr., Baltimore, MD 21210, USA
\and Johns Hopkins University, 3400 N. Charles Street, Baltimore, MD 21218, USA
 \and  INAF- Istituto di Radioastronomia, Via Gobetti 101, I-40129 Bologna, Italy
\and INAF - Osservatorio di Astrofisica e Scienza dello Spazio di Bologna, via Gobetti 93/3, 40129 Bologna, Italy
\and Department of Physics and Astronomy, University of Manitoba, Winnipeg, MB R3T 2N2, Canada
\and Leiden Observatory, Leiden University, PO Box 9513, NL-2300 RA, Leiden, the Netherlands
\and Harvard-Smithsonian Center for Astrophysics, 60 Garden St., Cambridge, MA 02138, USA
\and University of Maryland Baltimore County, 1000 Hilltop Circle, Baltimore, MD 21250, USA
\and SETI Institute, 189 N. Bernado Ave Mountain View,CA 94043
\and Instituto de Astrof\'isica, Facultad de F\'isica, Pontificia Universidad Cat\'olica  de Chile, Casilla 306, Santiago 22, Chile
}

\date{} 

\abstract{We present the final observations of a complete sample of 37
  radio galaxies from the Third Cambridge Catalogue (3C) with redshift
  $<$0.3 and declination $<$20$^{\circ}$ obtained with the VLT/MUSE
  optical integral field spectrograph. These data were obtained as
  part of the  MUse RAdio Loud Emission line Snapshot
(MURALES)  survey with the main goal of exploring the AGN feedback process in
  the most powerful radio sources. We present the data analysis and,
  for each source, the resulting emission line images and the 2D gas
  velocity field. Thanks to the unprecedented depth these
  observations reveal emission line regions (ELRs) extending several
  tens of kiloparsec in most objects. The gas velocity shows ordered
  rotation in 25 galaxies, but in several sources it is highly
  complex. We find that the 3C sources show a connection between radio
  morphology and emission line properties. In the ten FR~I sources the line
  emission region is generally compact, only a few kpc in size; only in one case does it exceed the size of the host. Conversely, all
  but two of the FR~II galaxies show large-scale structures of ionized
  gas. The median extent is 16 kpc with the maximum reaching a size of
  $\sim80$ kpc. There are no apparent differences in extent or
  strength between the ELR properties of the FR~II sources of high and low
  gas excitation. We confirm that the previous optical identification
  of 3C~258 is incorrect: this radio source is likely associated with
  a quasi-stellar object at $z\sim 1.54$.}


\titlerunning{The MURALES survey} 
\authorrunning{B. Balmaverde et al.}
 \maketitle

\section{Introduction}
\label{intro}
Radio galaxies (RGs), usually hosted by the brightest galaxies at
  the center of clusters or groups, are among the most energetic
  manifestations of   supermassive black holes (SMBHs) in the
  Universe. The study of these objects has become particularly
important due to their role in the  feedback process, which is
the exchange of matter and energy between AGN, their host galaxies,
and clusters of galaxies. The evidence of AGN feedback mode is often
witnessed in local RGs, showing the presence of cavities inflated by
the radio emitting gas in the X-ray images (e.g., kinetic feedback;
\citealt{birzan04,birzan20}). However, we still lack a comprehensive
view of the effects that the highly collimated jets and the nuclear
emission have on the host and its immediate environment. In
particular, it is still unclear how precisely the coupling between
radio jets and ionized gas occurs, and whether the jets are able to
accelerate the gas above the host escape velocity
\citep{mcnamara07}. We also lack clear observational evidence of the conditions under which jets enhance or quench star formation (positive or
negative feedback, e.g., \citealt{fabian12}). Furthermore, the
mechanical luminosity released by the AGN is not well-constrained
because it is estimated using indirect and model-dependent approaches 
(e.g., \citealt{blanton11,panagoulia14}). Finally, the radiative output
from the AGN can produce fast outflows of ionized gas, which might
also affect the properties of the ambient medium (e.g.,
\citealt{balmaverde16, wylezalek16,carniani16,cresci18}).

The AGN feedback action can be directly probed by studying the
  narrow-line region (NLR), a region of intense narrow emission lines
  that can usually   be traced up to few kiloparsec from the nucleus.
  The size of the NLR  correlates with the nuclear [O
    III] $\lambda $5007$\AA$ (hereafter [O III]) luminosity as $r
  \propto L_{\rm [O~III]} ^{0.5}$ \citep{bennert02}, suggesting that
  the gas is photoionized by the nucleus and that the ionization
  parameter scales with the distance.  However, in some quasi-stellar objects (QSOs), the
  [O~III] emission line has been detected up to ~20 kpc or more
  \citep{wampler75,stockton76,richstone77}.  This region,
  characterized by extended line emitting gas showing complex shape
  and kinematics, has been named the extended narrow-line region (ENLR).
The properties of the nuclear and the extended gas are related, and
there is a linear relation between the line and radio luminosities in
extended RGs
\citep{baum89a,baum89b,rawlings89,rawlings91,buttiglione10}. Moreover,
the luminosity and size of the ENLR also correlates  with the nuclear
H$\alpha$+[N~II] or [O~III] luminosity \citep{Mulchaey96}.  The
expected trend between the extended emission line luminosity and total
radio power, and between the size of the emission line region
and redshift, have been found by other studies using HST data
 \citep[e.g.,][]{tremblay09,baldi19,privon08}.  We can now re-explore these
relations using the higher sensitivity of MUSE.
 
In this context, the third Cambridge Catalogue of RGs (3C,
\citealt{spinrad85}) represents the best testbed to explore these
open problems. It covers a wide range of radio power; it contains all types
of sources from the point of view of their optical spectrum and radio
morphology, enabling us to perform statistically robust comparison
among the various classes. In the last two decades a superb suite of
ground- and space-based observations has been built for the 3C with
all major observing facilities from HST to Chandra, Spitzer, Herschel,
and JVLA.

In \citet{balmaverde18a} we presented a spectacular example of
  the capabilities of MUSE for this class of sources, discussing the
  observation of 3C~317, a radio galaxy located at the center of the
  Abell cluster A2052. We detected a complex network of emission lines
  filaments, enshrouding the northern cavity, extending out to $\sim$
  20 kpc. The emission line ratios in the filaments suggest ionization
  from slow shocks or from cosmic rays; we did not detect any star
  forming regions.  Modeling the global kinematics, we derived a
  velocity expansion of $\sim$ 250 km s$^{-1}$, a value  we
  used to   estimate  the cavity age.

The  MUse RAdio Loud Emission lines Snapshot (MURALES) project is a
program aimed at observing the 3C radio sources with the Multi-Unit Spectroscopic Explorer (MUSE) at the
Very Large Telescope (VLT) \citep{bacon10}. Our main goals are to
study the feedback process in a sample of the most powerful radio
sources, to constrain the coupling between the radio source and the
warm gas, to probe the fueling process, and to estimate the net effect
of the feedback on star formation. The MUSE data can be combined with
the unique  multi-band dataset available for these sources, produced
with all major observing facilities at all accessible wavelengths,
adding a key ingredient for our understanding of the radio-loud AGN
phenomenon. 

As a first step of  MURALES, we started with the MUSE observations
of the 3C sources with $z<0.3$, accessible from the VLT (i.e., $\delta
< 20^\circ$). We were recently awarded  MUSE observing time to extend the sample
to larger distances and radio power, observing a sample of 29 3C radio galaxies up to z=0.8. In \citet{balmaverde19} we presented the results of the
observations of the first 20 3C sources obtained in Period 99. Thanks
to their unprecedented depth (the median 3$\sigma$ surface brightness
limit in the emission line maps at the [O~III] wavelength is
6$\times$10$^{-18}$ erg s$^{-1}$ cm$^{-2}$ arcsec$^{-2}$), these
observations reveal extended emission line structures in most objects.
We found that in three of the four
\citet{fanaroff74} Class I radio galaxies (FR~Is), the line emission
regions are compact, $\sim$ 1 kpc in size; in all but one of the
FR~IIs, we detect large-scale structures of ionized gas reaching sizes
of up to $\sim$80 kpc.

The richness of the information stored in the MUSE datacubes enables us to investigate
many different scientific questions that cannot be exhausted here. For example,  we 
can obtain deep  emission line images
and compare them with the X-ray and radio structures; exploring the spatial
link between the hot and warm ionized ISM phases, and with the radio
outflows, could provide evidence of shocked regions of gas along the jets path and/or around the radio lobes.
We can also derive spatially resolved emission line ratio maps to
explore the gas physical conditions in order to detect star forming regions, in search of positive feedback.
In addition, the profile of the emission lines can reveal evidence of nuclear outflows affecting large-scale  environments.

We  complete our survey by presenting the MUSE data of the
remaining 17 sources, observed in Period 102. The paper is organized
as follows. In Sect. 2 we present the sample observed, provide an
observation log, and  describe the data reduction. In Sect. 3 we
describe the case of 3C~258. In Sect. 4 we present the properties of
the extended emission line region and in Sect. 5 we show the resulting emission line images,
line ratio maps, and 2D velocity fields. We also provide a description
of the individual sources. The results are summarized in Sect. 6.

In two forthcoming papers we will study the properties of the extended
emission line structures in order to explore their origin and their
connection with the radio jets. We will also look for nuclear outflows of
ionized gas.

We adopt the following set of cosmological parameters: $H_{\rm o}=69.7$
\kms\ Mpc$^{-1}$ and $\Omega_m$=0.286 \citep{bennett14}.

\begin{figure}  
\centering{ 
\includegraphics[width=7.5cm]{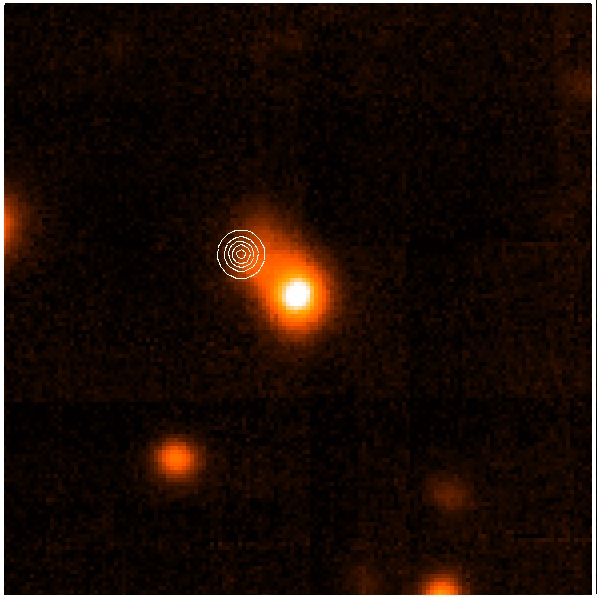}
\caption{Radio contours (white) of 3C~258 from \citet{neff95}
  superimposed on the optical continuum image (field of view $25\arcsec
  \times 25\arcsec$) derived from the MUSE data. The radio source is
  offset by 2\farcs8 to the NE with respect to its current (incorrect)
  optical identification, a galaxy at z=0.165. At the location of
  radio source we find a type I AGN at a tentative redshift z=1.54.}}
\label{3C258}
\end{figure}

\begin{landscape}
\begin{table*}
\caption{Main properties of the 3C subsample observed with MUSE, and the observations log}
\begin{tabular}{l l c l c r r r r r r r r r l r}
\hline
     Name      &  z    & FR & Class&L$_{178}$&       r  & L$_{\rm [NII],nuc}$ & L$_{\rm [NII],ext}$& L$_{\rm H\alpha,nuc}$ & L$_{\rm H\alpha,ext}$& L$_{\rm [OIII],nuc}$ & L$_{\rm [OIII],ext}$ &  Obs. date        & Seeing  & Sky    &   Depth\\
\hline                                                       
     3C~015    & 0.073 & I  & LEG  &  33.30 &      6.7 & 40.91 & 40.24    &40.23   &  38.16  &  40.51  &  38.16  &    Jun 30 2017   &  0.65   &  TN       &  5.6 \\ 
     3C~017    & 0.220 & II & BLO  &  34.44 &     11.6 & 41.95 & 41.25    &42.15   &  42.58  &  41.86  &  41.41  &    Jul~20 2017   &  0.49   &  TN/CL    &  5.9 \\  
     3C~018    & 0.188 & II & BLO  &  34.27 &     20.1 & 41.74 & 41.17    &41.61   &  41.17  &  42.14  &  41.46  &    Jun 30 2017   &  0.53   &  TN       &  5.9 \\ 
     3C~029    & 0.045 & I  & LEG  &  32.84 &   $<$1.4 & 40.39 & $<$39.91 &39.73   &  38.75  &  39.81  &  38.38  &    Jul~20 2017   &  0.51   &  TN/CL    &  3.2 \\ 
     3C~033    & 0.060 & II & HEG  &  33.65 &     10.7 & 41.04 & 41.02    &41.18   &  41.32  &  41.66  &  41.75  &    Jun 30 2017   &  0.63   &  TN       &  7.4 \\ 
     3C~040    & 0.018 & I  & LEG  &  32.29 &   $<$0.5 & 39.65 & $<$38.42 &39.38   &  38.22  &  39.0   &  37.13  &    Jul~22 2017   &  0.40   &  TN/CL    &  7.5 \\ 
     3C~063    & 0.175 & II & HEG  &  34.21 &     31.5 & 40.95 & 41.05    &41.31   &  41.48  &  41.41  &  40.98  &    Jul~21 2017   &  0.49   &  TN       &  7.5 \\ 
{\bf 3C~076.1} & 0.032 &  I & --   &  32.46 &      4.2 & 39.63 & 39.36    &38.97   &  37.46  &  39.84  &$<$38.22 &    Jan~05 2019   &  0.62   &  TN       &  1.9 \\   
{\bf 3C~078  } & 0.028 &  I & LEG  &  32.51 &   $<$0.9 & 40.34 & $<$39.84 &39.7    &$<$38.82 &  39.41  &$<$38.82 &    Nov~11 2018   &  0.53   &  TN       &  10.1 \\  
{\bf 3C~079  } & 0.256 & II & HEG  &  34.78 &     35.7 & 40.09 & 41.89    &42.49   &  42.59  &  42.93  &  42.91  &    Jan~05 2019   &  0.66   &  TN       &  4.8 \\  
{\bf 3C~088  } & 0.030 & II & LEG  &  32.49 &      3.7 & 40.38 & 39.75    &40.38   &  40.06  &  39.88  &  39.12  &    Nov~11 2018   &  0.59   &  TN/CL    &  1.7 \\  
{\bf 3C~089  } & 0.138 &  I & --   &  34.01 &      8.8 & 40.39 & 39.67    &39.47   &  38.62  &  40.52  &$<$39.76 &    Dec~29 2018   &  0.64   &  TN       &  3.1 \\  
{\bf 3C~098  } & 0.030 & II & HEG  &  32.99 &     16.3 & 40.20 & 40.03    &40.29   &  40.46  &  40.75  &  40.75  &    Nov~05 2018   &  0.66   &  TN       &  0.5 \\  
{\bf 3C~105  } & 0.089 & II & HEG  &  33.54 &   $<$3.6 & 40.85 & $<$40.15 &40.63   &  39.89  &  41.01  &  40.39  &    Dec~30 2018   &  0.71   &  TN/CL    &  8.4 \\  
{\bf 3C~135  } & 0.125 & II & HEG  &  33.84 &     43.7 & 41.38 & 40.78    &41.44   &  41.12  &  41.95  &  41.42  &    Dec~06 2018   &  0.52   &  TN       &  3.2 \\  
{\bf 3C~180  } & 0.220 & II & HEG  &  34.32 &     34.9 & 40.93 & 41.68    &41.82   &  42.14  &  42.35  &  42.66  &    Dec~03 2018   &  1.45   &  CL       &  3.6 \\  
{\bf 3C~196.1} & 0.198 & II & LEG  &  34.31 &     14.5 & 41.36 & 41.66    &41.22   &  41.6   &  41.3   &  41.05  &    Jan~13 2019   &  0.48   &  TN       &  1.1 \\  
{\bf 3C~198  } & 0.081 & II & SF   &  33.19 &     35.2 & 40.85 & 40.41    &41.38   &  41.14  &  41.07  &  41.32  &    Jan~14 2019   &  0.78   &  TN       &  1.1 \\  
{\bf 3C~227  } & 0.086 & II & BLO  &  33.74 &     49.6 & 41.15 & 41.27    &41.61   &  41.74  &  41.82  &  42.13  &    Jan~28 2019   &  0.91   &  TN/CL    &  1.1 \\  
{\bf 3C~264  } & 0.021 &  I & LEG  &  32.43 &   $<$1.1 & 40.01 & $<$39.28 &39.79   &  39.34  &  39.15  &  37.18  &    Jan~12 2019   &  0.85   &  TN       &  1.7 \\  
{\bf 3C~272.1} & 0.003 &  I & LEG  &  30.72 &      0.5 & 38.98 & 39.45    &38.78   &  39.44  &  38.45  &  39.27  &    Feb~03 2019   &  0.39   &  TN       &  17.5\\  
{\bf 3C~287.1} & 0.216 & II & BLO  &  34.04 &     27.0 & 41.47 & 41.94    &41.59   &  41.18  &  41.88  &  41.47  &    Jun~30 2018   &  0.65   &  TN       &  7.1 \\  
{\bf 3C~296  } & 0.024 &  I & LEG  &  32.22 &   $<$1.6 & 40.09 & $<$39.20 &39.67   &  38.44  &  39.58  &  38.37  &    Feb~11 2019   &  1.08   &  TN/CL    &  1.3 \\  
{\bf 3C~300  } & 0.270 & II & HEG  &  34.60 &     36.3 & 41.12 & 40.61    &41.87   &  42.37  &  42.3   &  42.78  &    Mar~11 2019   &  0.41   &  TN       &  4.1 \\
     3C~318.1  & 0.045 & -- &  --  &  32.72 &     18.1 & 39.91 & 40.71    &38.95   &  40.16  &  39.40  &$<$39.03 &    Jun~22 2017   &  1.38   &  CL       &  6.3 \\ 
     3C~327    & 0.105 & II & HEG  &  33.98 &     16.3 & 41.67 & 40.97    & 41.79  &  40.98  &  41.67  &  40.97  &    Jun~30 2017   &  0.70   &  TN       &  6.3 \\  
     3C~348    & 0.155 & I  & ELEG &  35.35 &     28.2 & 41.32 & 42.33    & 41.18  &  41.12  &  40.09  &  39.69  &    Jul~20 2017   &  1.76   &  CL       &  3.5 \\
     3C~353    & 0.030 & II & LEG  &  33.69 &     15.4 & 40.04 & 39.70    & 40.0   &  39.84  &  39.54  &  39.28  &    Jun~29 2017   &  1.30   &  CL       &  4.1 \\ 
     3C~386    & 0.017 & II &  --  &  32.18 &     10.1 & 39.51 & 40.26    & 39.22  &  40.25  &  40.22  &$<$38.23 &    Jun~03 2017   &  0.61   &  TN       &  7.3 \\
     3C~403    & 0.059 & II & HEG  &  33.16 &      8.7 & 40.82 & 41.69    & 40.87  &  40.83  &  41.33  &  41.71  &    Jun~30 2017   &  0.54   &  TN       &  5.5 \\   
     3C~403.1  & 0.055 & II & LEG  &  32.98 &      5.3 & 39.83 & 40.20    & 39.78  &  40.59  &  39.51  &  40.18  &    Jun~30 2017   &  0.80   &  TN       &  4.5 \\  
     3C~424    & 0.127 & II & LEG  &  33.78 &     29.0 & 40.70 & 41.45    & 40.75  &  41.67  &  40.27  &  41.0   &    Jul~01 2017   &  0.98   &  TN       &  1.2 \\  
     3C~442    & 0.026 & II & LEG  &  32.39 &      4.1 & 39.92 & 40.11    & 39.56  &  39.54  &  39.16  &  38.9   &    Jun~30 2017   &  0.61   &  TN       & 11.3 \\  
     3C~445    & 0.056 & II & BLO  &  33.26 &     15.7 & 42.20 & 41.12    & 42.5   &  42.02  &  42.2   &  41.87  &    Jul~01 2017   &  1.48   &  TN/CL    &  5.2 \\  
     3C~456    & 0.233 & II & HEG  &  34.23 &  $<14.3$ & 42.46 & $<$41.78 & 42.57  &  41.74  &  42.88  &  42.11  &    Jun~30 2017   &  1.27   &  TN       &  8.8 \\   
     3C~458    & 0.289 & II & HEG  &  34.58 &     81.2 & 40.96 & 40.84    & 41.12  &  41.67  &  41.64  &  42.09  &    Jul~22 2017   &  0.50   &  TN/CL    &  5.6 \\   
     3C~459    & 0.220 & II & BLO  &  34.55 &     59.6 & 42.39 & 41.87    & 42.15  &  41.83  &  41.83  &  42.02  &    Jul~22 2017   &  0.43   &  TN/CL    & 19.3 \\   
\hline
\end{tabular}

Column description: (1) Name: source name. The 17 newly observed sources are
shown in boldface; (2) z: redshift; (3 and 4) FR and excitation class
from \citet{buttiglione10,buttiglione11}. We used
their criterion for the HERG/LERG separation; some sources lack  a spectral
classification due to the non-detection of the key diagnostic emission
lines; (5) L$_{178}$: radio luminosity at 178 MHz in $\ergshz$, from
\citet{spinrad85}; (6) r: largest distance of emission line detection in
kpc units; (7 and 8) L$_{\rm [N~II],nuc}$ and L$_{\rm [N~II],ext}$: luminosity of the nuclear ELR
(within three times the seeing of the observations) and extended
components, both in $\ergs$ units; (9) Obs. date: date of the observation; (10) Seeing: mean seeing of the
observation; (11) Sky conditions: (TN) thin cirrus clouds, (CL) clear
night; (12) Depth: surface brightness limit of emission lines at 3$\sigma$ in
units of 10$^{-18}$ erg s$^{-1}$ cm$^{-2}$ arcsec$^{-2}$.
\label{tab1} 
\end{table*}
\end{landscape}

\section{Observations and data reduction}
\label{sample}

We observed a sample of 37 RGs with MUSE as part of the MURALES
survey. The sample is formed of all the 3C radio-sources limited to
$z<0.3$ and $\delta<20^\circ$. The information on the first subset of
20 objects was presented by \citet{balmaverde19}; in this paper we  complete the
analysis of the remaining 17 3C sources.  The main properties of the
galaxies are listed in Table \ref{tab1}. The full sample covers the
redshift range $ 0.003 < $z$ < 0.289$, with 21 sources located at
z$<0.1$. Their radio power spans more than four orders of magnitude,
from $\sim 10^{31}$ to $\sim 10^{35} \ergs$ at 178 MHz. Most of them
(26) are FR~II. RGs are also separated into spectroscopic classes
depending on the relative strength of the diagnostic optical emission
lines \citep{hine79,laing94};  we used the classification obtained by
\citet{buttiglione10,buttiglione11}. All classes (low excitation galaxies, LEGs;
high excitation galaxies, HEGs; and broad-line objects, BLOs) are
represented with an almost equal share of LEGs (including six
FR~II/LEGs) and HEGs or BLOs. We compare the redshift and radio power
distribution of our sample with that of the entire population of 114
3C RGs at z$<$0.3 presented by \citet{buttiglione09}. The mean
redshift and radio power are $z=0.09$ and $\log L_{178}=33.69$
$\ergs$, respectively, not dissimilar from the values measured for the
entire 3C subsample with $z<0.3$ ($z=0.13$ and $\log L_{178}= 33.61
\ergs$). The Kolmogorov--Smirnov test confirms that the two
distributions of $z$ and $L_{178}$ are not statistically
distinguishable. Our subsample can then be considered 
representative of the population of powerful, low redshift RGs.

\begin{figure}  
\centering{ 
  \includegraphics[width=0.9\columnwidth]{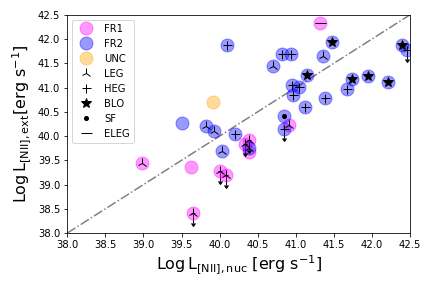}
  \includegraphics[width=0.85\columnwidth]{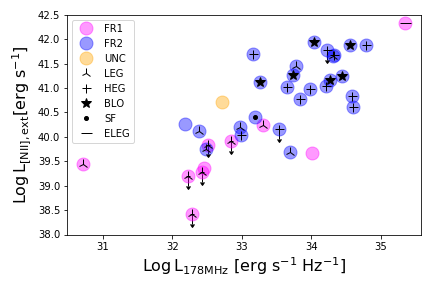}
  \includegraphics[width=0.95\columnwidth]{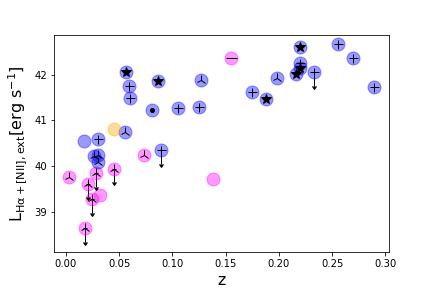}
  \includegraphics[width=0.9\columnwidth]{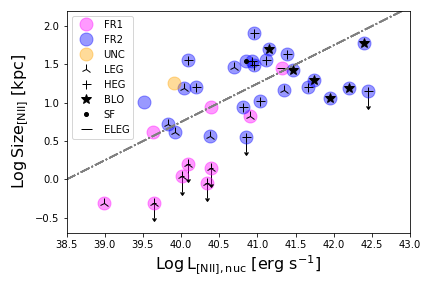}
 }
\caption{From top to bottom. First and second panels: Logarithm of the
  luminosities of the extended ELRs in the [N~II] line vs. nuclear
  line luminosity and radio luminosities (all quantities in
  $\ergs$). Sources are coded with symbols combining the optical
  spectroscopic classification and radio morphology. The dashed line
  is the bisectrix of the plane. Third panel: Luminosity of the
    H$\alpha$+[N~II] emission line vs. redshift. Fourth panel:
  Logarithm of the size of the extended ELR (in kpc) vs. nuclear
  line luminosity. The dashed line shows the $r \propto L_{\rm Ext}
  ^{0.5}$ relation.}
   \label{extended}
 
\end{figure}  

The observations presented here were obtained as part of the program
ID 0102.B-0048(A). Two exposures of 10 minutes each were obtained with
the VLT/MUSE spectrograph between November 5, 2018, and March 11,
2019 with the nominal wavelength range (4800 - 9300 \AA). The median
seeing of the observations is 0\farcs65, the same value of the first
set of observations. We split the total on-source exposure time  into two
sub-exposures, applying a 5\arcsec\ dithering pattern and a 90 degree
rotation between on-object exposures to reject cosmic rays.  We used
the ESO MUSE pipeline (version 1.6.2; \citealt{weilbacher20}) to
obtain fully reduced and calibrated data cubes. Summarizing, the
pipeline corrects for the bias, the flat, and the vignetting; removes
cosmic rays; calibrates each exposure in wavelength and in flux; and
subtracts the sky background. For the flux calibration of the image,
the pipeline uses a pre-processed spectrophotometric standard star
observation. A known issue of the MUSE calibration procedure is that
the astrometry in some cases is not precise. We used the position of
bright sources in the MUSE field observed by Pan-STARRs to correct for
any astrometric offset.

We followed the same strategy for the data analysis described in
\citet{balmaverde18}, producing flux, velocity, and velocity dispersion maps for all the emission lines in the spectrum. However, in
this paper, we focus on the
brightest and most spatially extended emission line (H$\alpha$, [N~II], or [O~III]).


\begin{figure*}  
\centering{ 
\includegraphics[width=2.0\columnwidth]{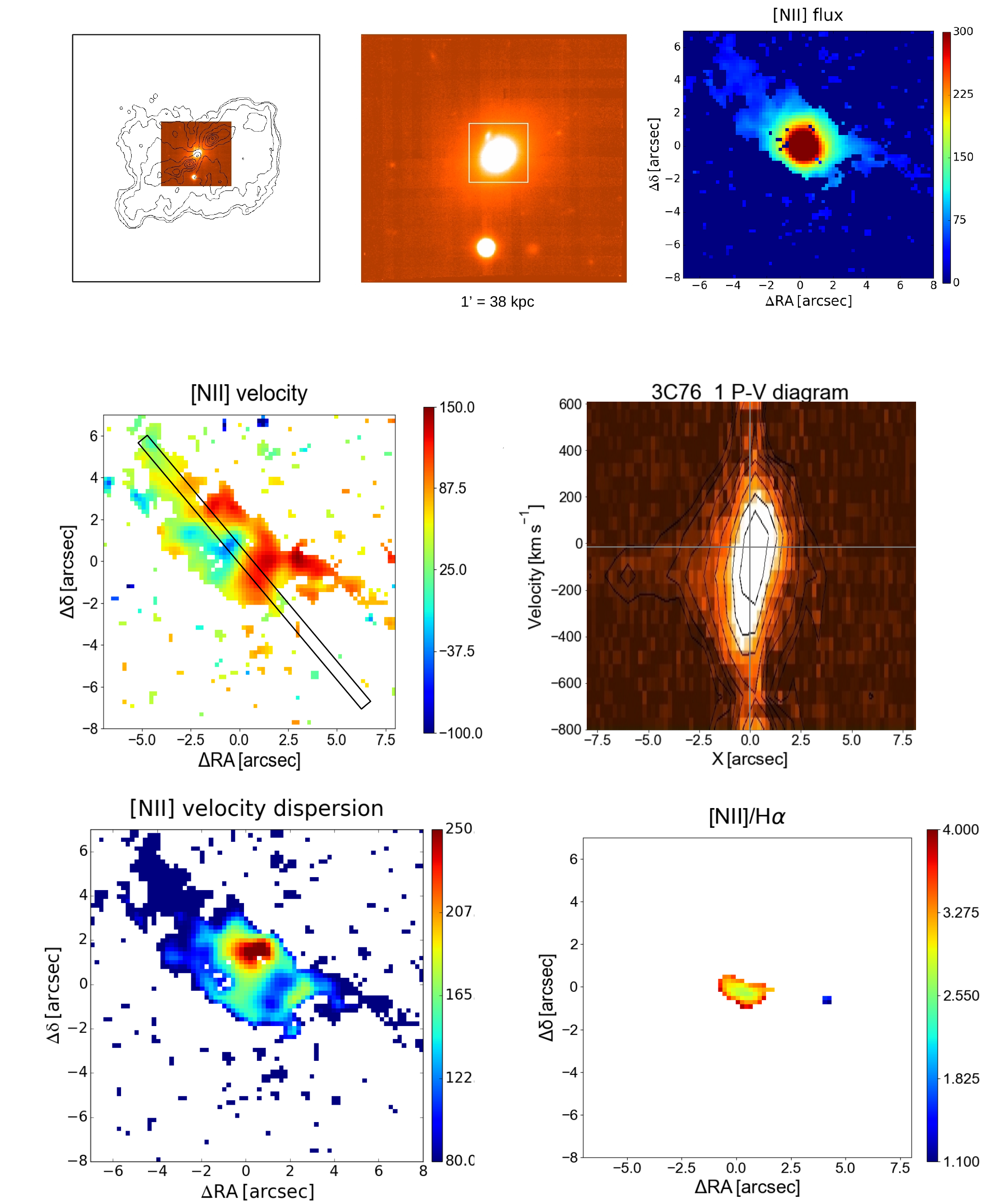}
\caption{3C~076.1: FR~I, 1$\arcsec$ = 0.64 kpc. Top left: Radio
  contours (black) overlaid onto the MUSE optical continuum image in
  the 5800--6250 \AA\ rest frame range. The size of the image is
  the whole MUSE field of view, 1$^\prime \times 1^\prime$. Top
  center: MUSE continuum image with superposed the region in which we
  explored the emission line properties (white square). Top right:
  [N~II] emission line image extracted from the white square  in
  the center panel. Surface brightness is  $10^{-18} {\rm
    erg}\,{\rm s}^{-1}\,{\rm cm}^{-2} {\rm arcsec}^{-2}$. Middle:
  Velocity field from the \nii\ line and position--velocity diagram
  extracted from the synthetic slit shown in the left panel (width
  1$\arcsec$, $PA=50$). Velocities are in units of \kms. Bottom:
  Velocity dispersion and \nii/\ha\ ratio map.} \label{3c076.1}}
\end{figure*}  

\begin{figure*}  

\centering{ 
\includegraphics[width=2.\columnwidth]{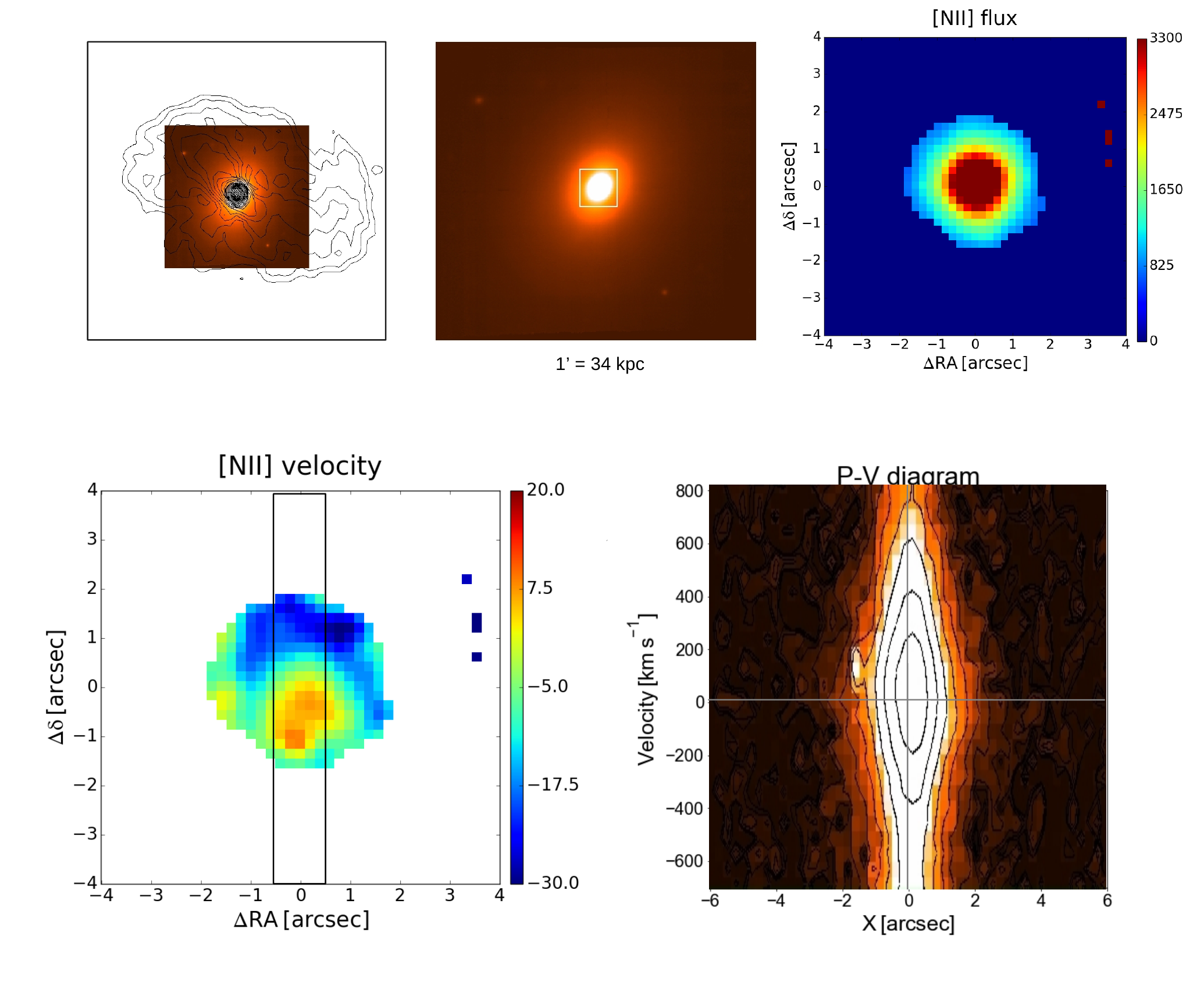}
\caption{3C~078: FR~I/LEG, 1$\arcsec$ = 0.56 kpc. Top left: Radio
  contours (black) overlaid onto the MUSE optical continuum image. The
  size of the image is the whole MUSE field of view, 1$^\prime \times
  1^\prime$. Top center: MUSE continuum image with superposed the
  region in which we explored the emission line properties (white
  square). Top right: [N~II] emission line image extracted from the
  white square  in the center panel. Surface brightness is 
  $10^{-18} {\rm erg}\,{\rm s}^{-1}\,{\rm cm}^{-2} {\rm
    arcsec}^{-2}$. Bottom panels: (left) Velocity field (in \kms) from
  the \nii\ line; (right) Position--velocity diagram extracted from the
  synthetic aperture shown in the left panel (width 0$\farcs$6,
  $PA=0^\circ$).} \label{3c078}}
\end{figure*}

\begin{figure*}  
\centering{ 
\includegraphics[width=2.\columnwidth]{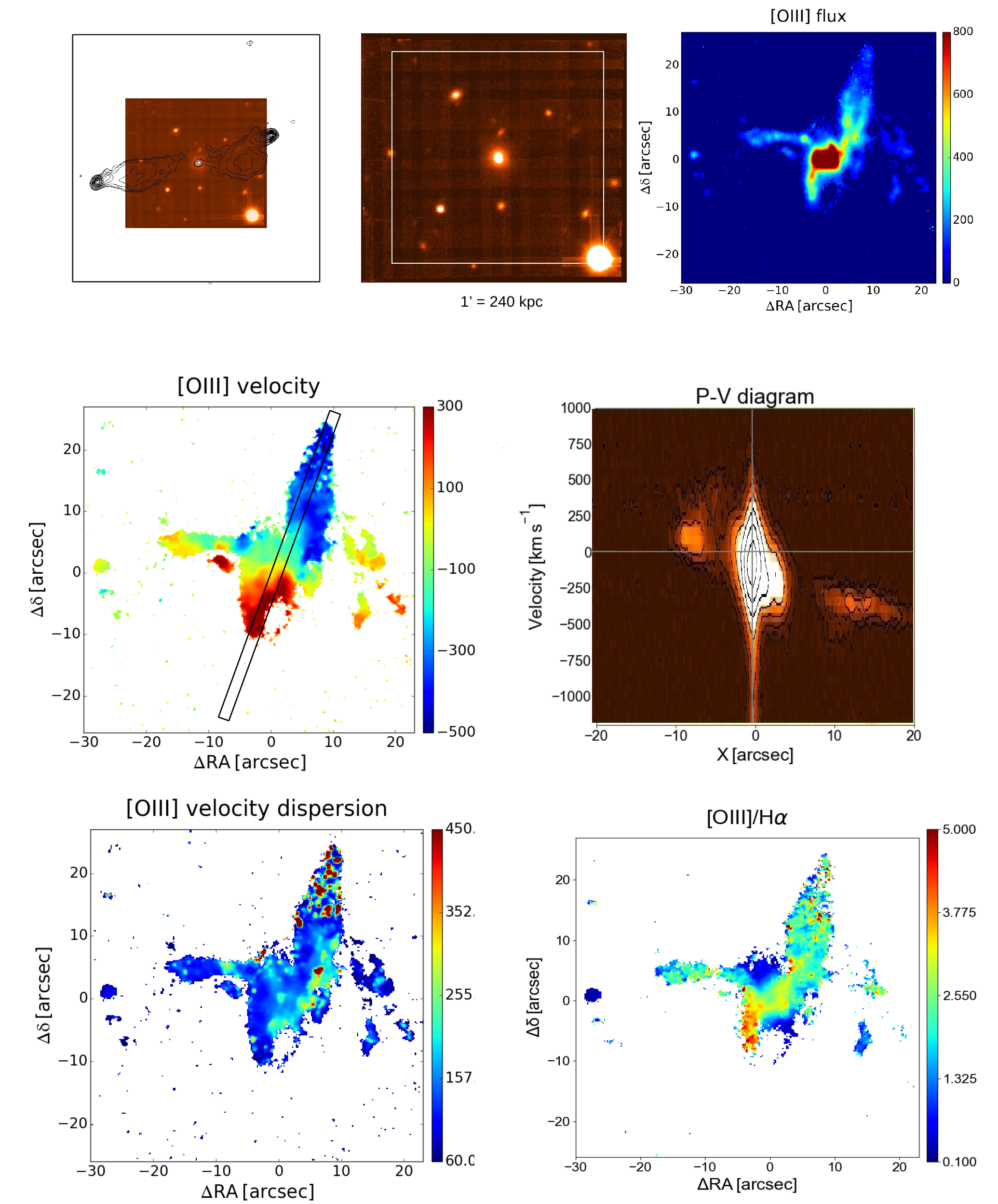}
\caption{3C~079: FR~II/HEG, 1$\arcsec$ = 4.0 kpc.Top left: Radio
  contours (black) overlaid onto the MUSE continuum image. The size of
  the image is the whole MUSE field of view, 1$^\prime \times
  1^\prime$. Top center: MUSE continuum image with superposed the
  region in which we explored the emission line properties (white
  square). Top right: \oiii\ emission line image extracted from the
  white square  in the center panel. Surface brightness is 
  $10^{-18} {\rm erg}\,{\rm s}^{-1}\,{\rm cm}^{-2} {\rm
    arcsec}^{-2}$. Central panels: (left) Velocity field (in \kms)
  from the \oiii\ line; (right) position--velocity diagram extracted
  from the synthetic aperture shown in the left panel. Bottom panels:
  (left) Velocity dispersion distribution (width 1$\arcsec$,
  $PA=-20^\circ$) and (right) \oiii/\ha\ map. }\label{3c079} }
\end{figure*}  

\begin{figure*}  
\centering{ 
\includegraphics[width=2.\columnwidth]{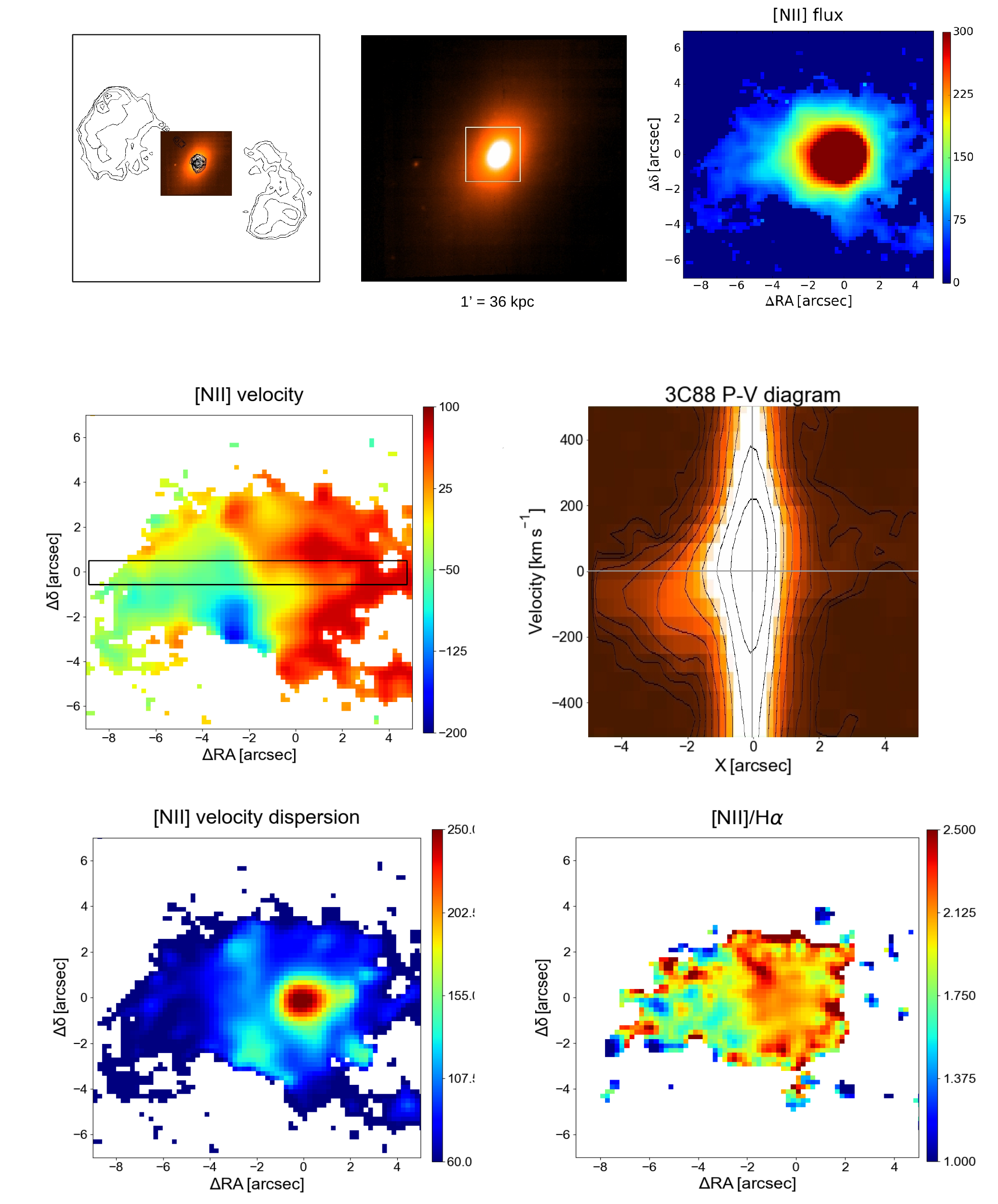}
\caption{3C~088: FR~II/LEG, 1$\arcsec$ = 0.60 kpc. Top left: Radio
  contours (black) overlaid onto the MUSE continuum. The size of the
  image is the whole MUSE field of view, 1$^\prime \times
  1^\prime$. Top center: MUSE continuum image with superposed the
  region in which we explored the emission line properties (white
  square). Top right: \oiii\ emission line image extracted from the
  white square   in the center panel. Surface brightness is 
  $10^{-18} {\rm erg}\,{\rm s}^{-1}\,{\rm cm}^{-2} {\rm arcsec}^{-2}$.
  Central panels: (left) Velocity field (in \kms) from the
  \oiii\ line; (right) position--velocity diagram extracted from the
  synthetic aperture shown in the left panel (width 0$\farcs$6,
  $PA=90^\circ$). Bottom panels: (left) Velocity dispersion
  distribution and (right) \oiii/\ha\ map.} \label{3c088}}
\end{figure*}  

\begin{figure*}  
\centering{ 
\includegraphics[width=2.\columnwidth]{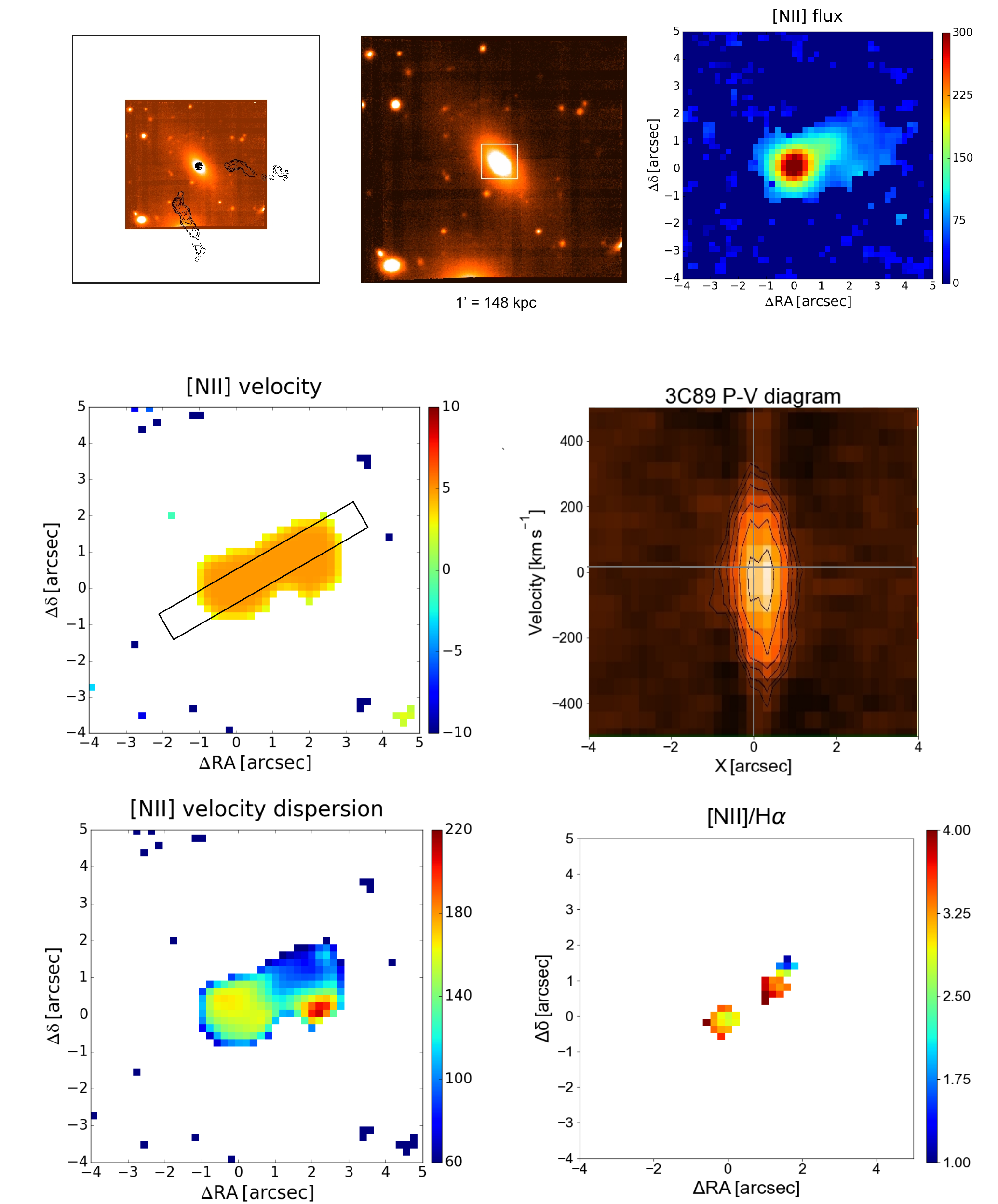}
\caption{3C~089: FR~I, 1$\arcsec$ = 2.46 kpc. Top left: Radio contours
  (black) overlaid onto the MUSE continuum image. The size of the
  image is the whole MUSE field of view, 1$^\prime \times
  1^\prime$. Top center: MUSE continuum image with superposed the
  region in which we explored the emission line properties (white
  square). Top right: [N~II] emission line image extracted from the
  white square  in the central panel. Surface brightness is 
  $10^{-18} {\rm erg}\,{\rm s}^{-1}\,{\rm cm}^{-2} {\rm arcsec}^{-2}$.
  Central panels: (left) Velocity field (in \kms) from the \nii\ line;
  (right) position--velocity diagram extracted from the synthetic
  aperture shown in the left panel (width 1$\arcsec$,
  $PA=30^\circ$). Bottom panels: (left) Velocity dispersion
  distribution and (right) \nii/\ha\ map. } \label{3c089}}
\end{figure*}  

\begin{figure*}  
\centering{ 
\includegraphics[width=2.\columnwidth]{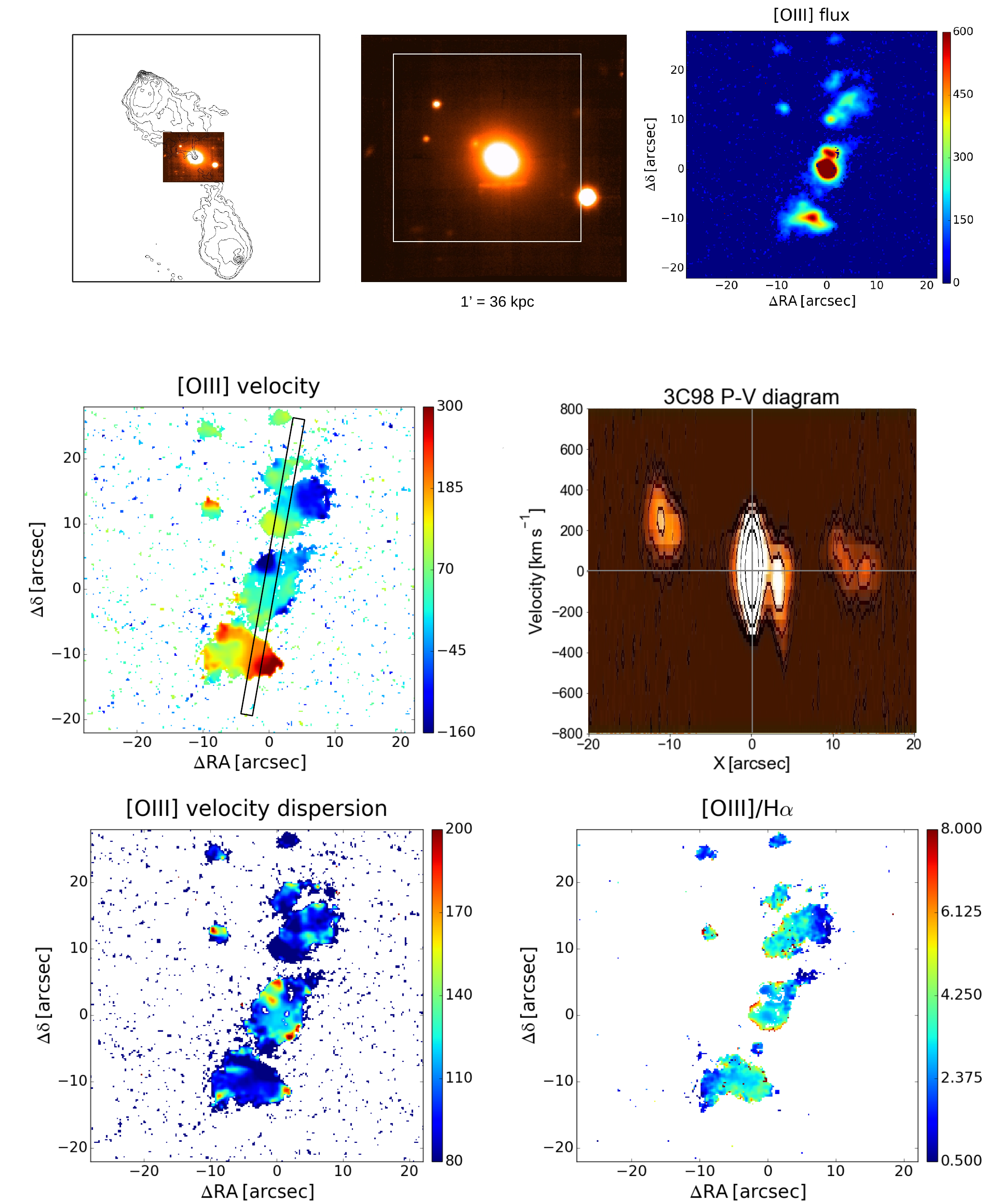}
\caption{3C~098: FR~II/HEG, 1$\arcsec$ = 0.60 kpc.Top left: Radio
  contours (black) overlaid onto the MUSE continuum image. The size of
  the image is the whole MUSE field of view, 1$^\prime \times
  1^\prime$. Top center: MUSE continuum image with superposed the
  region in which we explored the emission line properties (white
  square). Top right: \oiii\ emission line image extracted from the
  white square  in the central panel. Surface brightness  is  
  $10^{-18} {\rm erg}\,{\rm s}^{-1}\,{\rm cm}^{-2} {\rm arcsec}^{-2}$.
  Central panels: (left) Velocity field (in \kms) from the
  \oiii\ line; (right) position--velocity   diagram extracted from the
  synthetic aperture shown  in the left
  panel (width 1$\arcsec$, $PA=-10^\circ$). Bottom panels: (left) Velocity dispersion distribution and
  (right) \oiii/\ha\ map. } 
\label{3c098}}
\end{figure*}  

\begin{figure*}  
\centering{ 
\includegraphics[width=2.\columnwidth]{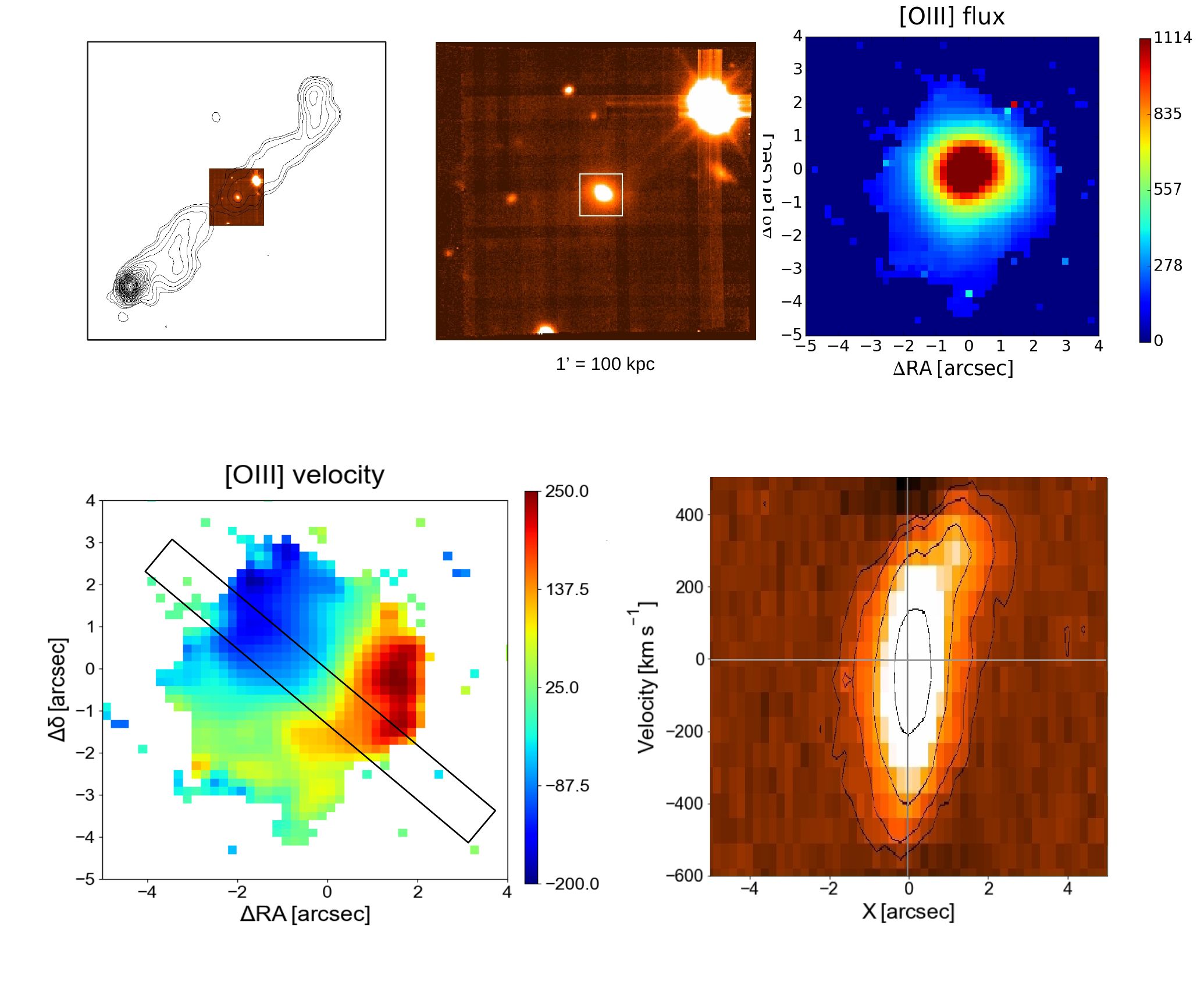}
\caption{3C~105: FR~II/HEG, 1$\arcsec$ = 1.67 kpc. Top left: Radio
  contours (black) overlaid onto the MUSE continuum image. The size of
  the image is the whole MUSE field of view, 1$^\prime \times
  1^\prime$. Top center: MUSE continuum image with superposed the
  region in which we explored the emission line properties (white
  square). Top right: \oiii\ emission line image extracted from the
  white square  in the central panel. Surface brightness is 
  $10^{-18} {\rm erg}\,{\rm s}^{-1}\,{\rm cm}^{-2} {\rm arcsec}^{-2}$.
  Bottom panels: (left) Velocity field (in \kms) from the \nii\ line;
  (right) position--velocity diagram extracted from the synthetic
  aperture shown  in the left panel (width 1$\arcsec$, $PA=50^\circ$). }
\label{3c105}}
\end{figure*}  

\section{The case of 3C~258}
\label{3c258}

The radio source 3C~258 was identified as a galaxy at redshift
$z=0.165$ \citep{wyndham66}. This value was challenged by
\citet{dey94} who noted the likely presence of a background quasar,
but, due to poor seeing of the observations, no firm claims can be
made. The NICMOS/HST images by \citet{floyd08} are consistent with the
higher redshift as they show a very compact host galaxy, much fainter
than   expected for an object at z=0.165.  The superposition of
the radio image from \citet{neff95} on the MUSE continuum image
indicates that what had been identified as the optical counter part is
actually offset by 2\farcs8 at $PA = -20^\circ$ (see
Fig. \ref{3C258}). A point-like object is seen at the location of the
radio source, also in the optical HST \citep{dekoff96} and Chandra
images \citep{massaro12}. The MUSE spectrum of this object shows a
single emission line, with a broad profile, at $\lambda = 7116
\AA$. The most likely identification of this line is Mg II at
$\lambda$ 2798 \AA, confirming the \citeauthor{dey94} result and
suggesting a tentative association with a type I AGN at $z\sim 1.54$:
at this redshift the MUSE spectrum covers the rest frame range $\sim
1900 - 3600 $\AA, a spectral region devoid of strong emission
lines. We will not  consider this object further in our analysis.

\section{Luminosities and sizes of the ELR}
\label{elr}

In Table \ref{tab1} we list the measurements of sizes and luminosities
of the  ELR measured in the 37 sources of the sample. We consider
separately the central (i.e., within a radius of 3 times the seeing of
the observations) and the extended component. The size of the ELR is
measured as the largest radius at which emission lines are detected
above the 3$\sigma$ level. ELRs are considered to be not resolved when
they do not extend beyond a radius of 3 times the seeing: this occurs
in five FR~Is and in two FR~IIs.

The top panel of  Fig. \ref{extended} compares the luminosities
of the nucleus and of the extended ELR: they are linked by a strong
correlation with a slope consistent with unity and a median ratio of
$L_{\rm Nuc}/L_{\rm Ext} \sim 2$.  We estimated the Pearson
  partial and semi-partial correlation coefficient, finding that the
  correlation is significant ($>$ 95\%); we obtained a null
  probability value of 0.037 (partial correlation probability) and
  0.024 (semi-partial probability value) after removing the common
  dependence of luminosities on redshift. We verified that a
  similarly tight correlation is also present considering the nuclear
  and extended line fluxes.  An apparently similar connection is
found between the radio power and the extended line luminosity (see
the second panel), reminiscent of the correlation between the
extended and nuclear ELR luminosities just discussed and the well-known link between the line and radio luminosities in extended RGs
\citep{baum89a,baum89b,rawlings89,rawlings91,buttiglione10}. 
  However, by applying the statistical tests just described, we found
  that this relation is mainly driven by the common dependence 
  on redshift, and it is not significant. In the third panel of
  Fig. \ref{extended} we explore the dependence on redshift of the
  extended luminosity H$\alpha$+[N~II]. The expected positive trend (because going to higher redshifts we select more powerful sources
  able to ionize much more gas) is indeed observed. The bottom panel
  shows a positive trend, albeit with a large spread, between
  the size and the luminosity of the ELRs. The distribution of the
  sources in this plane could be consistent with the behavior
  expected from photoionization of a central source (i.e., $r \propto
  L_{\rm Ext} ^{0.5}$), as already found in Seyfert galaxies and QSOs
  (e.g., \citealt{bennert02,husemann14,sun18,chen19}).

 Overall, the ten FR~Is are (with the exception of 3C~348, or
  Hercules A, a source with a hybrid FRI/FRII morphology) associated
with the less luminous (and smaller) ELRs. When compared to the
FR~IIs, they show a large deficit in  luminosity (a factor
$\gtrsim$ 10) and in size (a factor $\gtrsim$ 3) of the extended ELR
at given nuclear luminosity.

Conversely, there is no clear separation in terms of size and
luminosity of the extended ELRs between the various spectroscopic
classes among the FR~IIs. Those of the BLO subclass have the more
luminous central ELR, but the size and luminosity of their 
EELRs are similar to those seen in the FR~IIs/HEGs and LEGs. This is
likely  due to the effects of the circumnuclear selective
obscuration on the NLR \citep{baldi13,baum10}. The comparison
of the HEG/BLO\footnote{These two classes, although they differ for the
  presence or lack of broad emission lines, have similar high ionization
  properties to those of  their NLR} and LEG classes is limited
by the small number of the LEG/FR~IIs (six objects) and we cannot
perform a robust statistical analysis. Nonetheless, HEG-BLO and
LEG/FR~II cover the same ranges in power and size of the extended
ELRs. In particular, we find two LEG/FR~IIs (3C~196.1 and 3C~424), in
which the ELR extends for $\sim$ 15- 20 kpc, with a luminosity of
$\sim 3 \times 10^{41} \ergs$, values similar to the brightest and
most extended ELRs in HEGs. In addition, the only two FR~IIs where no
extended line emission is detected (3C~105 and 3C~456) belong to the
HEG class.

\section{Results for the individual sources}
\label{results}

\begin{figure*}  
\centering{ \includegraphics[width=2.\columnwidth]{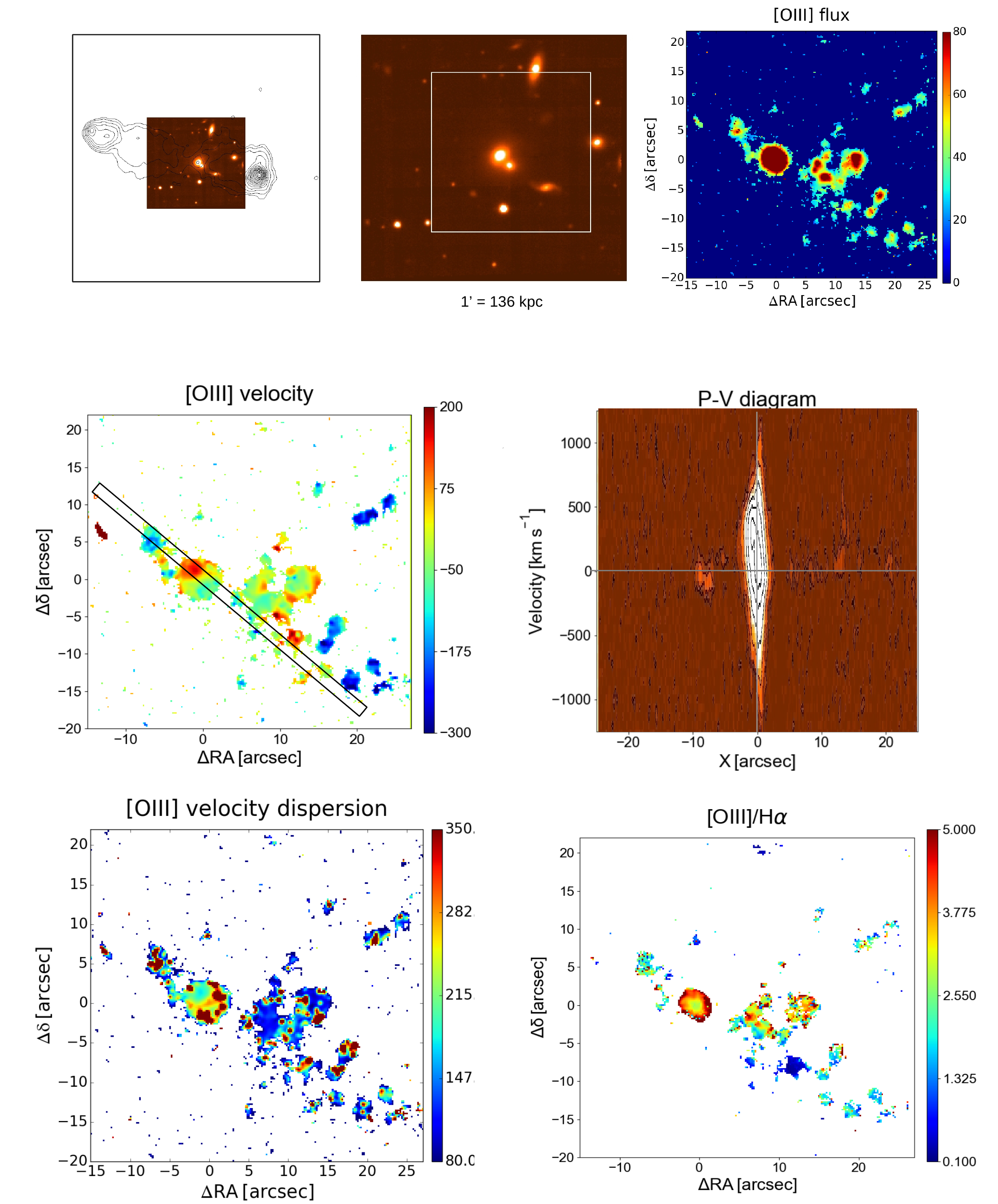}
\caption{3C~135: FR~II/HEG: 1$\arcsec$ = 2.26 kpc.Top left: radio
  contours (black) overlaid onto the MUSE continuum image. The size of
  the image is the whole MUSE field of view, 1$^\prime \times
  1^\prime$. Top center: MUSE continuum image with superposed the
  region in which we explored the emission line properties (white
  square). Top right: \oiii\ emission line image extracted from the
  white square  in the central panel. Surface brightness is 
  $10^{-18} {\rm erg}\,{\rm s}^{-1}\,{\rm cm}^{-2} {\rm
    arcsec}^{-2}$. Central panels: (left) velocity field (in \kms)
  from the \oiii\ line; (right) position--velocity diagram extracted
  from the synthetic aperture shown in the left panel(width
  1$\arcsec$, $PA=50^\circ$). Bottom panels: (left) velocity
  dispersion distribution and (right) \oiii/\ha\ map.  }
\label{3c135}}
\end{figure*}

\begin{figure*}  
\centering{ 
\includegraphics[width=2.\columnwidth]{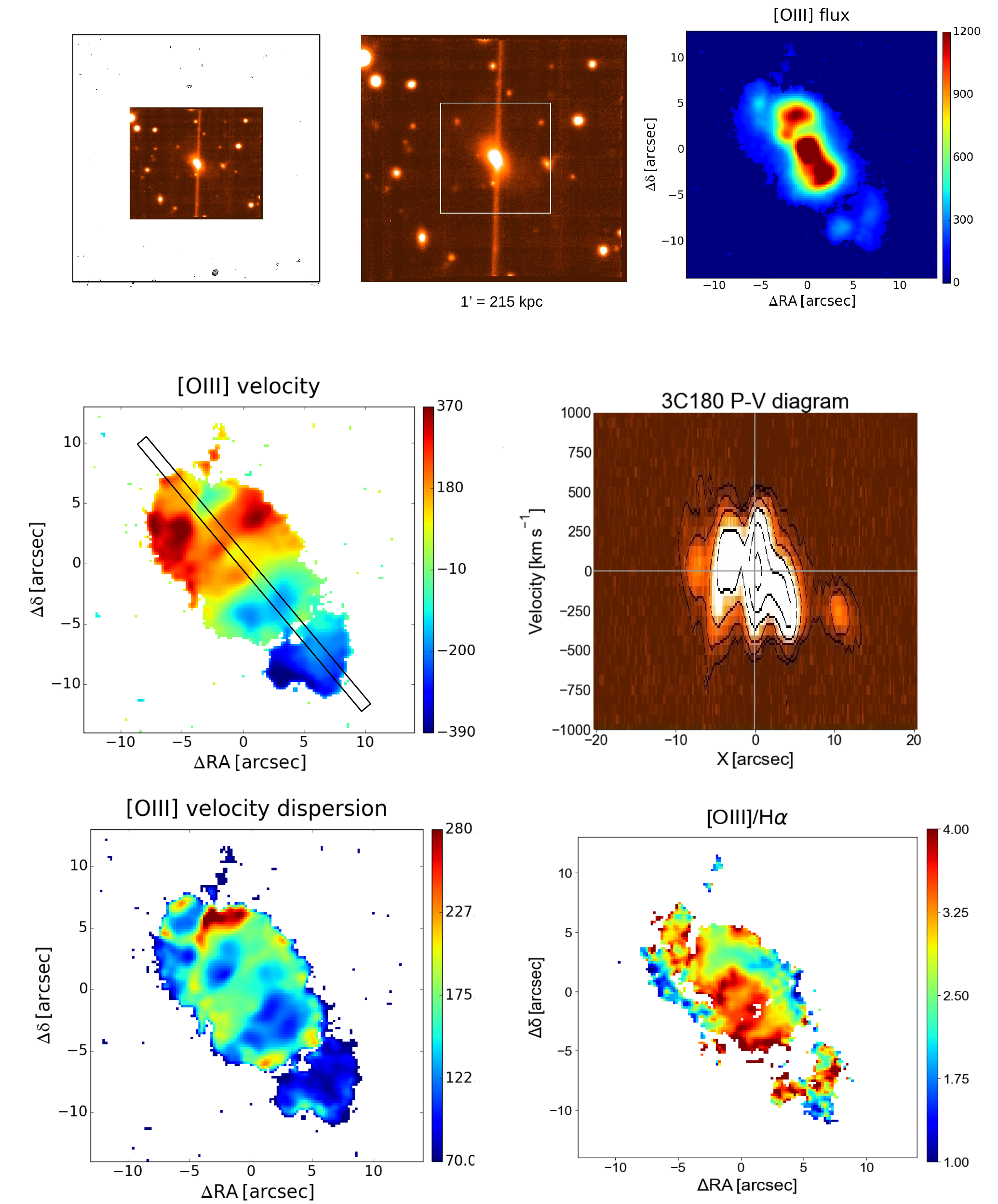}
\caption{3C~180: FR~II/HEG, 1$\arcsec$ = 3.58 kpc.Top left: radio
  contours (black) overlaid onto the MUSE continuum image. The size of
  the image is the whole MUSE field of view, 1$^\prime \times
  1^\prime$. Top center: MUSE continuum image with superposed the
  region in which we explored the emission line properties (white
  square). Top right: \oiii\ emission line image extracted from the
  white square  in the central panel. Surface brightness is 
  $10^{-18} {\rm erg}\,{\rm s}^{-1}\,{\rm cm}^{-2} {\rm
    arcsec}^{-2}$. Central panels: (left) velocity field (in \kms)
  from the \oiii\ line; (right) position--velocity diagram extracted
  from the synthetic aperture shown in the left panel (width
  1$\arcsec$, $PA=40^\circ$) . Bottom panels: (left) velocity
  dispersion distribution and (right) \oiii/\ha\ map.  }
\label{3c180}}
\end{figure*}

\begin{figure*}  
\centering{ 
\includegraphics[width=1.95\columnwidth]{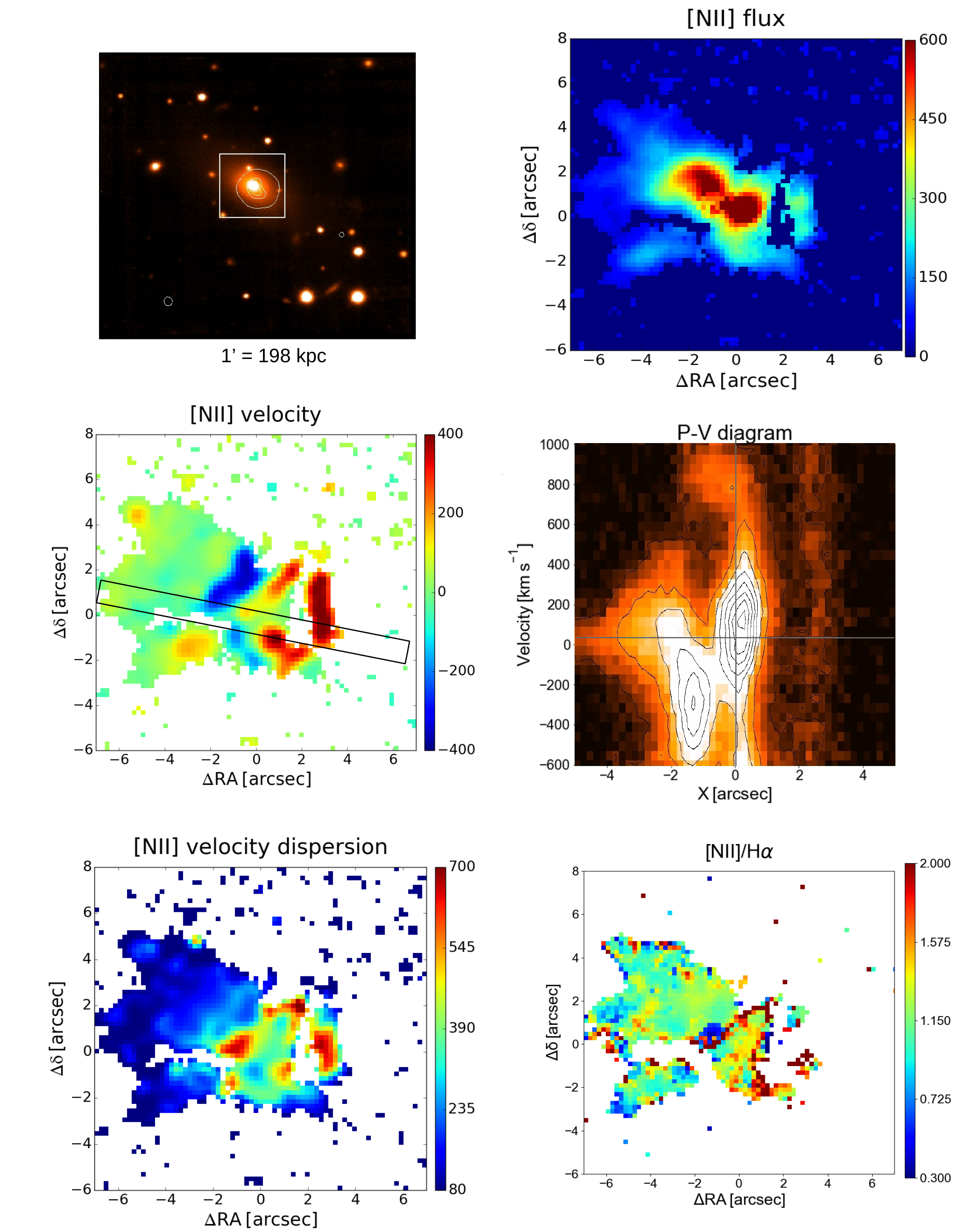}
\caption{3C~196.1: FR~II/LEG, 1$\arcsec$ = 3.30 kpc. Top left: radio
  contours (black) overlaid onto the MUSE continuum image. The size of
  the image is the whole MUSE field of view, 1$^\prime \times
  1^\prime$. The white square marks the region in which we explored
  the emission line properties. Top right: [N~II] emission line image
  extracted from the white square  in the central panel. Surface
  brightness is  $10^{-18} {\rm erg}\,{\rm s}^{-1}\,{\rm cm}^{-2}
  {\rm arcsec}^{-2}$.  Central panels: (left) velocity field (in \kms)
  from the \nii\ line; (right) position--velocity diagram extracted
  from the synthetic aperture shown in the left panel (width
  1\arcsec, $PA=30^\circ$). Bottom panels: (left) velocity
  dispersion distribution and (right) \nii/\ha\ map.  }
\label{3c196.1}}
\end{figure*}

\begin{figure*}  
\centering{ 
\includegraphics[width=2.\columnwidth]{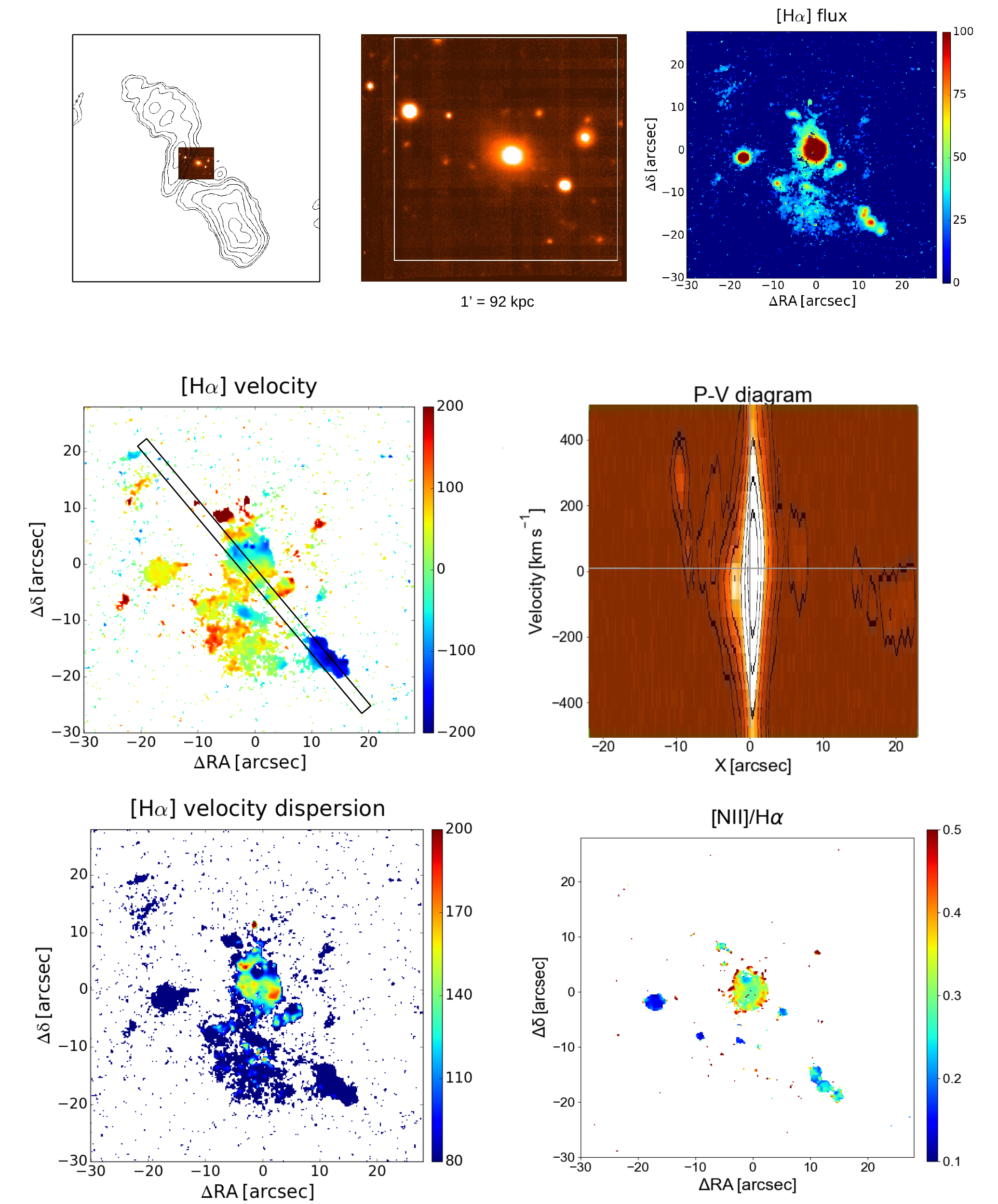}
\caption{3C~198: FR~II/SF, 1$\arcsec$ = 1.54 kpc. Top left: Radio
  contours (black) overlaid onto the MUSE continuum image. The size of
  the image is the whole MUSE field of view, 1$^\prime \times
  1^\prime$. Top center: MUSE continuum image with superposed the
  region in which we explored the emission line properties (white
  square). Top right: \ha\ emission line image extracted from the
  white square  in the central panel. Surface brightness is 
  $10^{-18} {\rm erg}\,{\rm s}^{-1}\,{\rm cm}^{-2} {\rm arcsec}^{-2}$.
  Central panels: (left) Velocity field (in \kms) from the \ha\ line;
  (right) position--velocity diagram extracted from the synthetic
  aperture shown in the left panel (width 2$\arcsec$,
  $PA=30^\circ$). Bottom panels: (left) Velocity dispersion
  distribution and (right) \nii/\ha\ map.  }
\label{3c198}}
\end{figure*}

\begin{figure*}  
\centering{ 
\includegraphics[width=2.\columnwidth]{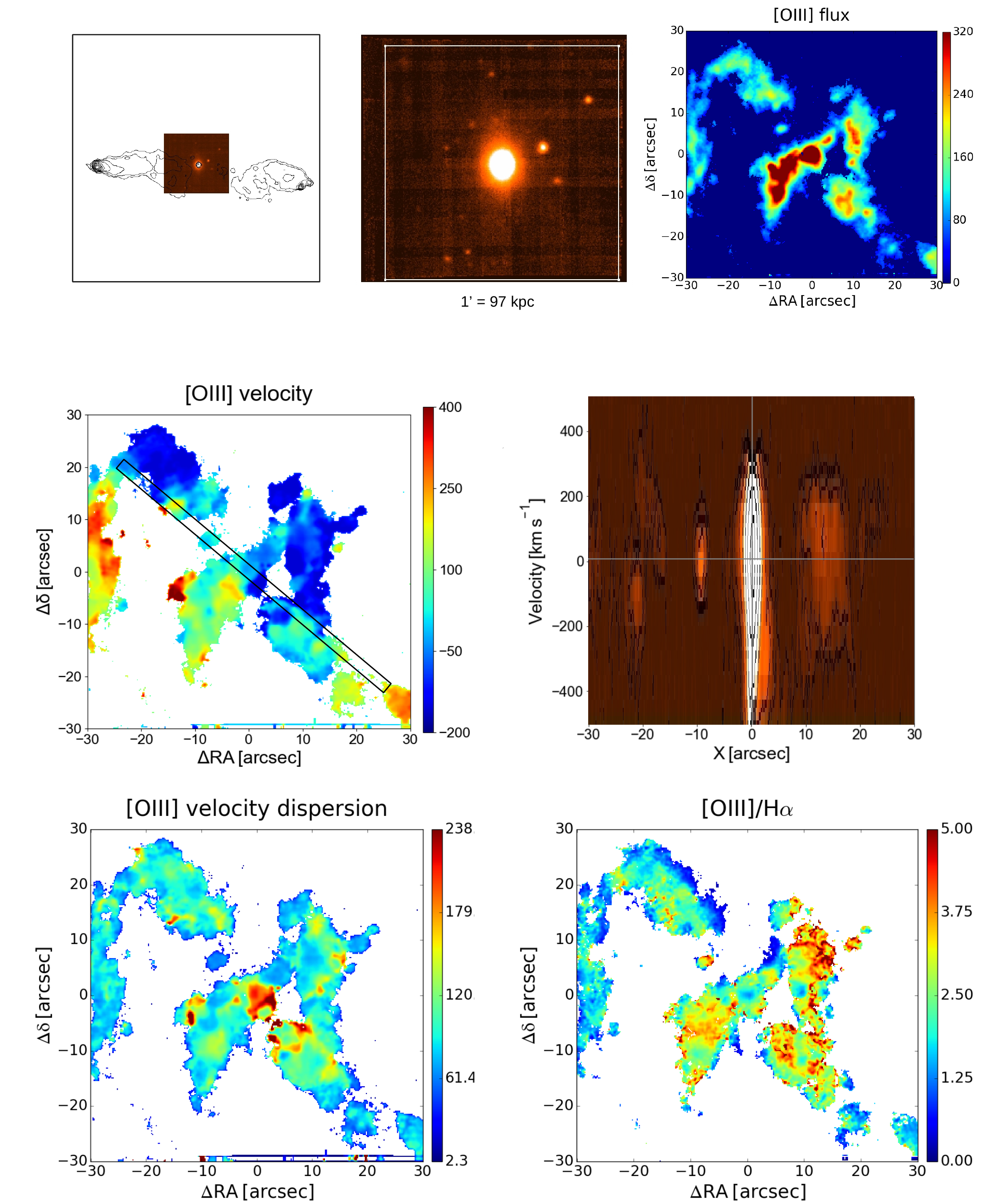}
\caption{3C~227: FR~II/BLO, 1$\arcsec$ = 1.62 kpc. Top left: Radio
  contours (black) overlaid onto the MUSE continuum image. The size of
  the image is the whole MUSE field of view, 1$^\prime \times
  1^\prime$. Top center: MUSE continuum image with superposed the
  region in which we explored the emission line properties (white
  square). Top right: \oiii\ emission line image extracted from the
  white square  in the central panel. Surface brightness is 
  $10^{-18} {\rm erg}\,{\rm s}^{-1}\,{\rm cm}^{-2} {\rm
    arcsec}^{-2}$. Central panels: (left) Velocity field (in \kms)
  from the \oiii\ line; (right) position--velocity diagram extracted
  from the synthetic aperture shown in the left panel (width
  1$\arcsec$, $PA=50^\circ$). Bottom panels: (left) Velocity
  dispersion distribution and (right) \oiii/\ha\ map.  }
\label{3c227}}
\end{figure*}

\begin{figure*}  
\centering{ 
\includegraphics[width=2.\columnwidth]{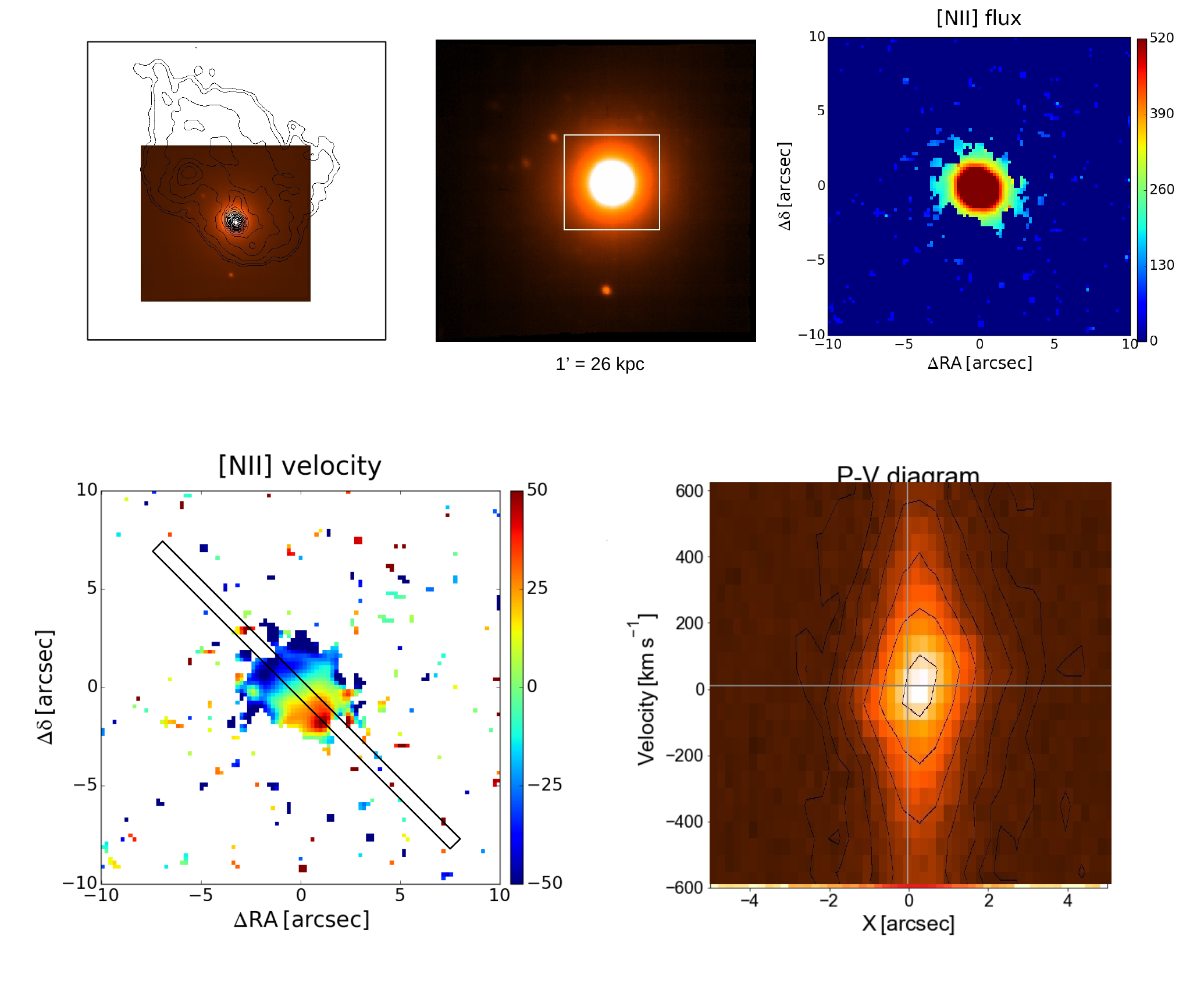}
\caption{3C~264: FR~I/LEG, 1$\arcsec$ = 0.43 kpc. Top left: Radio
  contours (black) overlaid onto the MUSE continuum image. The size of
  the image is the whole MUSE field of view, 1$^\prime \times
  1^\prime$. Top center: MUSE continuum image with superposed the
  region in which we explored the emission line properties (white
  square). Top right: [N~II] emission line image extracted from the
  white square  in the central panel. Surface brightness is 
  $10^{-18} {\rm erg}\,{\rm s}^{-1}\,{\rm cm}^{-2} {\rm arcsec}^{-2}$.
  Bottom panels: (left) Velocity field (in \kms) from the \nii\ line;
  (right) position--velocity diagram extracted from the synthetic
  aperture shown in the left panel (width 0$\farcs$6, $PA=45^\circ$).
}
\label{3c264}}
\end{figure*}

\begin{figure*}  
\centering{ 
\includegraphics[width=2.\columnwidth]{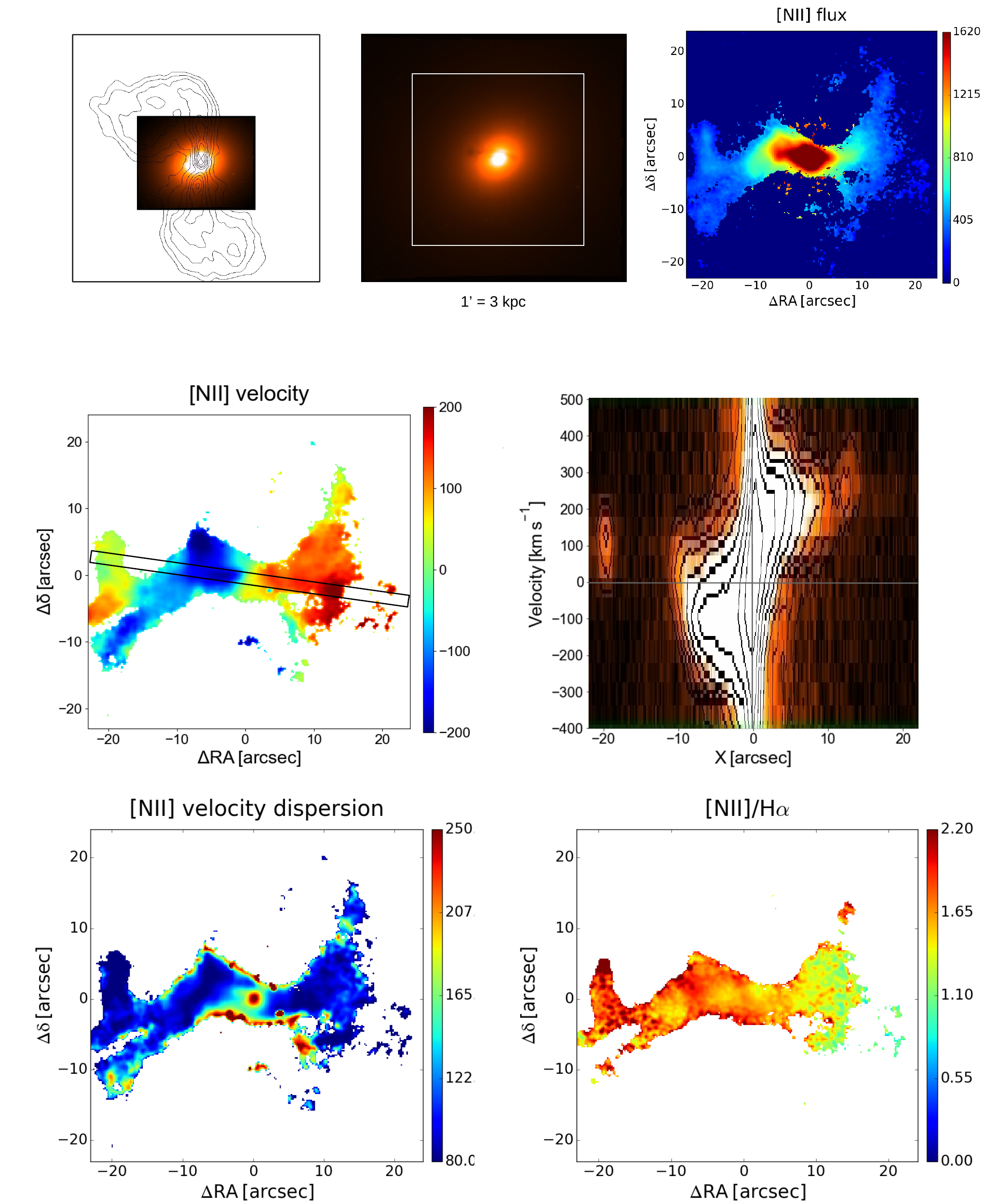}
\caption{3C~272.1: FR~I/LEG, 1$\arcsec$ = 0.06 kpc. Top left: Radio contours (black) overlaid onto the MUSE 
continuum image. The size
of the image is the whole MUSE field of view, 1$^\prime \times
1^\prime$. Top center: MUSE continuum image with superposed the region
in which we explored the emission line properties (white square). Top
right: [N~II] emission line image extracted from the white square
 in the central panel. Surface brightness is  $10^{-18} {\rm
erg}\,{\rm s}^{-1}\,{\rm cm}^{-2} {\rm arcsec}^{-2}$.  Central panels: (left) Velocity field (in \kms) from
the \nii\ line; (right) position--velocity diagram extracted from the
synthetic aperture shown  in the left
panel (width 1$\arcsec$, $PA=80^\circ$). Bottom panels: (left) Velocity dispersion distribution and
(right) \nii/\ha\ map.}
\label{3c272.1}}
\end{figure*}

\begin{figure*}  
\centering{ 
\includegraphics[width=2.\columnwidth]{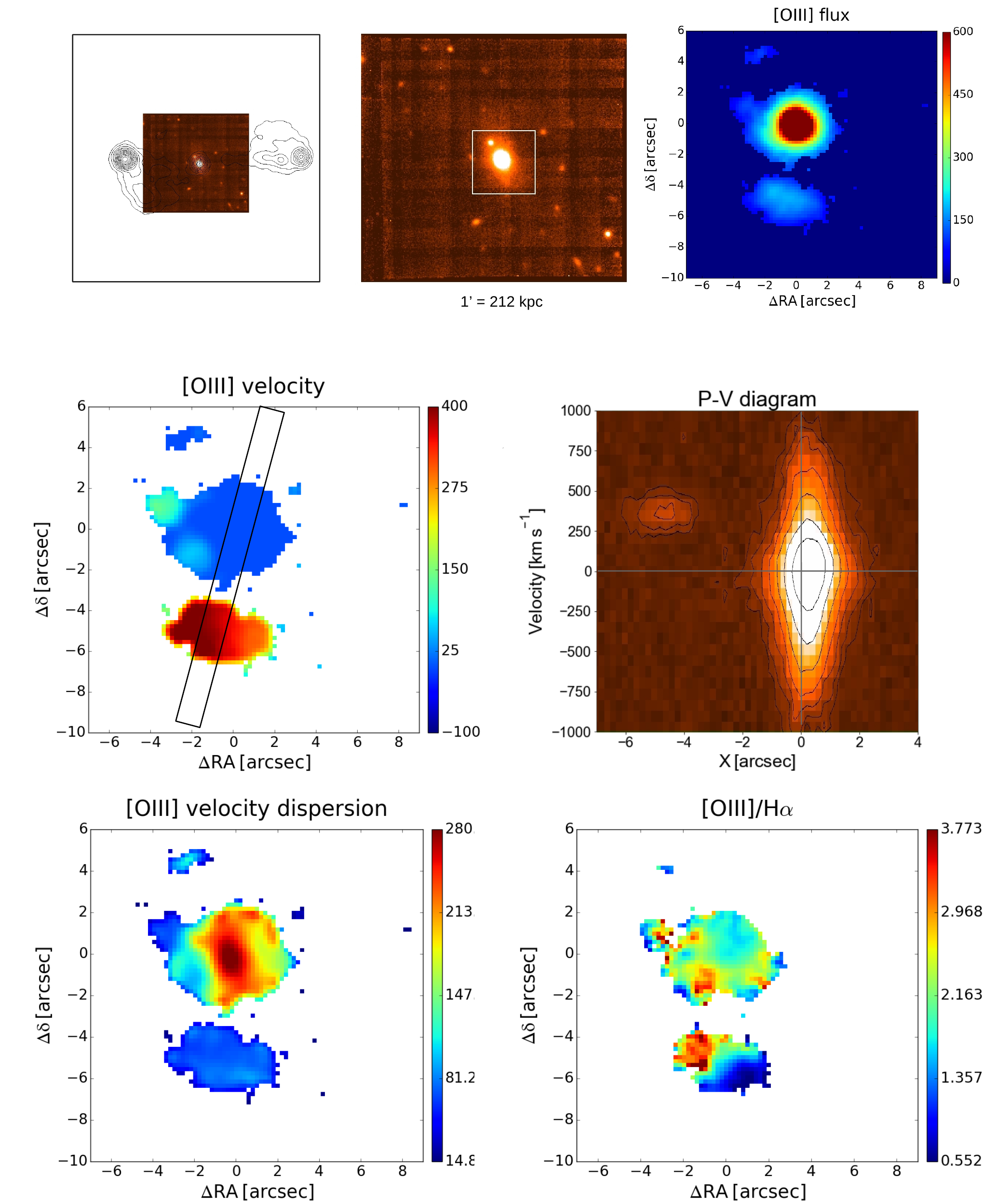}
\caption{3C~287.1: FR~II/BLO, 1$\arcsec$ = 3.53 kpc. Top left: Radio contours (black) overlaid onto the MUSE 
continuum image. The size
of the image is the whole MUSE field of view, 1$^\prime \times
1^\prime$. Top center: MUSE continuum image with superposed the region
in which we explored the emission line properties (white square). Top
right: \oiii\ emission line image extracted from the white square
 in the central panel. Surface brightness is  $10^{-18} {\rm
erg}\,{\rm s}^{-1}\,{\rm cm}^{-2} {\rm arcsec}^{-2}$. Central panels: (left) Velocity field (in \kms) from
the \oiii\ line; (right) position--velocity diagram extracted from the
synthetic aperture shown  in the left
panel (width 1$\arcsec$, $PA=-20^\circ$). Bottom panels: (left) Velocity dispersion distribution and
(right) \oiii/\ha\ map.}
\label{3c287.1}}
\end{figure*}

\begin{figure*}  
\centering{ 
\includegraphics[width=2.\columnwidth]{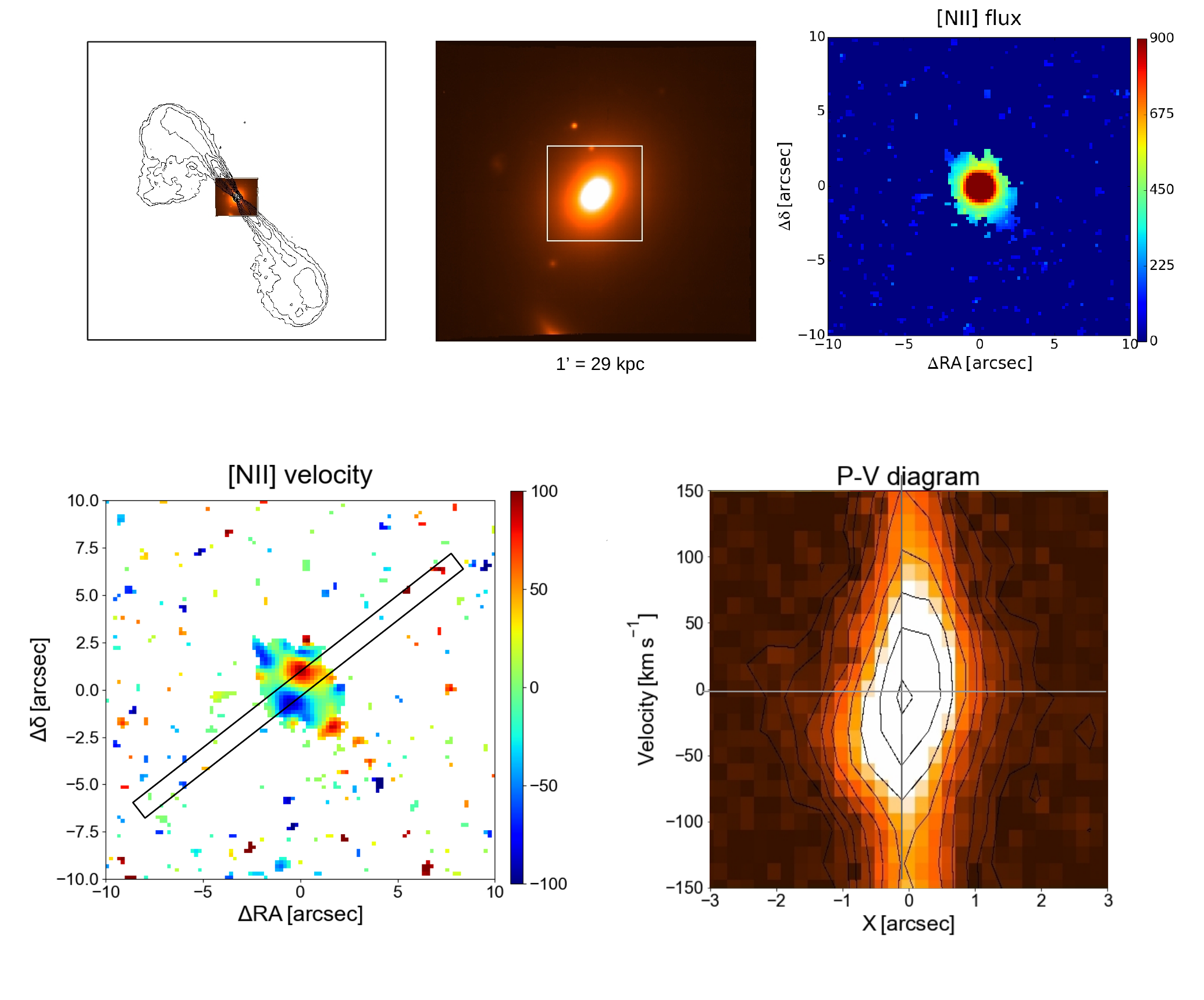}
\caption{3C~296: FR~I/LEG, 1$\arcsec$ = 0.49 kpc. Top left: Radio
  contours (black) overlaid onto the MUSE continuum image. The size of the image is
  the whole MUSE field of view, 1$^\prime \times 1^\prime$. Top
  center: MUSE continuum image with superposed the region in which we
  explored the emission line properties (white square). Top right:
  [N~II] emission line image extracted from the white square  in
  the central panel. Surface brightness is  $10^{-18} {\rm
    erg}\,{\rm s}^{-1}\,{\rm cm}^{-2} {\rm arcsec}^{-2}$.  Bottom panels: (left) Velocity field (in \kms) from
  the \nii\ line; (right) position--velocity diagram extracted from the
  synthetic aperture shown  in the left panel (width 1$\arcsec$, $PA=-50^\circ$).   }
\label{3c296}}
\end{figure*}

\begin{figure*}  
\centering{ 
\includegraphics[width=2.\columnwidth]{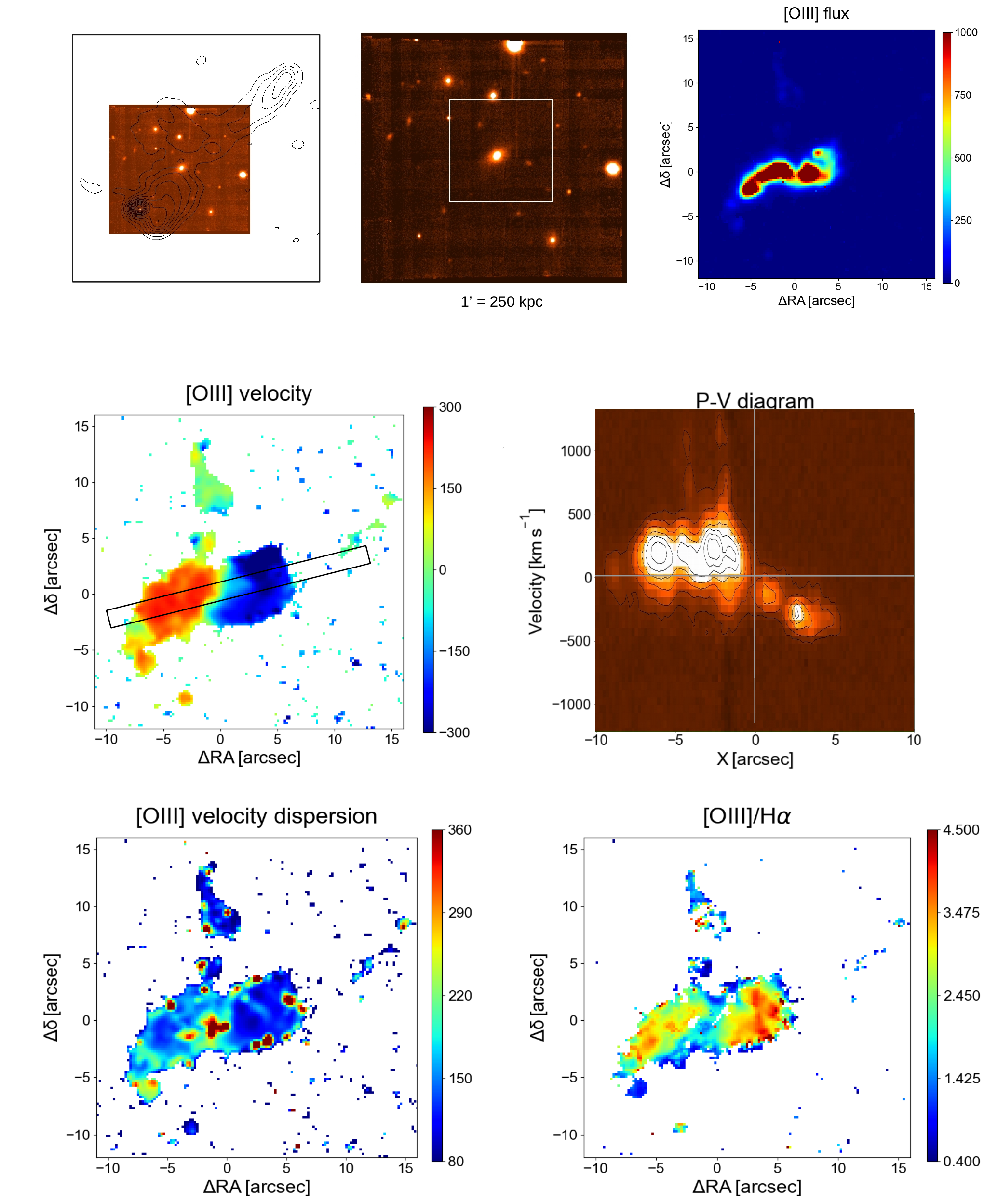}
\caption{3C~300: FR~II/HEG, 1$\arcsec$ = 4.17 kpc. Top left: Radio contours (black) overlaid onto the MUSE 
continuum image. The size
of the image is the whole MUSE field of view, 1$^\prime \times
1^\prime$. Top center: MUSE continuum image with superposed the region
in which we explored the emission line properties (white square). Top
right: \oiii\ emission line image extracted from the white square
 in the central panel. Surface brightness is  $10^{-18} {\rm
erg}\,{\rm s}^{-1}\,{\rm cm}^{-2} {\rm arcsec}^{-2}$. Central panels: (left) Velocity field (in \kms) from
the \oiii\ line; (right) position--velocity diagram extracted from the
synthetic aperture shown  in the left
panel (width 1$\arcsec$, $PA=-20^\circ$). Bottom panels: (left) Velocity dispersion distribution and
(right) \oiii/\ha\ map. }
\label{3c300}}
\end{figure*}

In this section we  present the main results for the individual sources.
In Figures  \ref{3c076.1} to \ref{3c300} we show the continuum
image with superposed the radio contours (taken from the NRAO VLA
Archive Survey Images or from the NASA/IPAC Extragalactic Database),
the image of the brightest emission lines, and the corresponding
velocity field and velocity dispersion. We also produced a
position--velocity diagram by extracting the spectra in a rectangular
synthetic aperture along the direction of the largest line extent, as well
as a ratio image between the brightest line ([O~III] or [N~II]) and
\ha. We presented only spaxels with an emission line signal-to-noise ratio greater than 3.

The emission line ratio maps provide important clues 
to the excitation mechanism of the gas. According to the location in the
emission line ratio diagnostic plane defined by \citet{kewley06}, we can separate Seyfert galaxies from starburst or LINER galaxies.
In particular, assuming an intrinsic value of Ha/Hb  of 3.1 \citep{Ferland85},
 a value of ([O~III]/H$\alpha$) emission line ratio greater than $\sim$3 is  indicative of a Seyfert-like nebular activity.  Similarly, a [N~II]/H$\alpha$ value can be used instead to separate LINERs from starburst galaxies. We will investigate the spatially resolved 
 emission line maps in comparison with the radio and X-ray emission in a future paper.

The definition of ordered rotation is based on the fit of the 2D
gas velocity field obtained by using the {\sl kinemetry} software
\citep{krajnovic06} that reproduces the moments of the line-of-sight
velocity distributions. The results from {\sl kinemetry} obtained for
all sources are presented in the Appendix.

\smallskip
\noindent
{\bf 3C~076.1:} FR~I, no optical classification, z=0.032, 1$\arcsec$ =
0.64 kpc (see Fig. \ref{3c076.1}). Two diffuse emission line regions
emerge from the compact central source, extending $\sim4$ kpc on each
side. The central ELR shows regular rotation with an amplitude
of $\sim 300$ \kms. The velocity dispersion is, except on the nucleus,
consistent with the MUSE spectral resolution.

\smallskip
\noindent
{\bf 3C~078 (NGC~1218):} FR~I/LEG, z=0.028, 1$\arcsec$ = 0.56 kpc (see
Fig. \ref{3c078}). The central ELR is only marginally resolved. 

\smallskip
\noindent {\bf 3C~079:} FR~II/HEG, z=0.256, 1$\arcsec$ = 4.0 kpc (see
Fig. \ref{3c079}). The ELR has an X-shaped morphology. The two
brightest and most extended structures ($\sim 80$ kpc of total size)
are oriented perpendicularly to the radio axis, and show ordered
rotation with an amplitude of $\sim$800 \kms. In addition, there are
several filaments of line emission of similar size but of lower
brightness extending perpendicularly to the main emission line
structure. The brightest regions of the ELR have higher
\oiii/\ha\ values than those of lower brightness; no clear trend with
distance is seen.

\smallskip
\noindent {\bf 3C~088 (UGC~2748):}  FR~II/LEG, z=0.030, 1$\arcsec$ = 0.60 kpc (see
Fig. \ref{3c088}). A diffuse halo of line emission, $\sim 3$ kpc of
radius, surrounds the central ELR.  It shows significant rotation,
with an amplitude of $\sim 170$ \kms.  The velocity dispersion reaches
$\sim 250$ \kms\ on the nucleus but elsewhere is rather low.
The
\nii/\ha\ ratio is higher in the central region than in the diffuse
halo.

\smallskip
\noindent {\bf 3C~089:} FR~I, no optical classification, z=0.138,
1$\arcsec$ = 2.46 kpc (see Fig. \ref{3c089}). A linear feature of line
emission emerges along the $PA$ of the western radio jet and reaches a
distance of $\sim$ 10 kpc. The velocity field is remarkably flat.

\smallskip
\noindent {\bf 3C~098:} FR~II/HEG, z=0.030, 1$\arcsec$ = 0.60 kpc (see
Fig. \ref{3c098}). A series of bright knots extends on both sides of
the nucleus to a distance of $\sim 15$ kpc, oriented at $PA \sim
-10^{\circ}$, $\sim 35^{\circ}$ away from the radio axis. The velocity
field is quite complex, but a general rotation is seen with the S side
of the ELR receding from us and the N side approaching. The line width
exceeds the instrumental value only in the central component of the
ELR. The \oiii/\ha\ ratio has a radial structure, decreasing at larger
angles from the radio axis, while no trend with distance is
  observed.

\smallskip
\noindent {\bf 3C~105:} FR~II/HEG, z=0.089, 1$\arcsec$ = 1.67 kpc (see
Fig. \ref{3c105}). The ELR is compact, with a size of a few kpc. It
shows rotation around $PA \sim 70^{\circ}$ with an amplitude of $\sim
400$ \kms.

\smallskip
\noindent {\bf 3C~135:} FR~II/HEG, z=0.125, 1$\arcsec$ = 2.26 kpc (see
Fig. \ref{3c135}). An irregular series of emission line knots form a
conical structure, mainly toward the SW, reaching a distance of $\sim
77$ kpc and tracing the edge of the western radio lobe. The ELR shows
rapid rotation in the central regions, while the velocity gradients
sharply decreases in the outer diffuse structures. The
\oiii/\ha\ ratio shows a general decreasing trend toward larger radii.

\smallskip
\noindent {\bf 3C~180:} FR~II/HEG, z=0.220, 1$\arcsec$ = 3.58 kpc (see
Fig. \ref{3c180}). A series of arc-like features is seen on both sides
of the nucleus, confirming the structure already seen in the HST
images \citep{baldi19}. The ELR extends by $\sim 45$ kpc on both sides
and it is oriented along a similar angle of the radio axis (the radio
image used, the only available high resolution image from the NRAO VLA
Archive Survey Images, is at a frequency of 8.4 GHz and only shows the
two hot spots). In general, the ELR shows  a rotating structure with a
full amplitude of $\sim 700 $\kms, but with local velocity fluctuations
on the larger scales. The velocity dispersion shows an increase at the
edges of the brightest arcs. The \oiii/\ha\ has a complex structure.

\smallskip
\noindent {\bf 3C~196.1:} FR~II/LEG, z=0.198, 1$\arcsec$ = 3.30 kpc
(see Fig. \ref{3c196.1}). Its radio structure is best seen in the 5
GHz image from \citet{neff95}: it is a double source with a total size
of 5$\arcsec$ ($\sim 16$ kpc) oriented along $PA \sim 40$. The ELR has
a similar size and orientation and its SW side enshrouds the radio
lobe. In this region we detect very high recession velocities,
exceeding $\sim 500$ \kms, and large line widths. These results
suggest that we are seeing a cavity of ionized gas inflated by the
radio outflow. The opposite NE side is more diffuse, with no
significant velocity gradient and a small velocity dispersion.
The SW line lobe has a higher \nii/\ha\  than the opposite side.

\smallskip
\noindent {\bf 3C~198:} FR~II/SF, z=0.081, 1$\arcsec$ = 1.54 kpc (see
Fig. \ref{3c198}). This is the only 3C source at $z<0.3$ where the
emission lines are dominated by the presence of star forming regions
\citep{buttiglione10}. The ELR shows both compact knots and diffuse
emission, mainly on its SW side where it reaches a distance of $\sim
45$ kpc. While the bright knot on the E side is cospatial with a
nearby galaxy, the remaining compact line features do not have
continuum counterparts. All these structures have a spectrum
characteristic of star formation, similarly to what is observed on the
nucleus, suggestive of jet-induced star formation.

\smallskip
\noindent {\bf 3C~227:} FR~II/BLO, z=0.086, 1$\arcsec$ = 1.62 kpc (see
Fig. \ref{3c227}). The ELR in this source covers the whole MUSE field
of view, reaching distances of $\sim 50 $ kpc. It is dominated by a
bright linear structure, oriented at $PA \sim 120^{\circ}$, showing
evidence of ordered rotation out to a radius of $\sim 5 \arcsec$. A
second structure of lower surface brightness gives  an
overall X-shaped structure to the ELR, similar to that seen in 3C~079. On the NE
side this features wraps toward higher values of $PA$, and it also
shows a large velocity gradient and a higher \oiii/\ha.

\smallskip
\noindent {\bf 3C~258:} As already mentioned above, we confirm that
the identification of this radio source with a galaxy at redshift
$z=0.165$ is incorrect. We confirm the suggestion of \citet{dey94} and
\citet{floyd08} that it is instead   a QSO at $z\sim
1.54$, based on the presence of a single broad emission line $\lambda
= 7116 \AA$, likely Mg II at $\lambda$ = 2798 \AA\ in rest frame.

\smallskip
\noindent {\bf 3C~264 (NGC~3862):} FR~I/LEG, z=0.021, 1$\arcsec$ = 0.43 kpc (see
Fig. \ref{3c264}).  The central ELR is only marginally resolved, but it
shows ordered rotation with an amplitude of $\sim$100 \kms. A dusty disk or ring seen in HST data has been 
reported by \citet{baum97}.

\smallskip
\noindent {\bf 3C~272.1 (M~84):} FR~I/LEG, z=0.003, 1$\arcsec$ = 0.068 kpc (see
Fig. \ref{3c272.1}). This is the nearest source of our sample. The ELR
extends for $\sim 2.5$ kpc in the EW direction, perpendicular to the
radio jets and cospatial with the dust lane crossing the nuclear
regions \citep{martel99}. At the largest radii its linear structure
opens in a fan-like shape, similarly to the dust distribution. It
shows an ordered rotation with a full amplitude of $\sim 400$
\kms. The line width is always consistent with the MUSE instrumental
resolution, except  the central $\sim 2$ \arcsec, where it reaches
$\sim 250$ \kms. There is a general trend of decreasing
\nii/\ha\  moving from  E to  W.
  
\smallskip
\noindent {\bf 3C~287.1:} FR~II/BLO, z=0.216, 1$\arcsec$ = 3.53 kpc
(see Fig. \ref{3c287.1}).  The central ELR is only marginally
resolved. At a distance of $\sim 5$\arcsec\ toward the S, in a
direction almost perpendicular to the radio jet, a detached diffuse
ELR is also seen, with a large velocity offset of $\sim$400 \kms\ with
respect to the nuclear component. The \oiii/\ha\ ratio has a complex
structure. In particular the southern component of the ELR shows a
strong gradient in  \oiii/\ha.

\smallskip
\noindent {\bf 3C~296 (NGC~5532):} FR~I/LEG, z=0.024, 1$\arcsec$ = 0.49 kpc (see
Fig. \ref{3c296}). The central ELR is only marginally resolved.

\smallskip
\noindent {\bf 3C~300:} FR~II/HEG, z=0.270, 1$\arcsec$ = 4.17 kpc (see
Fig. \ref{3c300}).  The ELR has an S-shaped structure extending $\sim
60$ kpc in the EW direction and showing well-ordered rotation along
its major axis. More diffuse emission is found on the N and W
side out to radii of more than $\sim 60$ kpc. The \oiii/\ha\ ratio is
higher in the off-nuclear regions than in the center.

\section{Summary and conclusions}
We completed the presentation of the observations of the 37 RGs from
the 3C catalog with redshift $<$0.3 and declination $<20^{\circ}$
obtained with the VLT/MUSE optical integral field spectroscoph by
showing the results obtained for the second set of 17 sources; the
first 20 were presented in \citet{balmaverde19}. These data were
obtained as part of the MURALES survey, a project whose aim is to 
explore the distribution and kinematics of the ionized gas in the
most powerful radio sources in order to better understand the AGN
feedback.

We presented the data analysis and, for each source, the resulting
emission line images, the 2D gas velocity field, and an emission line
ratio map indicative of the gas excitation origin. Thanks to their
unprecedented depth these observations reveal ELRs extending several
tens of kiloparsec in most objects.

All but 2 of the 26 FR~IIs show large-scale structures of ionized
gas with a median extent of 16 kpc, but reaching sizes $\gtrsim 80$
kpc. The central gas component is often (in 21 sources) quite regular
and characteristic of rotation. On larger scales the gas kinematics is
usually more complex. The gas excitation (derived from either \oiii/\ha\ or 
\nii/\ha) shows a quite complex behavior. In some sources
(e.g., 3C~300) it increases with distance, but the opposite behavior is
also seen (e.g., 3C~088) and we also see  objects in which no trend
with distance is found (e.g., 3C~079). We also observed sources with a
radial structure of the gas excitation, with higher ionization gas
located along the radio axis (e.g., 3C~098 in this paper and 3C~033 in
Paper II). 
This  scenario suggests that different gas excitation mechanisms are concurrent in the various sources, 
and that the gas ionization status likely depends on many complex factors.
 We will investigate this issue in detail in a following paper.

There are
no apparent differences between the ELR properties of the FR~IIs of
high and low gas excitation regarding luminosities or sizes. This is
unexpected because it has been suggested that LEGs and HEGs are
related to a different accretion process: hot versus cold gas
\citep{baum95,hardcastle07,buttiglione10}. Nonetheless, in LEGs we
detected large reservoirs of warm emitting line gas, similar in
extent and luminosity to those found in HEGs. This finding leaves open
the possibility that the two classes are connected in a evolutionary
scenario, i.e., that a transition between HEGs and LEGs (or
vice versa) occurs (see also \citealt{macconi20}).

Conversely, there is a clear connection between radio morphology, and
in particular between the Fanaroff--Riley class and emission line
properties. In the ten FR~Is observed the line emission regions are
generally compact, only a few kpc in size;  only   one case
(3C~348)   exceeds the size of the host reaching a radius of $\sim
40$ kpc. The ELRs in FR~Is are smaller and less luminous than those
seen in the FR~IIs. This deficit is found even comparing sources at
the same level of line nuclear luminosity, a robust proxy for the
strength of the nuclear ionizing continuum. This implies that what
distinguishes objects belonging to the two radio morphological classes
is a different content of gas in the warm or cold phase on scales larger
than the host galaxy, the FR~Is being generally gas poor. This result
adds to similar conclusions obtained from the properties of the
interstellar medium of RGs (see, e.g., \citealt{dekoff00,ocana10}). 

We also confirmed that the optical identification of 3C~258 is
incorrect. The currently identified optical counterpart is offset by
2\farcs8 with respect to the compact radio source. Conversely, at the
location of the radio emission both the HST and Chandra images show
the presence of a point-like object. The MUSE spectrum of this object
shows a single emission line, with a broad profile, at $\lambda = 7116
\AA$, the most likely identification of this line being Mg II at
$\lambda$ 2798 \AA.  This indicates an association with a type I AGN
at a tentative redshift $z\sim 1.54$.

As already discussed in the Introduction, AGN feedback in radio-loud
sources can manifest itself in two different ways: radio and radiative
modes. They correspond to a transfer of energy to the ambient medium
from the relativistic jets and from nuclear outflows, respectively. In
two companion papers we will address in detail these issues by
studying the origin of the extended emission line structures, and in
particular their connection with the radio emission, and we will
explore the properties of nuclear outflows of ionized gas.

\begin{appendix}
  \appendix
  \section{Analysis of the gas kinematics}
We fit the 2D gas velocity field in the innermost regions with the
{\sl kinemetry} software \citep{krajnovic06}, a generalization of
surface photometry that reproduces the moments of the line-of-sight
velocity distributions.  The parameters of  interest to us returned by
this software are 1) the kinematical $PA$, 2) the coefficient of the
harmonic expansion $k1$ (from which the rotation curve is derived),
and 3) the ratio of the fifth to the first coefficient $k5/k1$ and its error $\sigma_{k5/k1}$
(which quantifies the deviations from purely ordered rotation).  In this
Appendix we show the results obtained for all sources,  excluding
only 3C~089 where the velocity field is remarkably flat.

We considered that a source shows ordered rotation when the
kinematical parameters are defined at a radius larger than 2 times the
seeing of the observations and where $(k5/k1$+$\sigma_{k5/k1})<$0.2.

\begin{figure*}  
\centering{ 
\includegraphics[width=8.5cm]{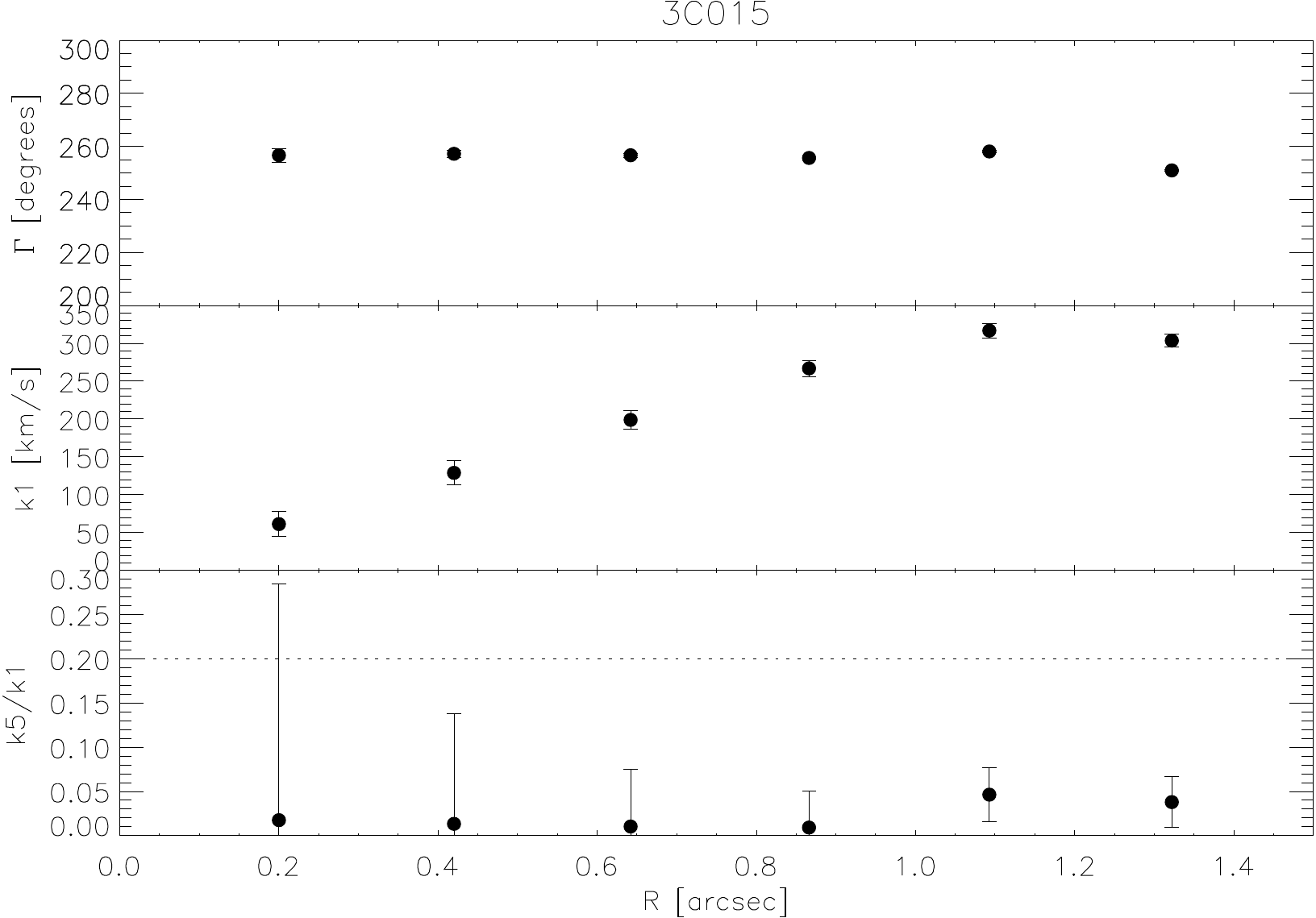}   
\includegraphics[width=8.5cm]{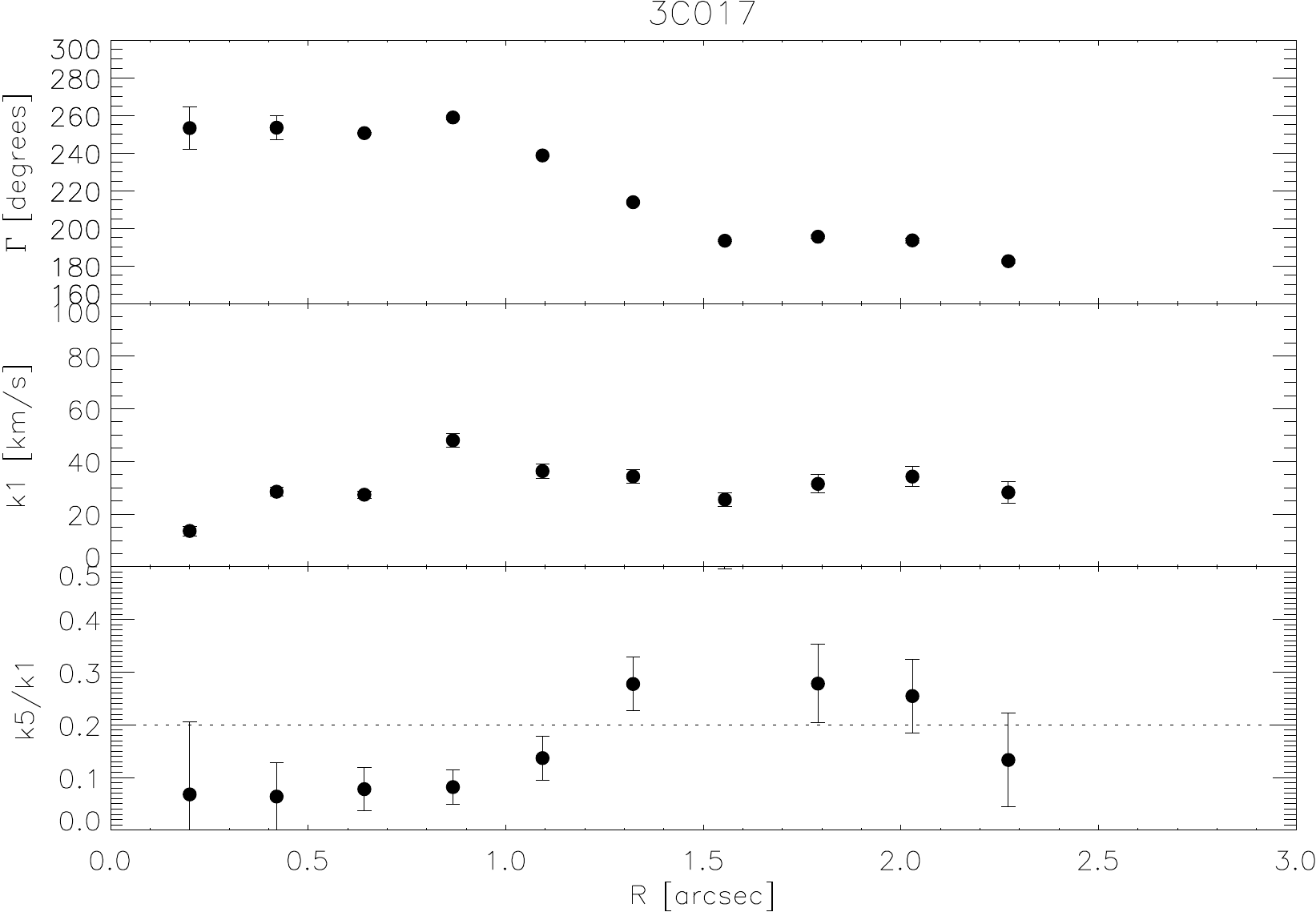}
\includegraphics[width=8.5cm]{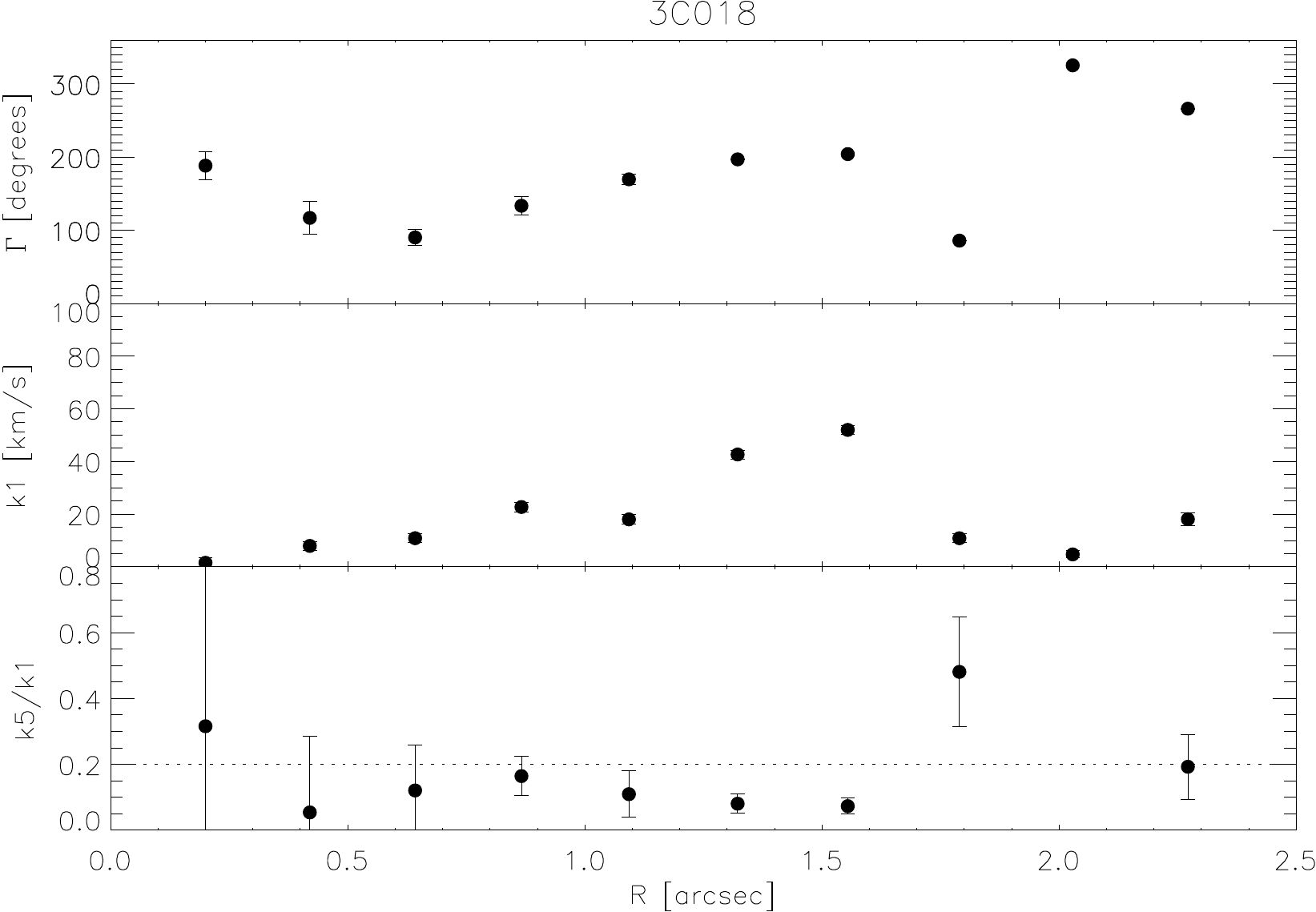}
\includegraphics[width=8.5cm]{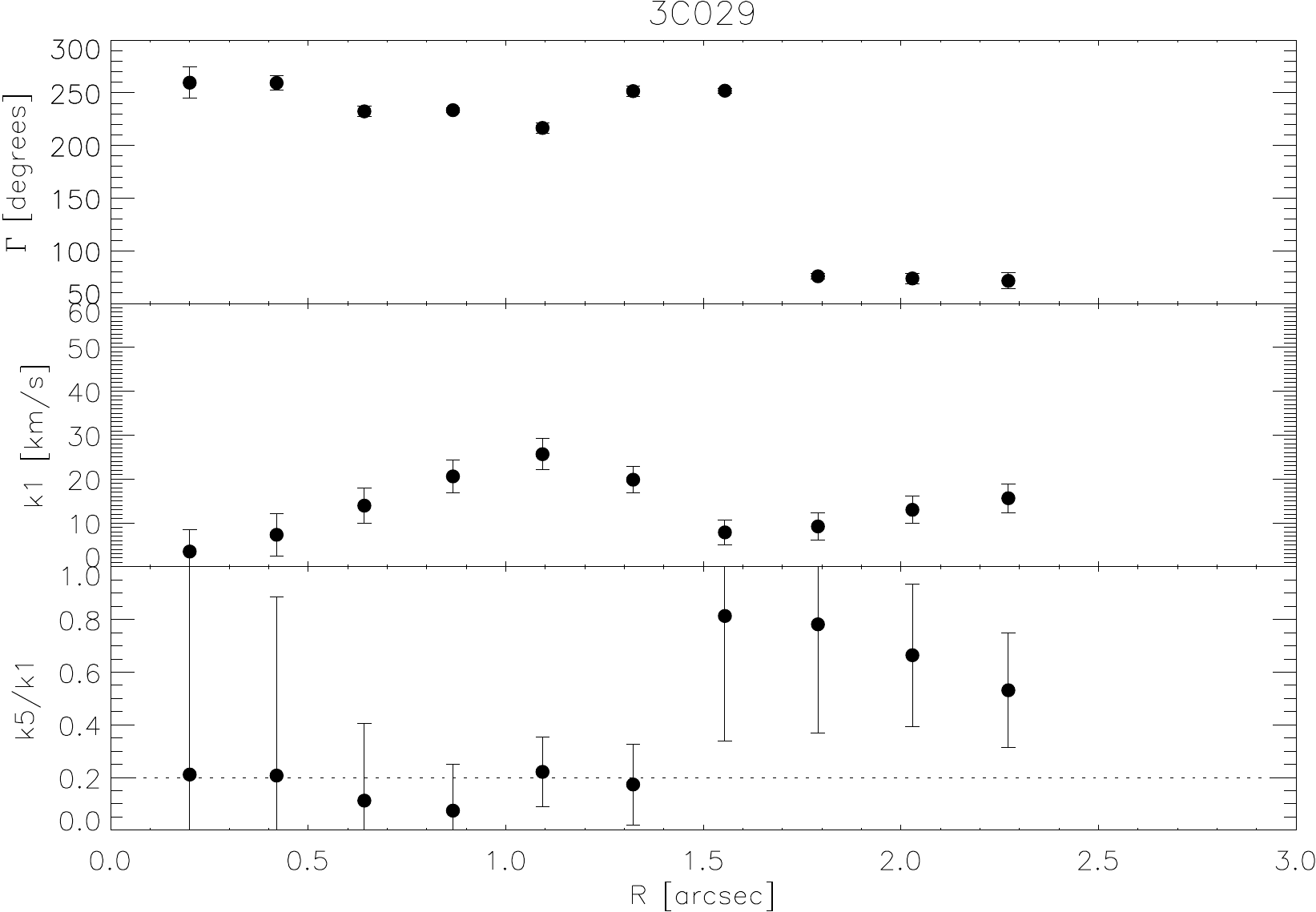}
\includegraphics[width=8.5cm]{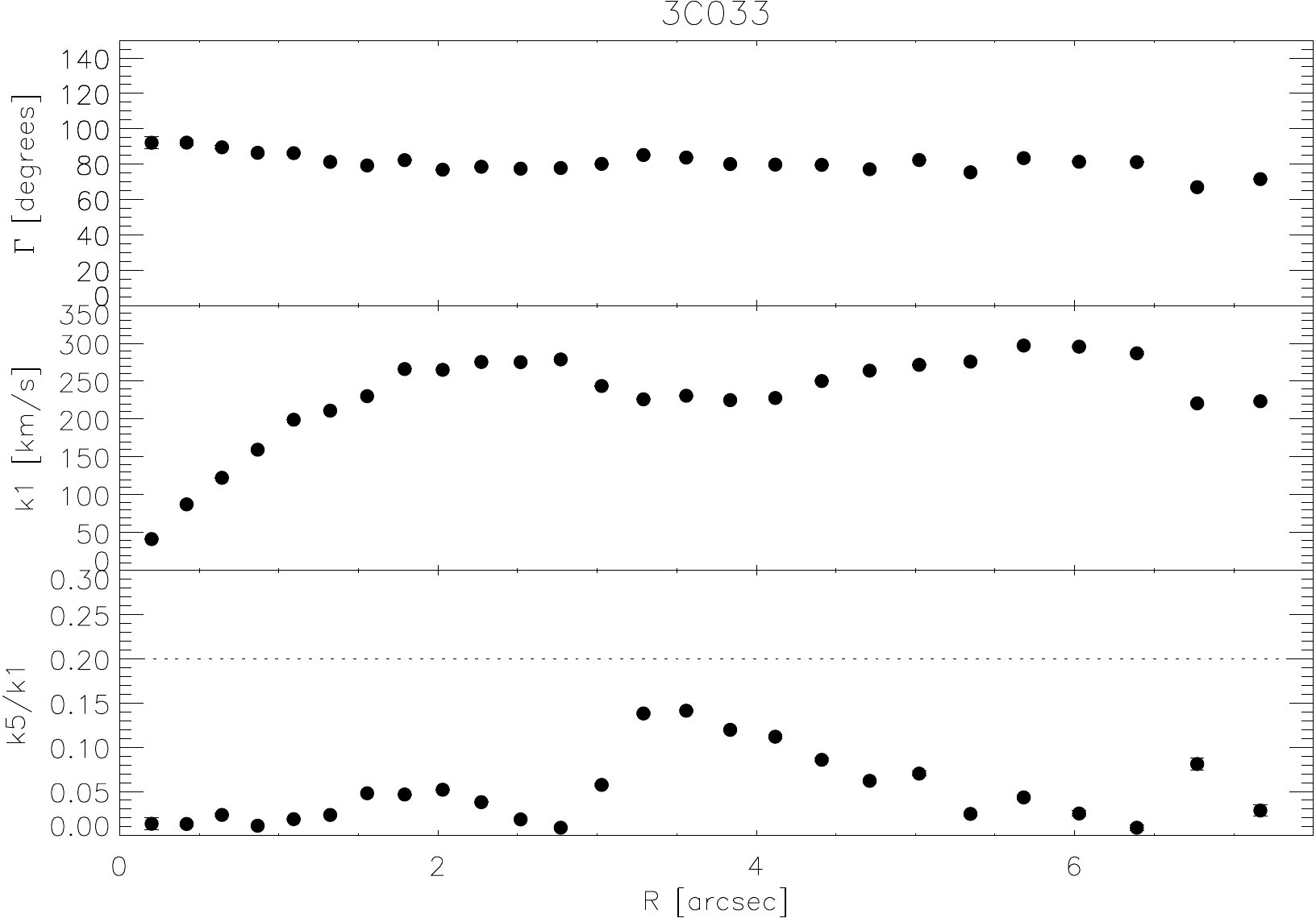}
\includegraphics[width=8.5cm]{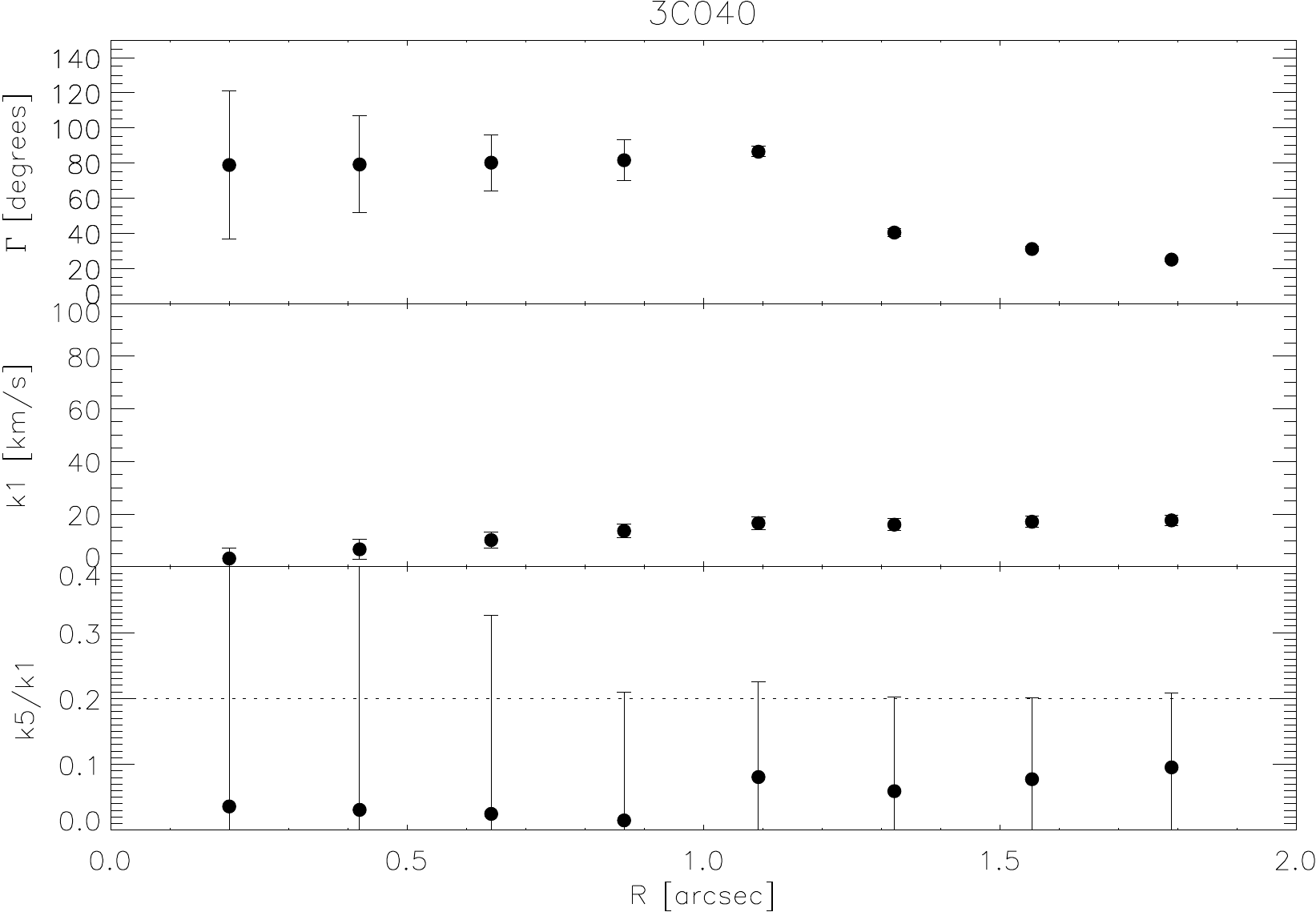}
\includegraphics[width=8.5cm]{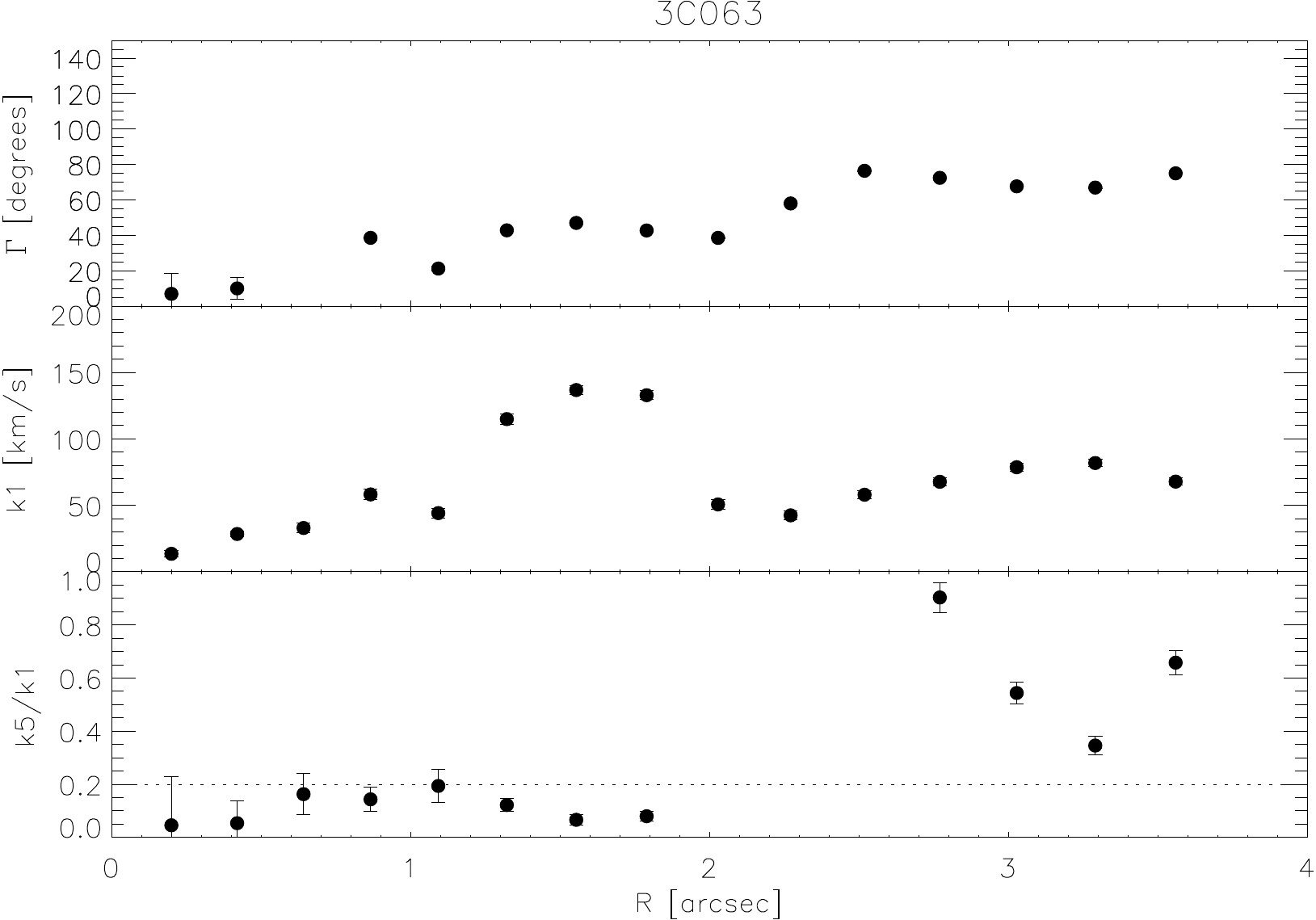}
\includegraphics[width=8.5cm]{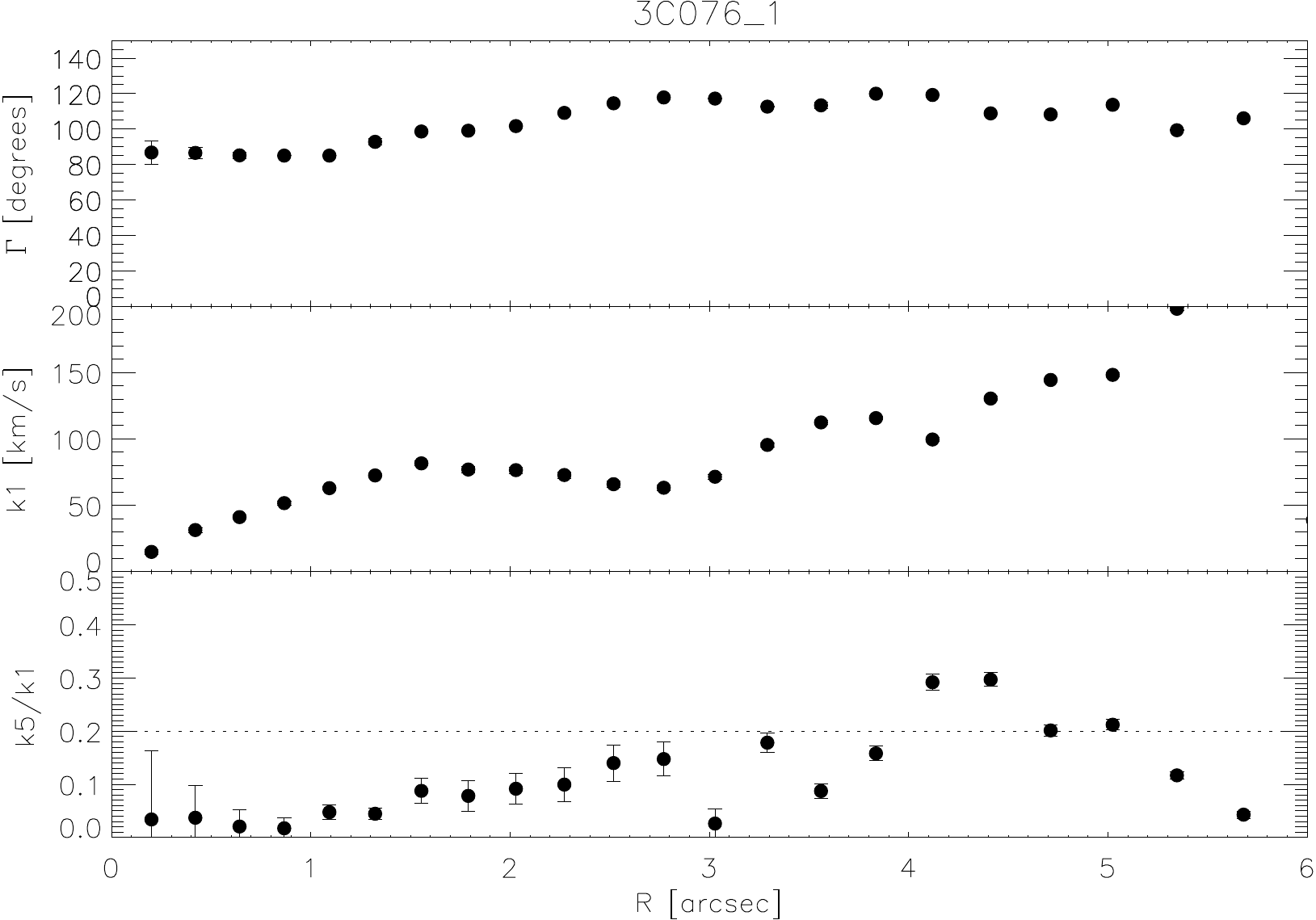}
}
\caption{Results obtained for 36 radio galaxies (all but 3C~089) with
  the {\sl kinemetry} software. From top to bottom in each panel: Kinematic position angle $PA$ (in degrees), amplitude of the
  rotation curve (in \kms), and the ratio of the fifth to the  first
  coefficient $k5/k1$, which quantifies the deviations from simple
  rotation.  }
\end{figure*}

\addtocounter{figure}{-1}
\begin{figure*}  
\centering{ 
\includegraphics[width=8.5cm]{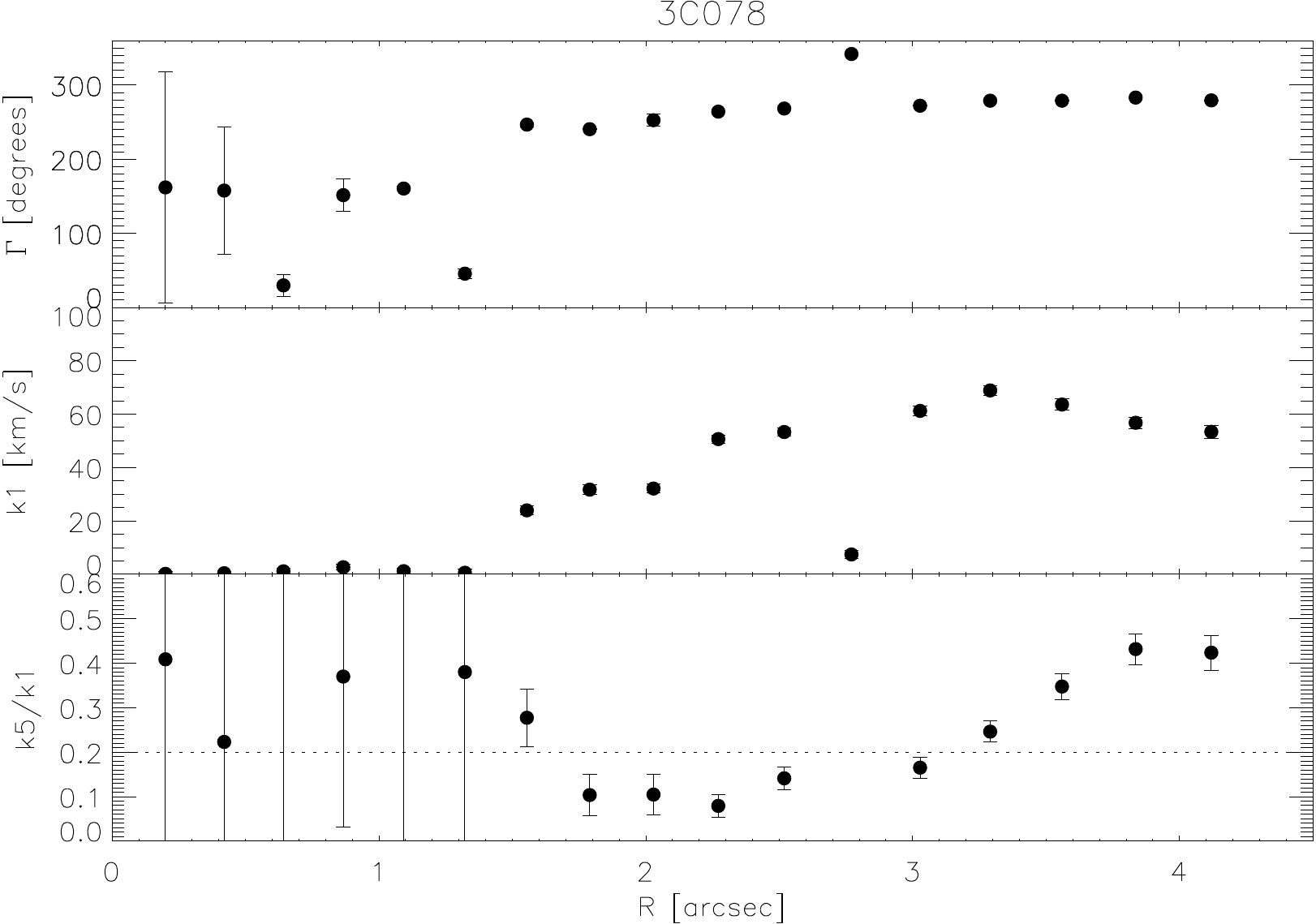}
\includegraphics[width=8.5cm]{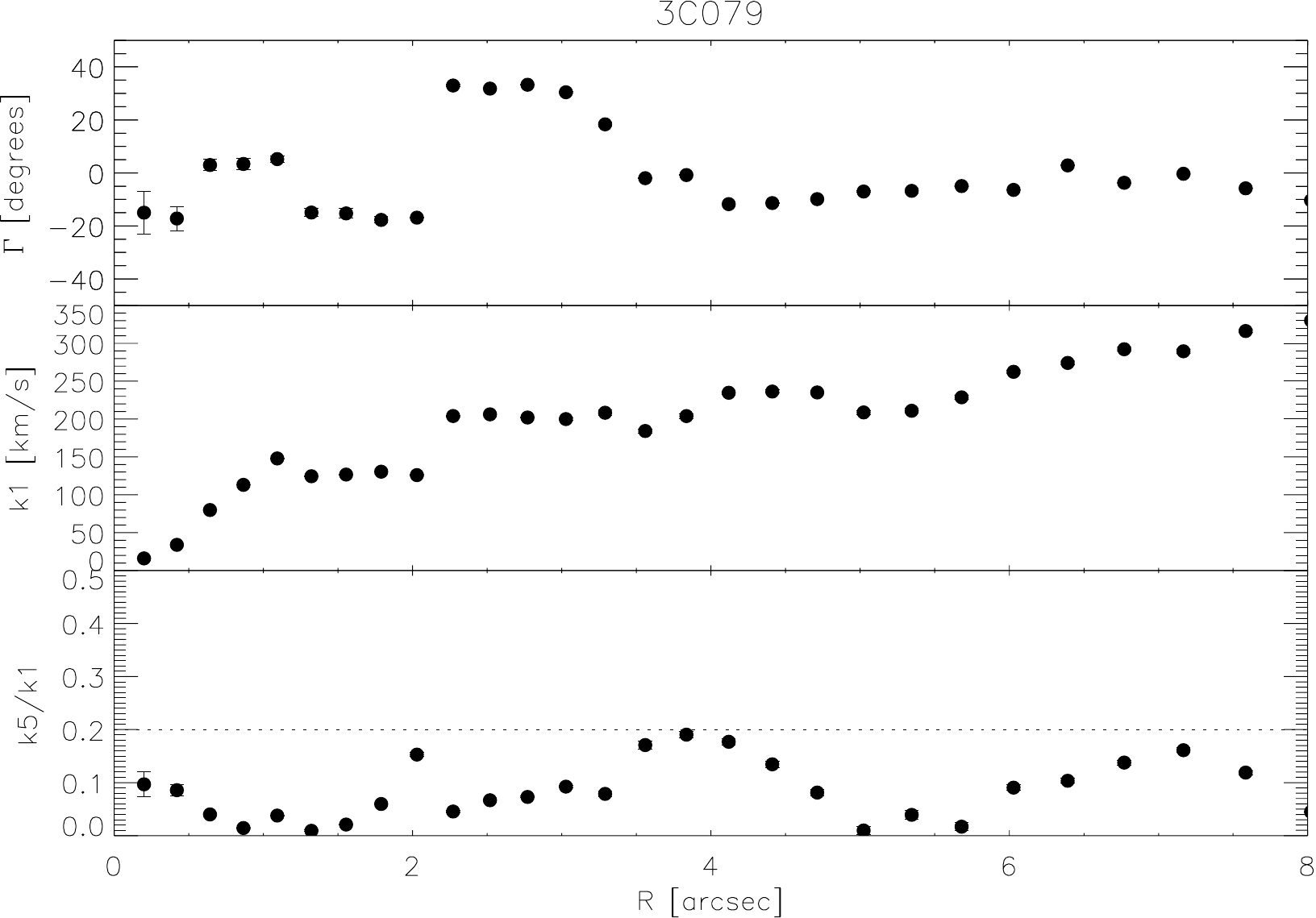}
\includegraphics[width=8.5cm]{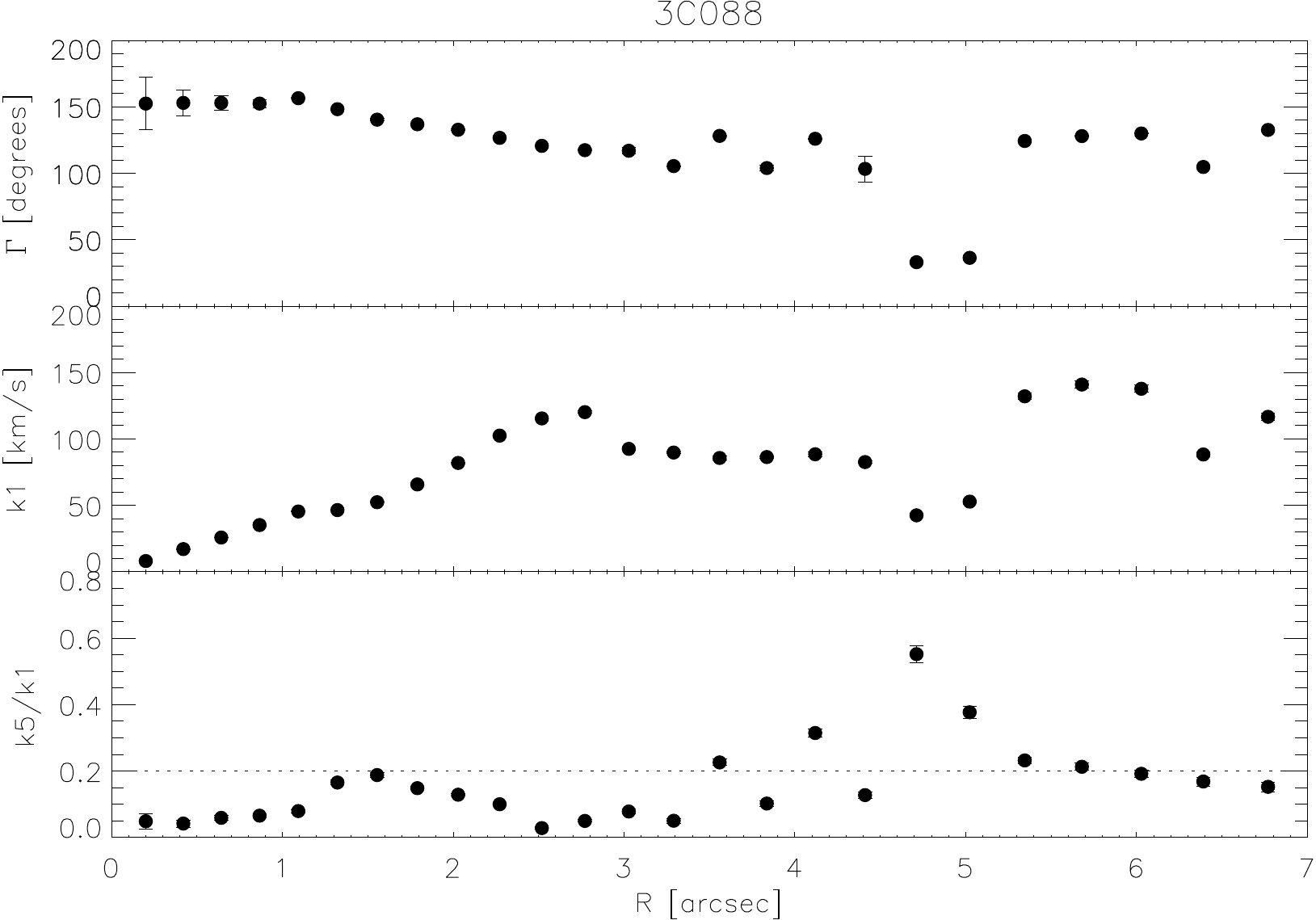}
\includegraphics[width=8.5cm]{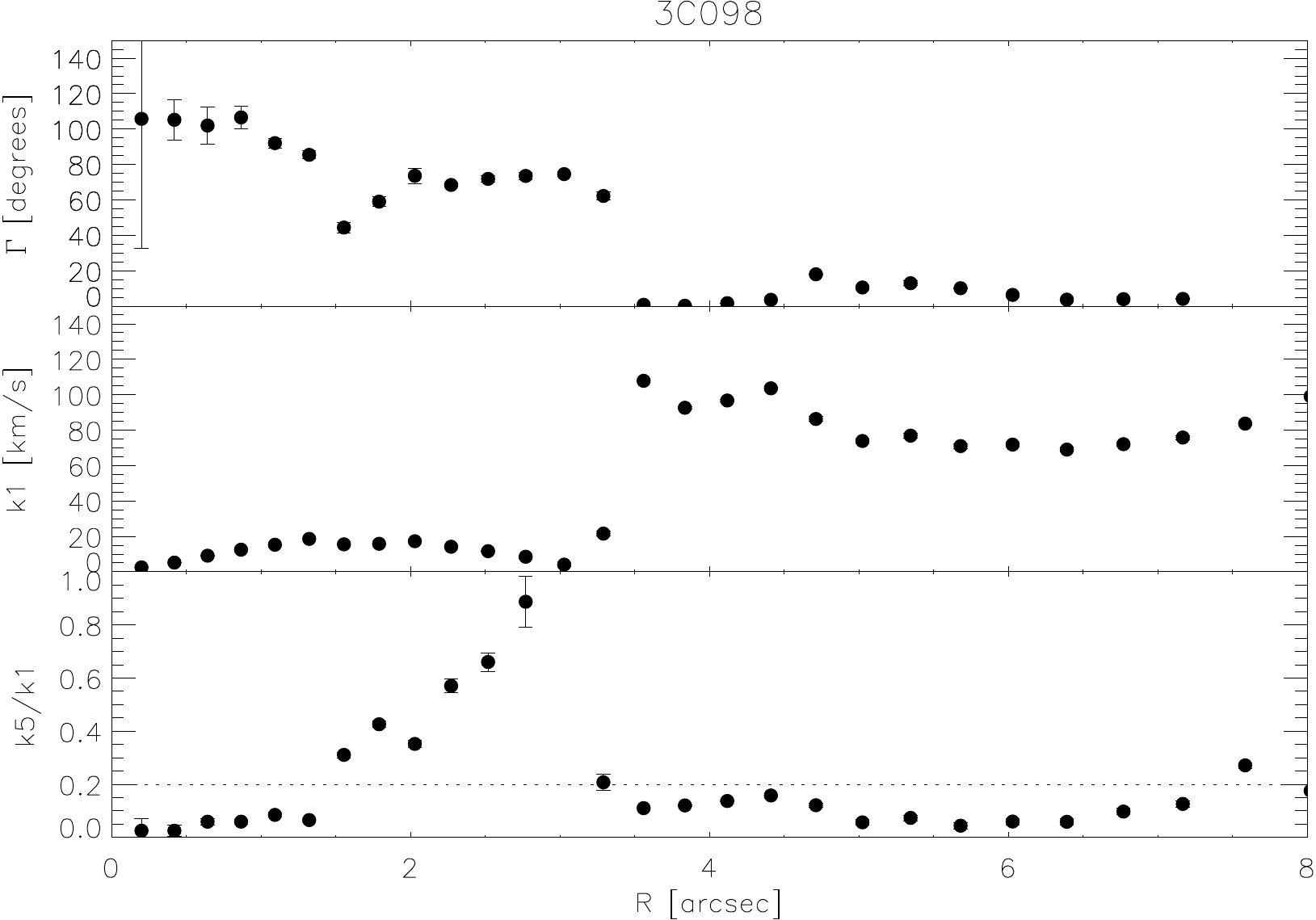}
\includegraphics[width=8.5cm]{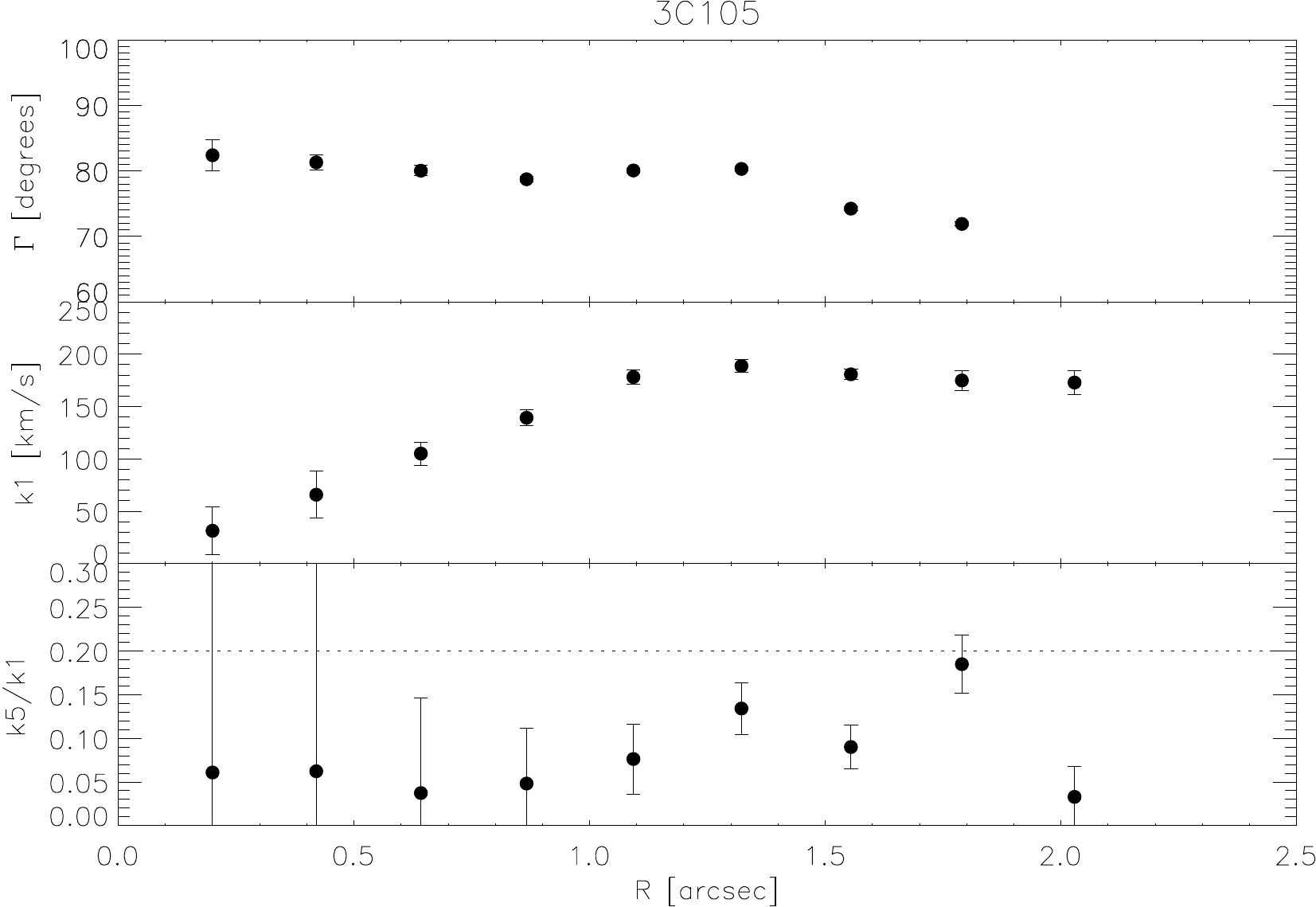}
\includegraphics[width=8.5cm]{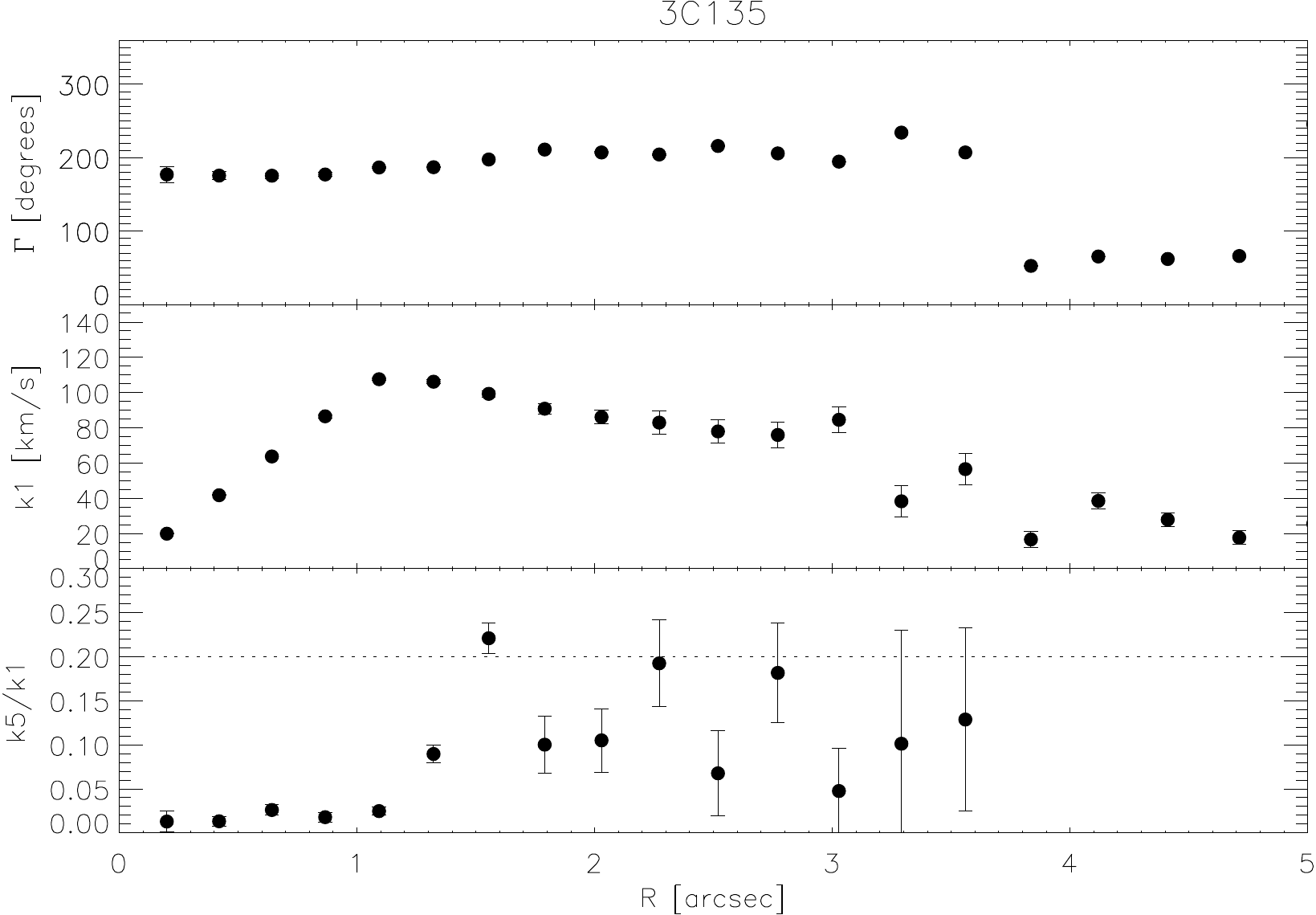}
\includegraphics[width=8.5cm]{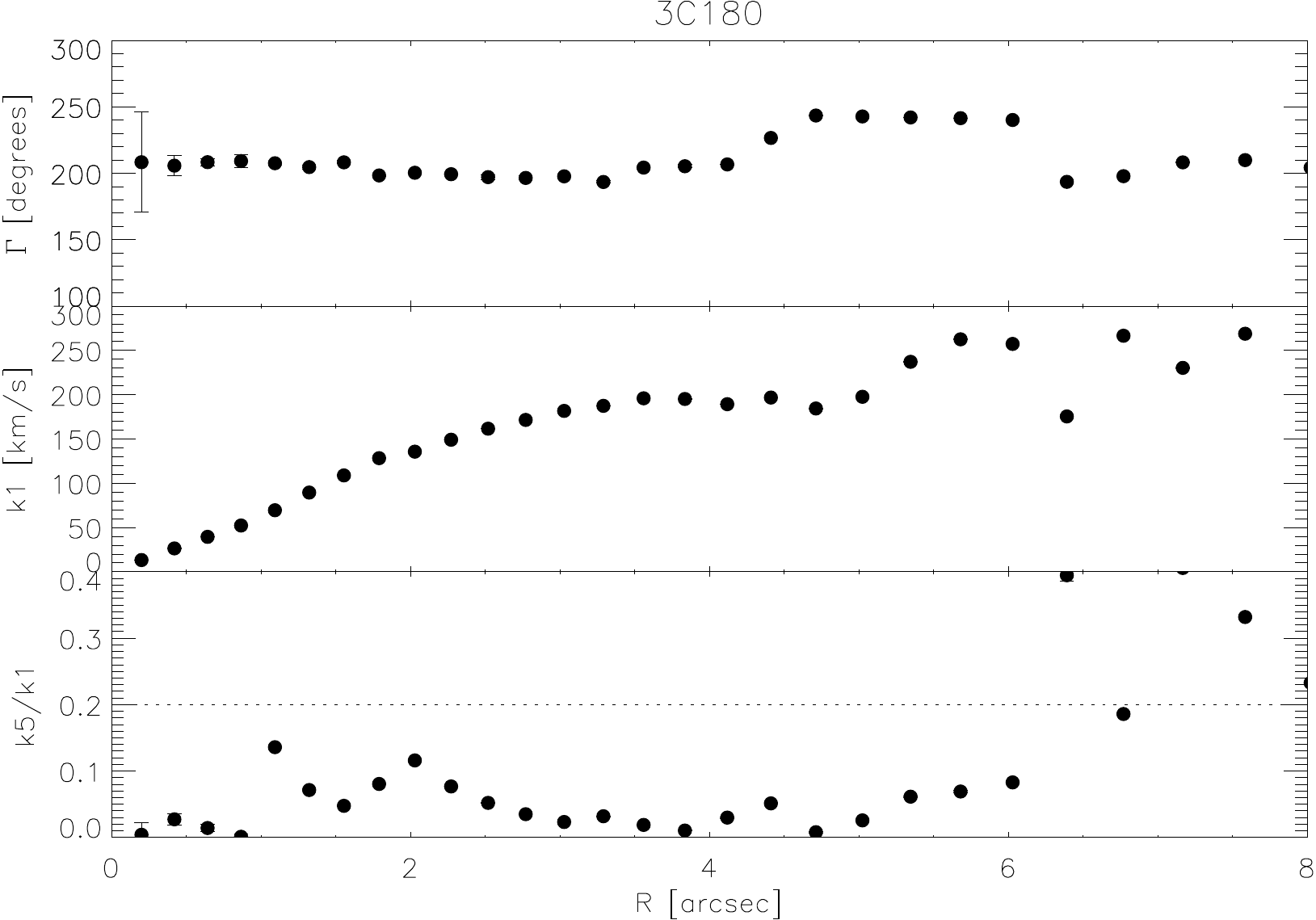}
\includegraphics[width=8.5cm]{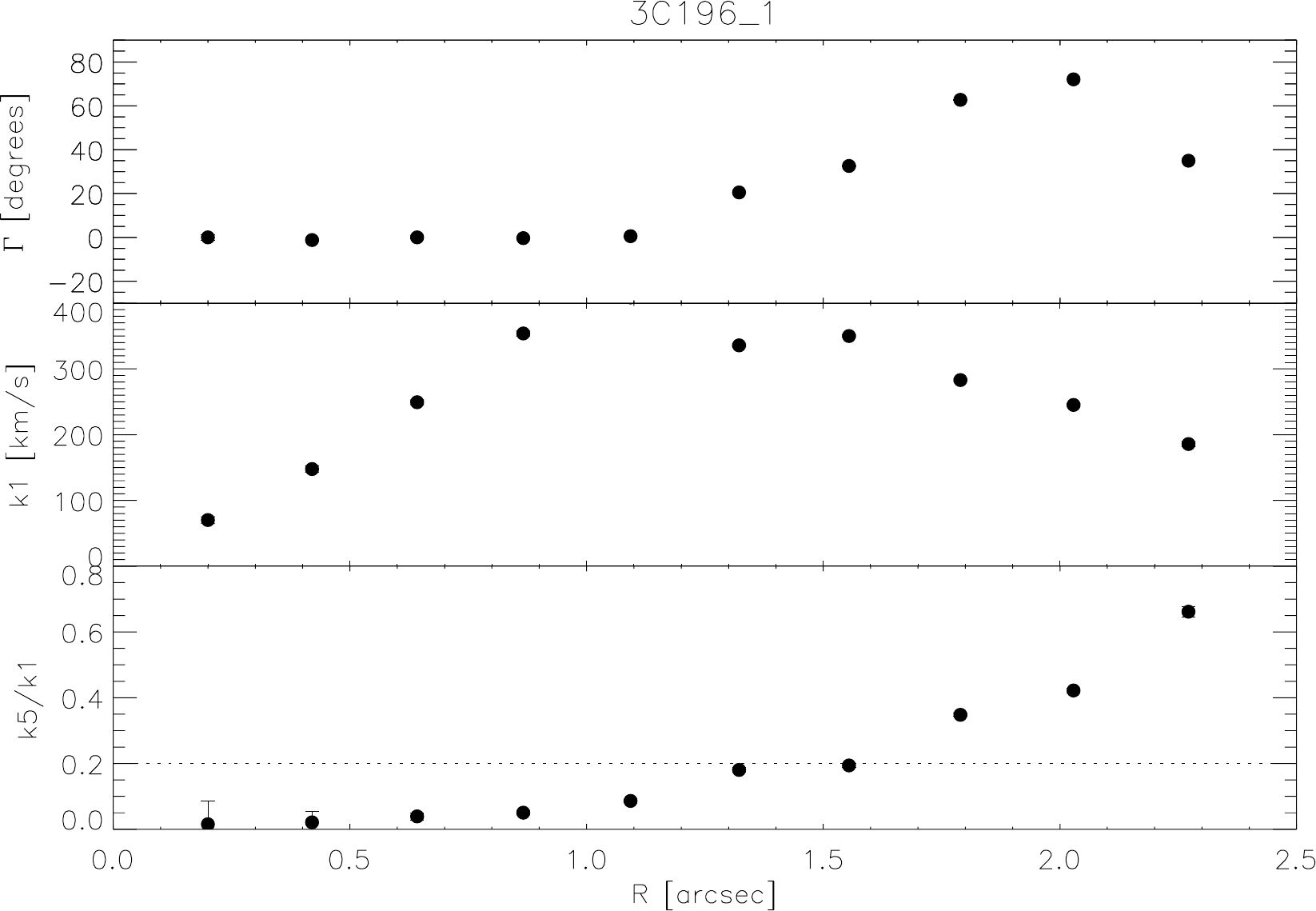}
}
\caption{- continued.}
\end{figure*}   

\addtocounter{figure}{-1}
\begin{figure*}  
\centering{ 
\includegraphics[width=8.5cm]{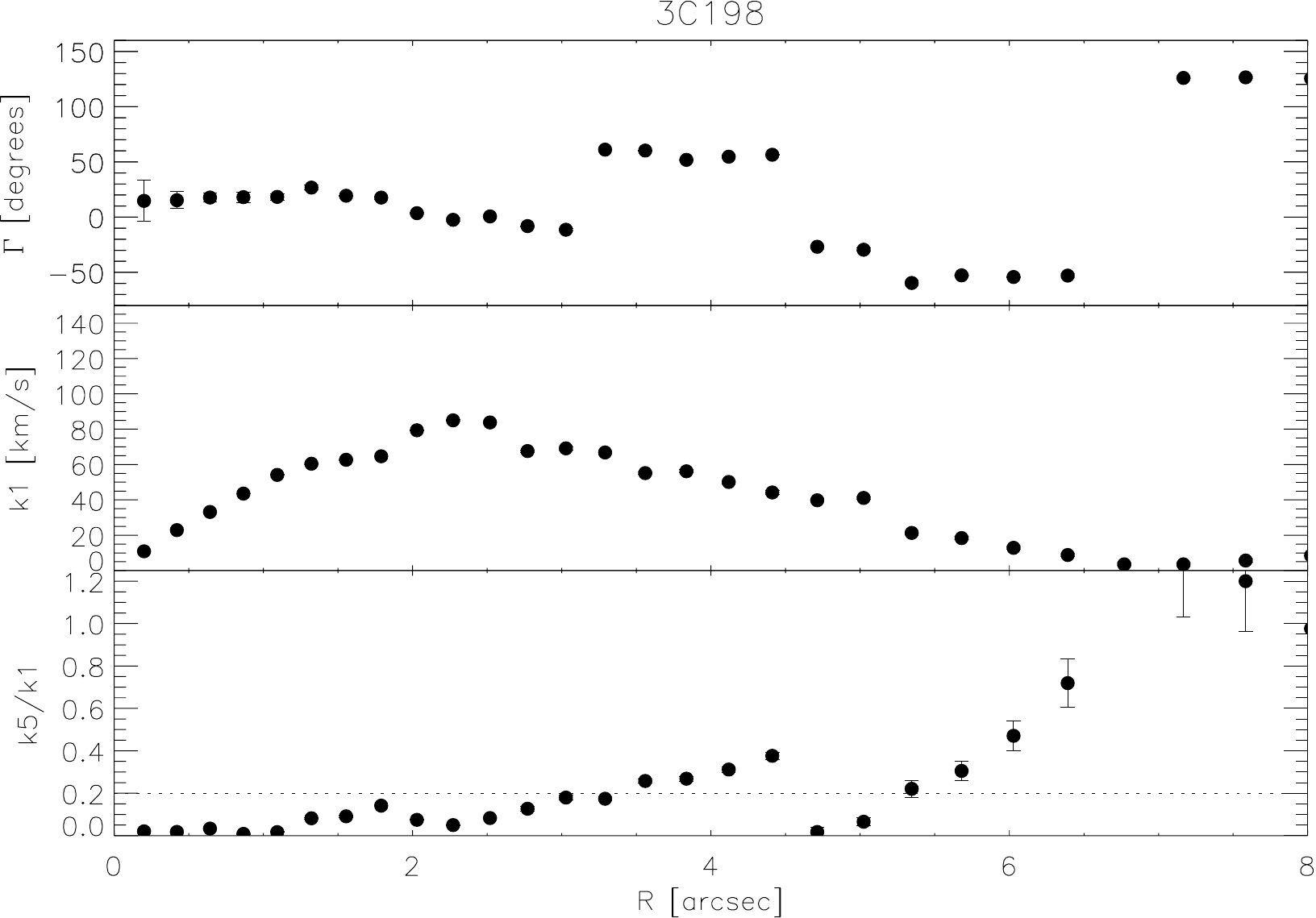}
\includegraphics[width=8.5cm]{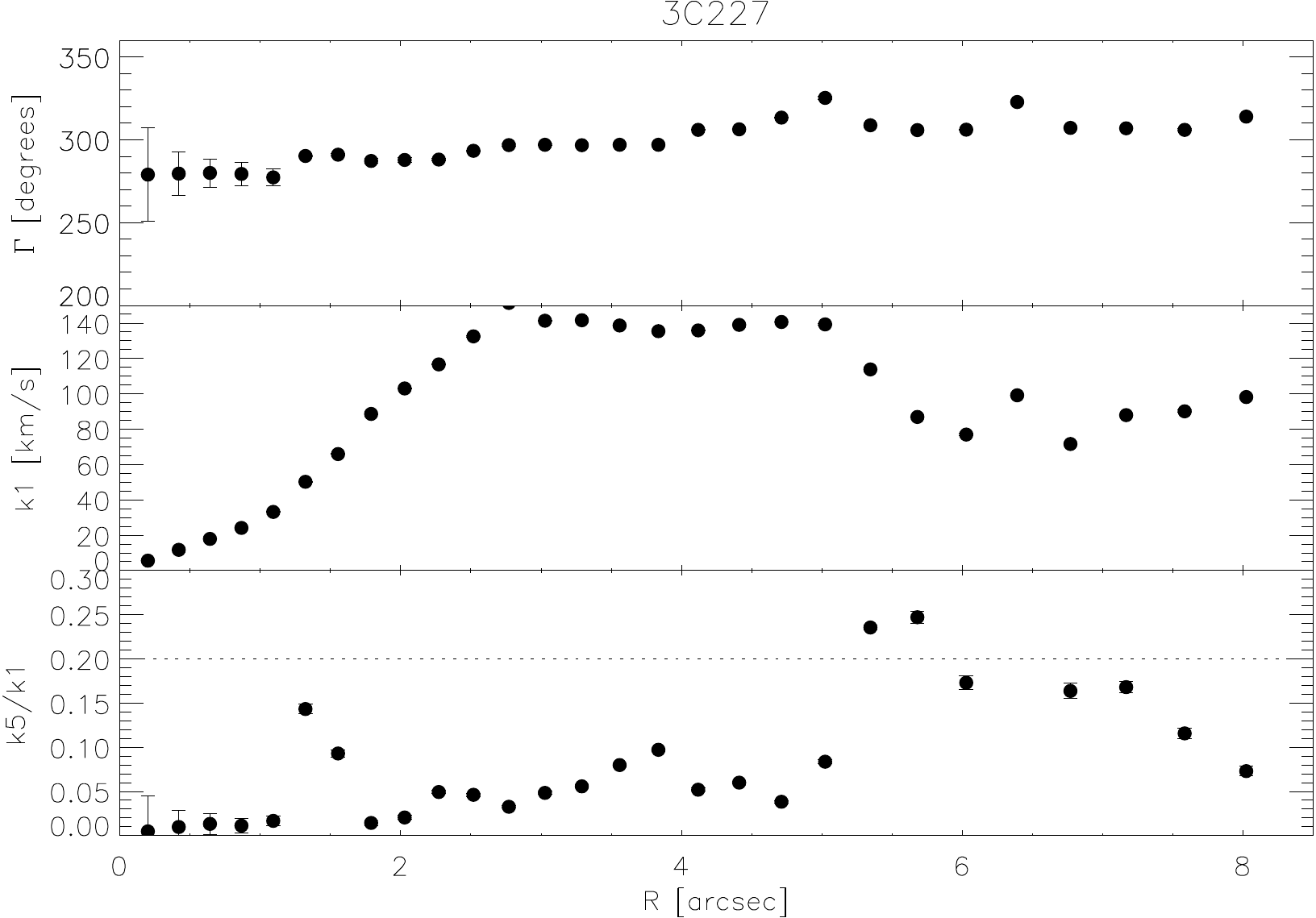}
\includegraphics[width=8.5cm]{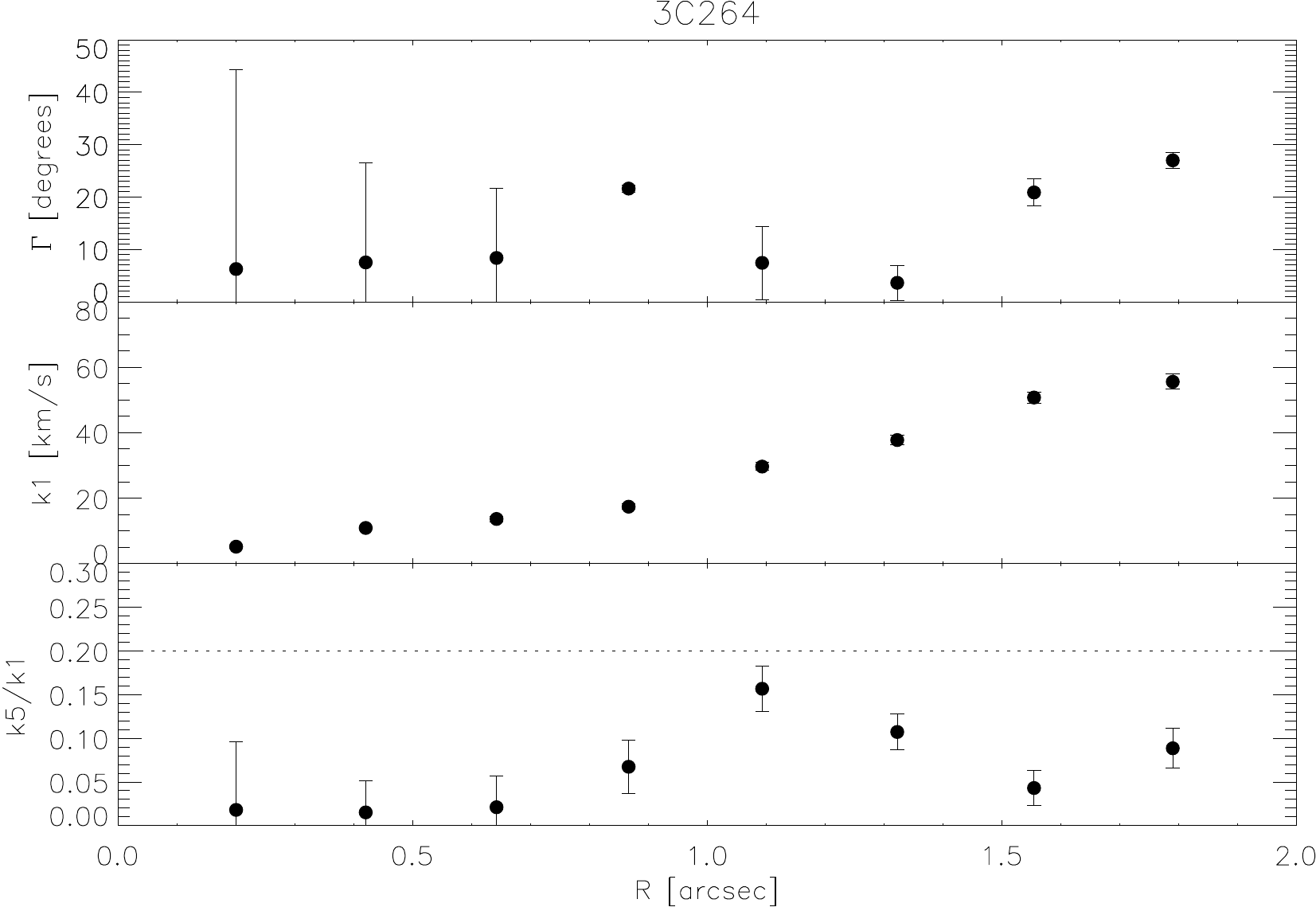}
\includegraphics[width=8.5cm]{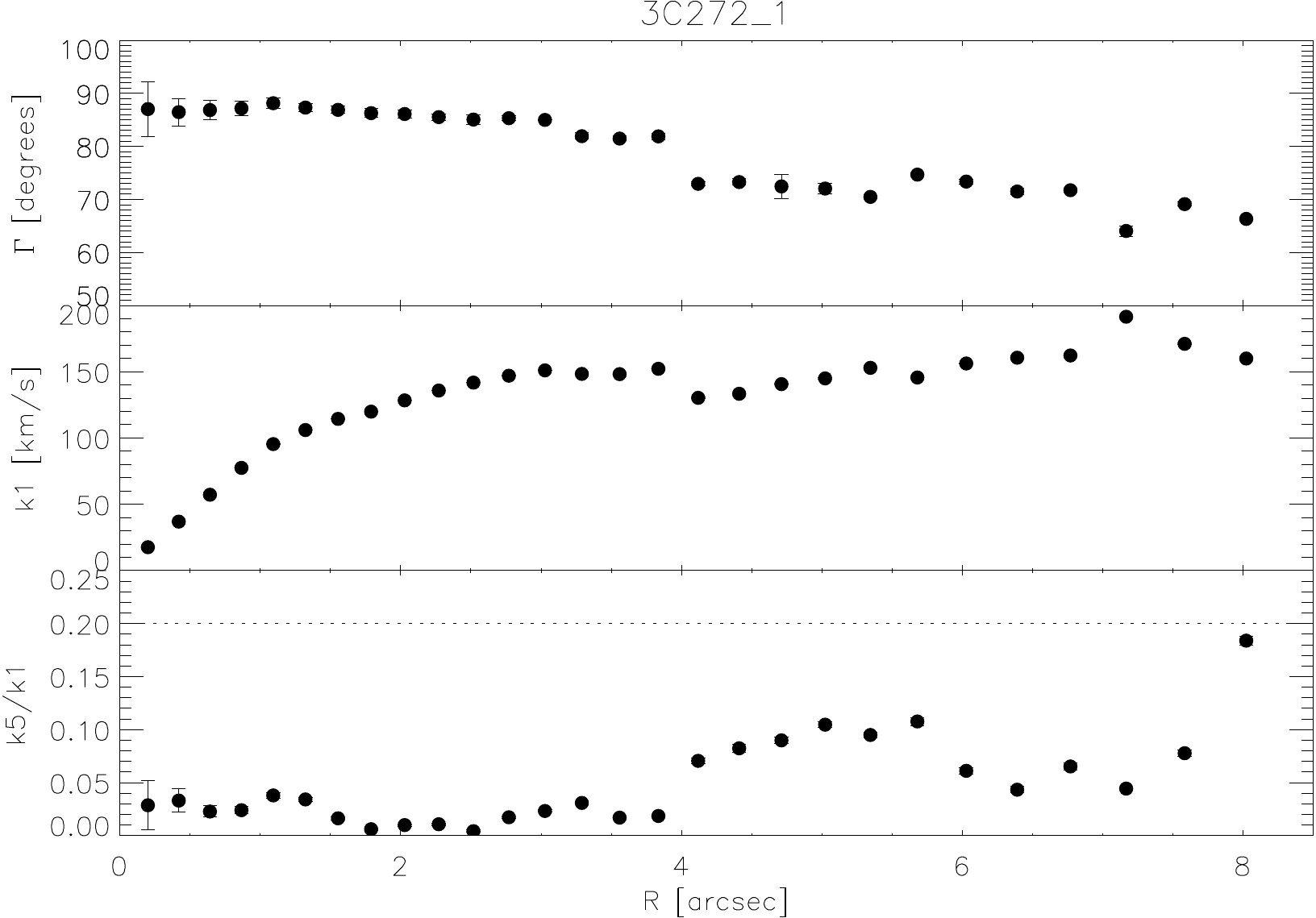}
\includegraphics[width=8.5cm]{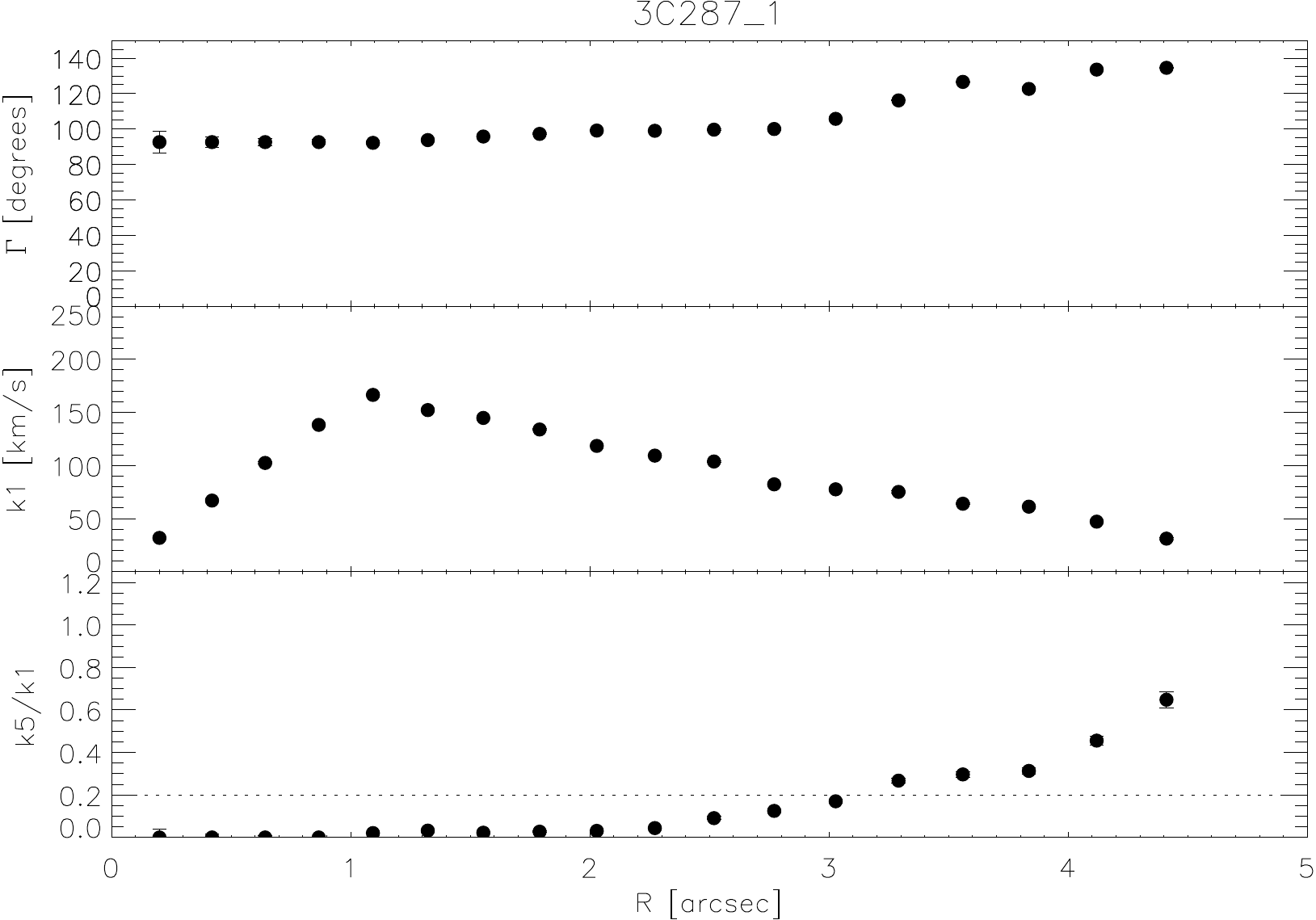}
\includegraphics[width=8.5cm]{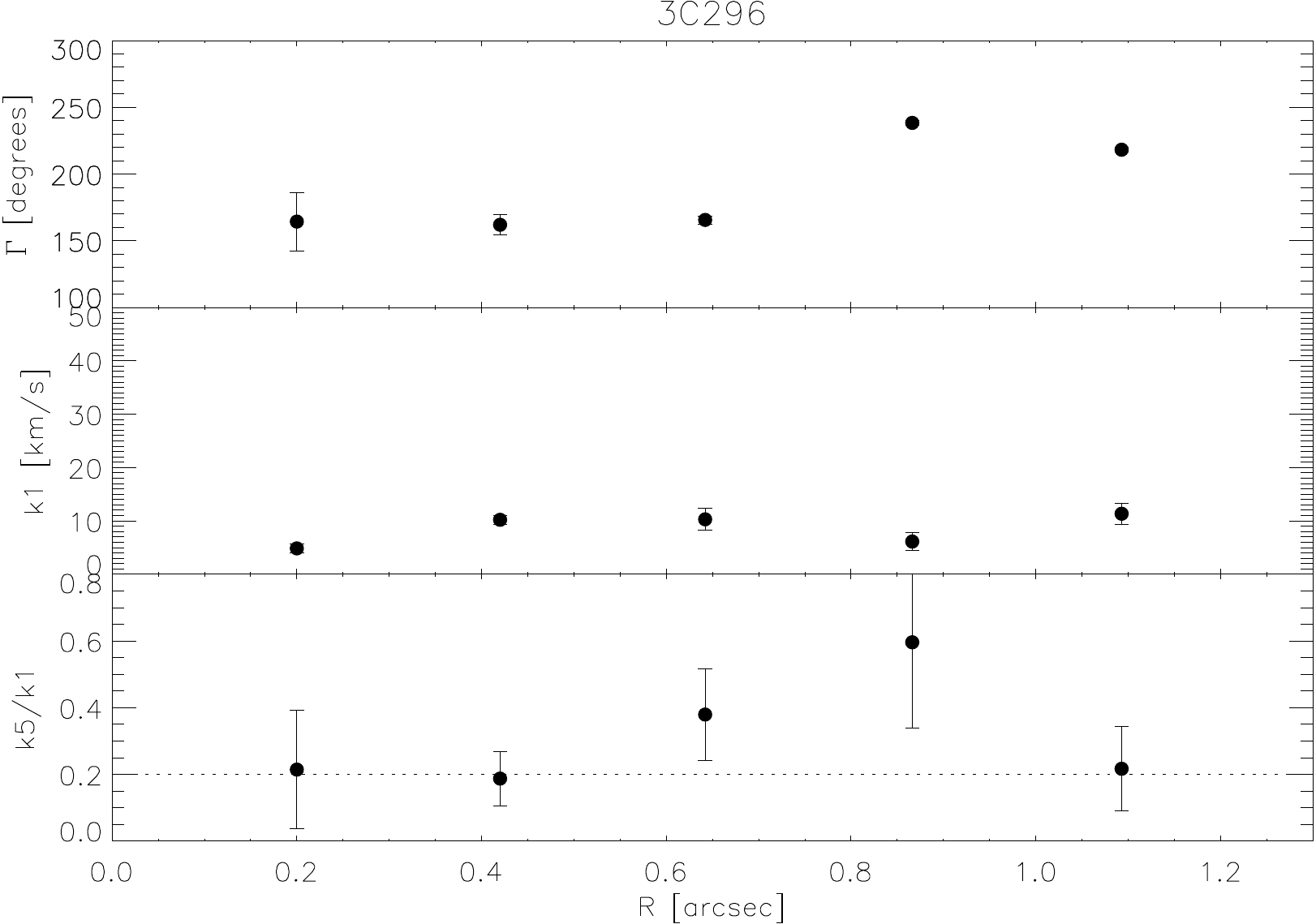}
\includegraphics[width=8.5cm]{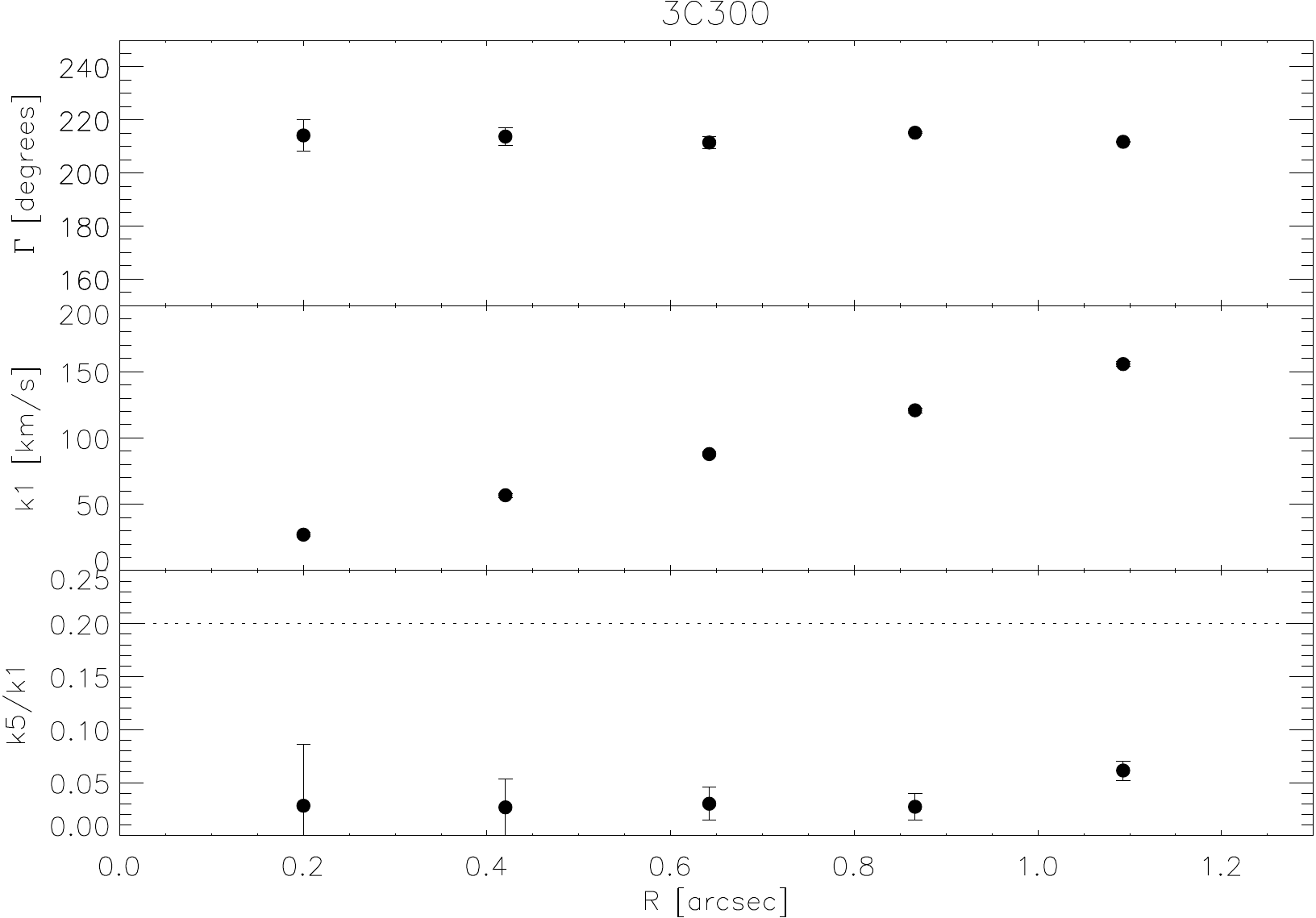}
\includegraphics[width=8.5cm]{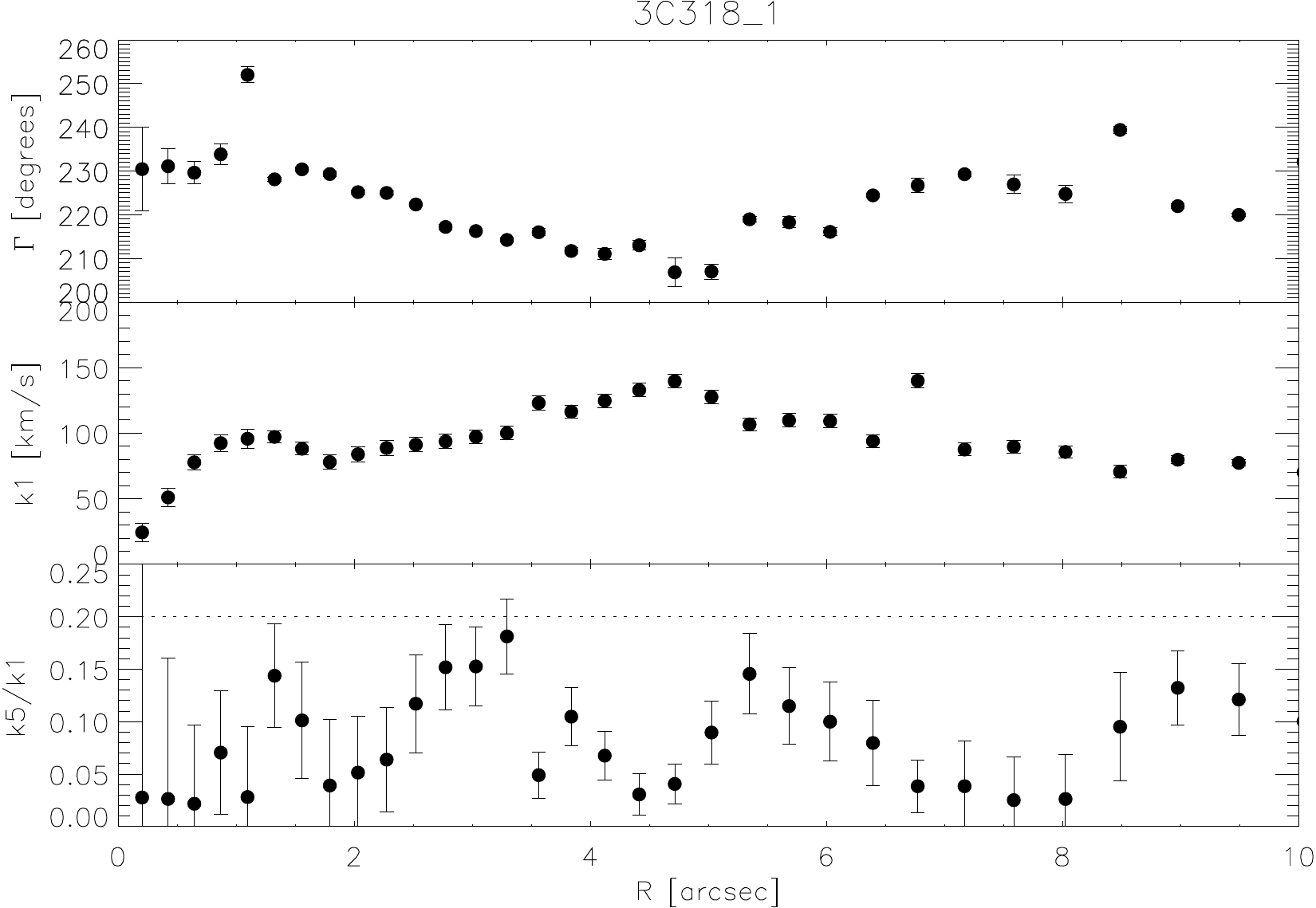}
}
\caption{- continued.}
\end{figure*}

\addtocounter{figure}{-1}
\begin{figure*}  
\centering{ 
\includegraphics[width=8.5cm]{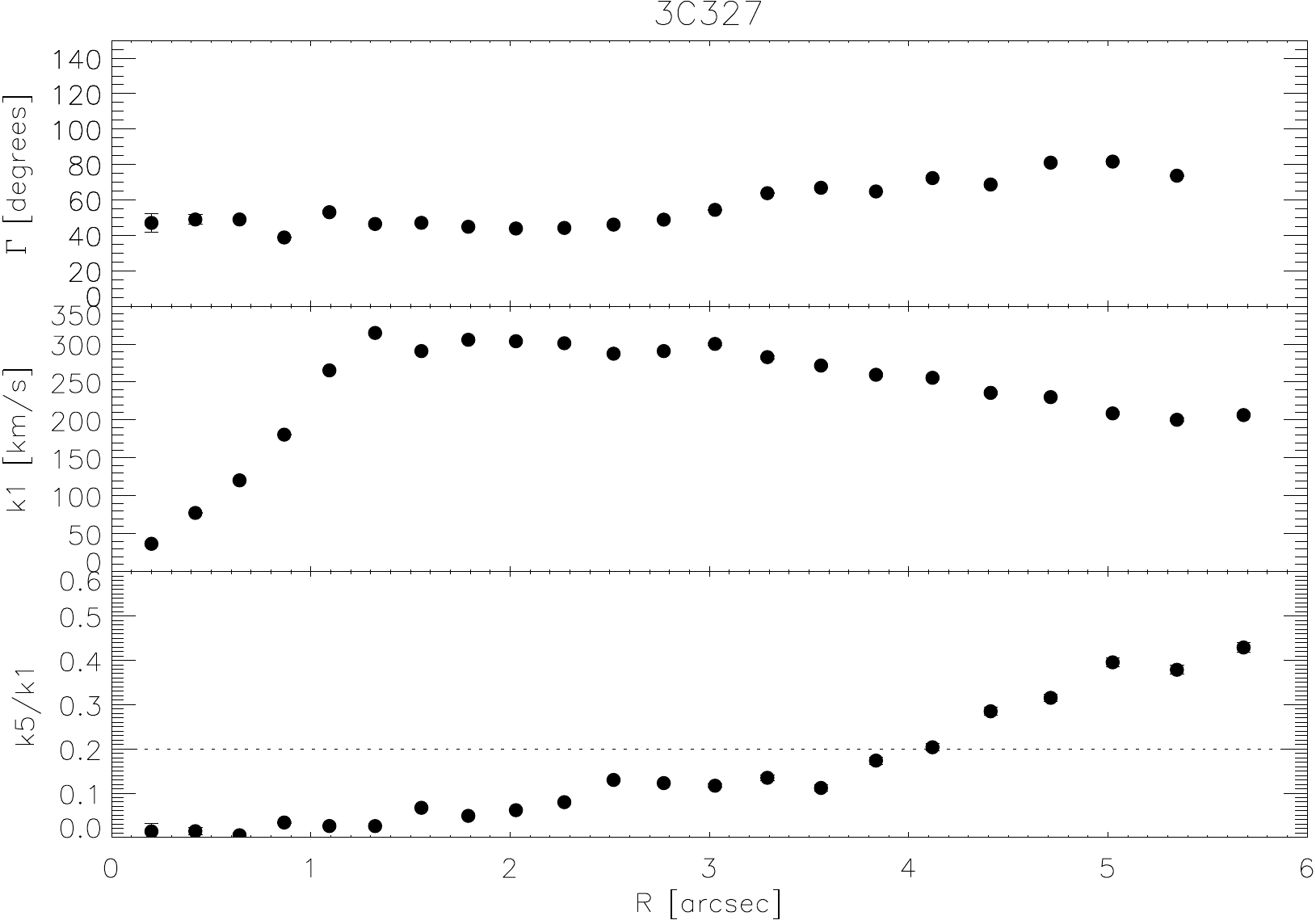}
\includegraphics[width=8.5cm]{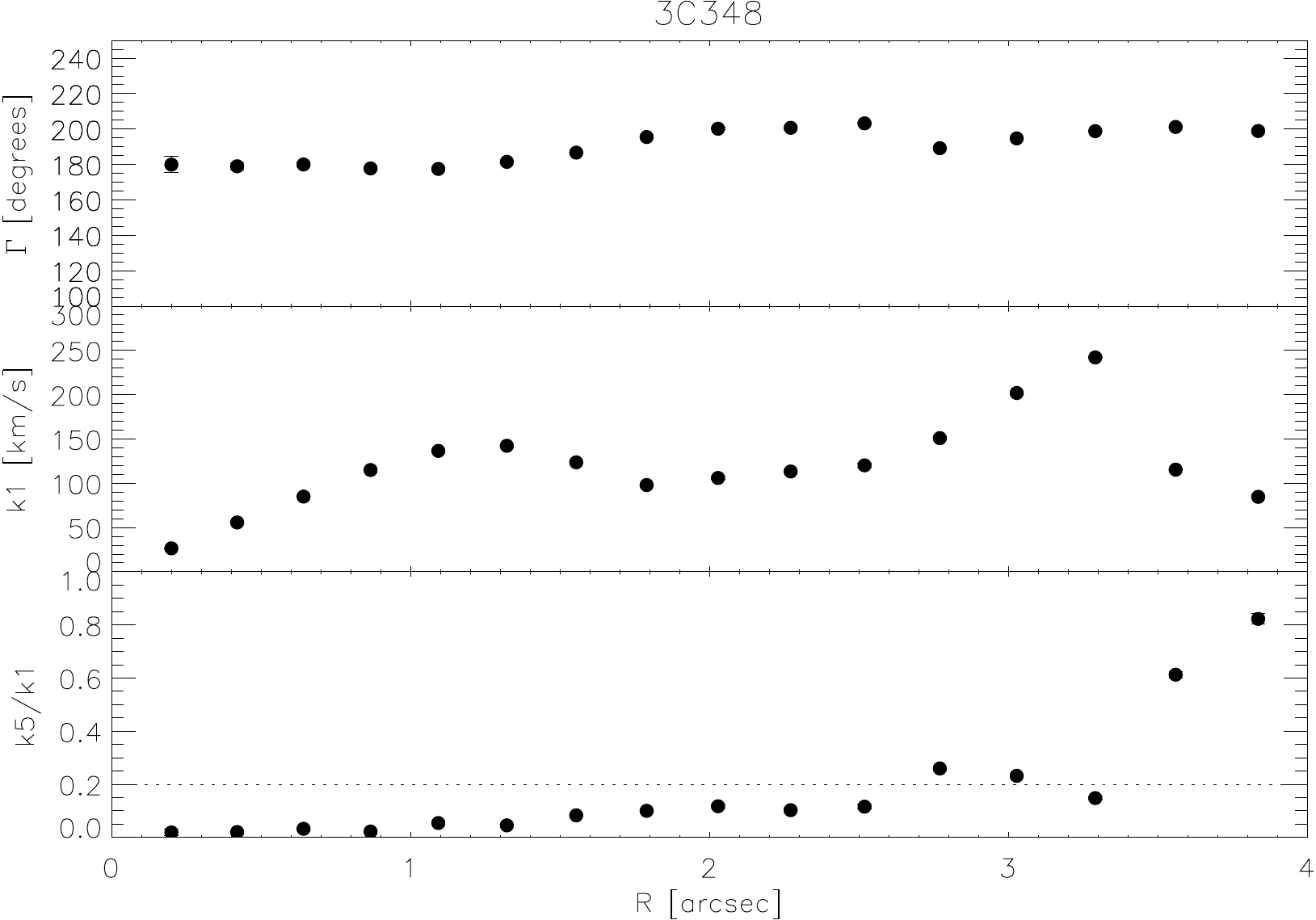}
\includegraphics[width=8.5cm]{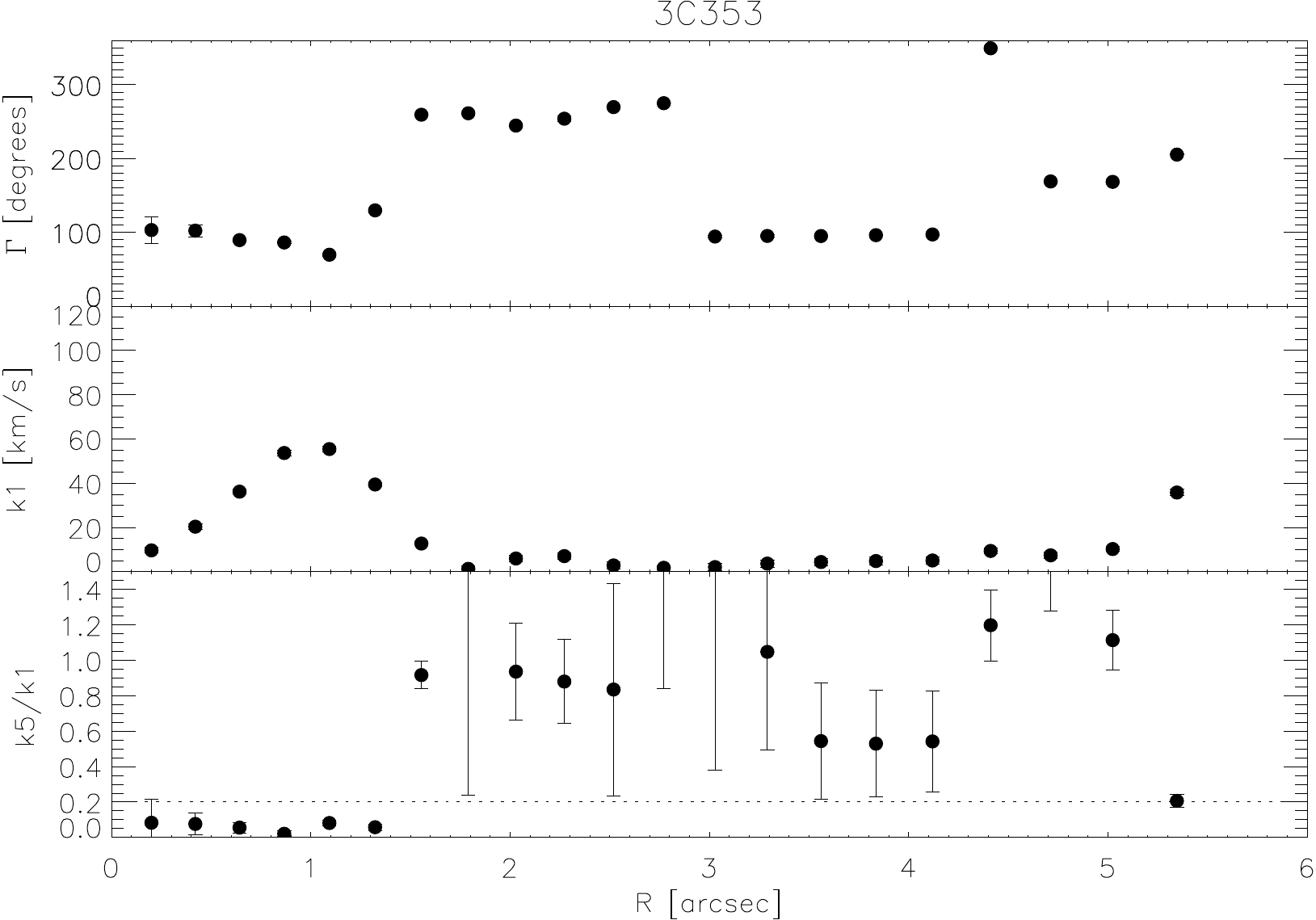}
\includegraphics[width=8.5cm]{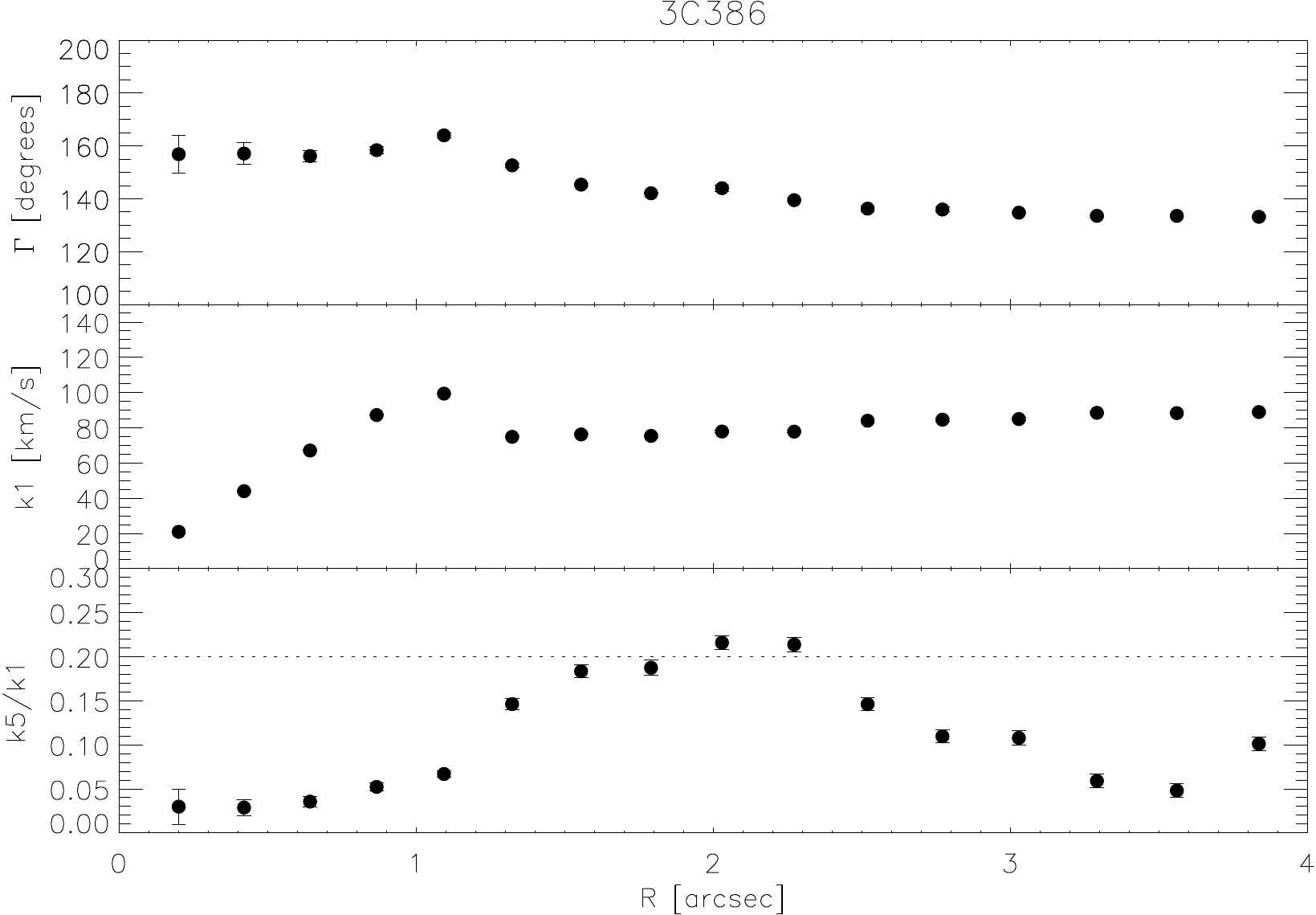}
\includegraphics[width=8.5cm]{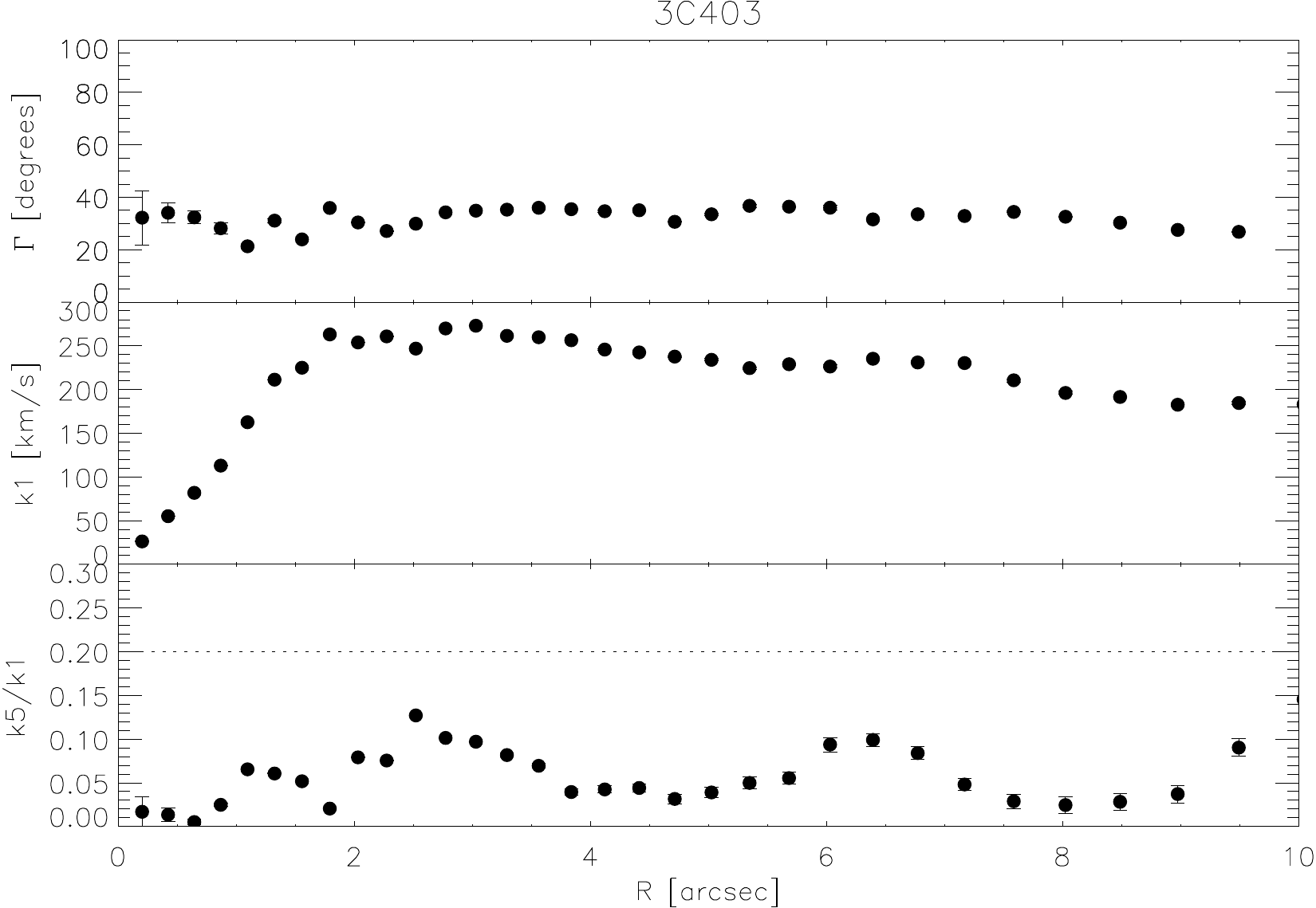}
\includegraphics[width=8.5cm]{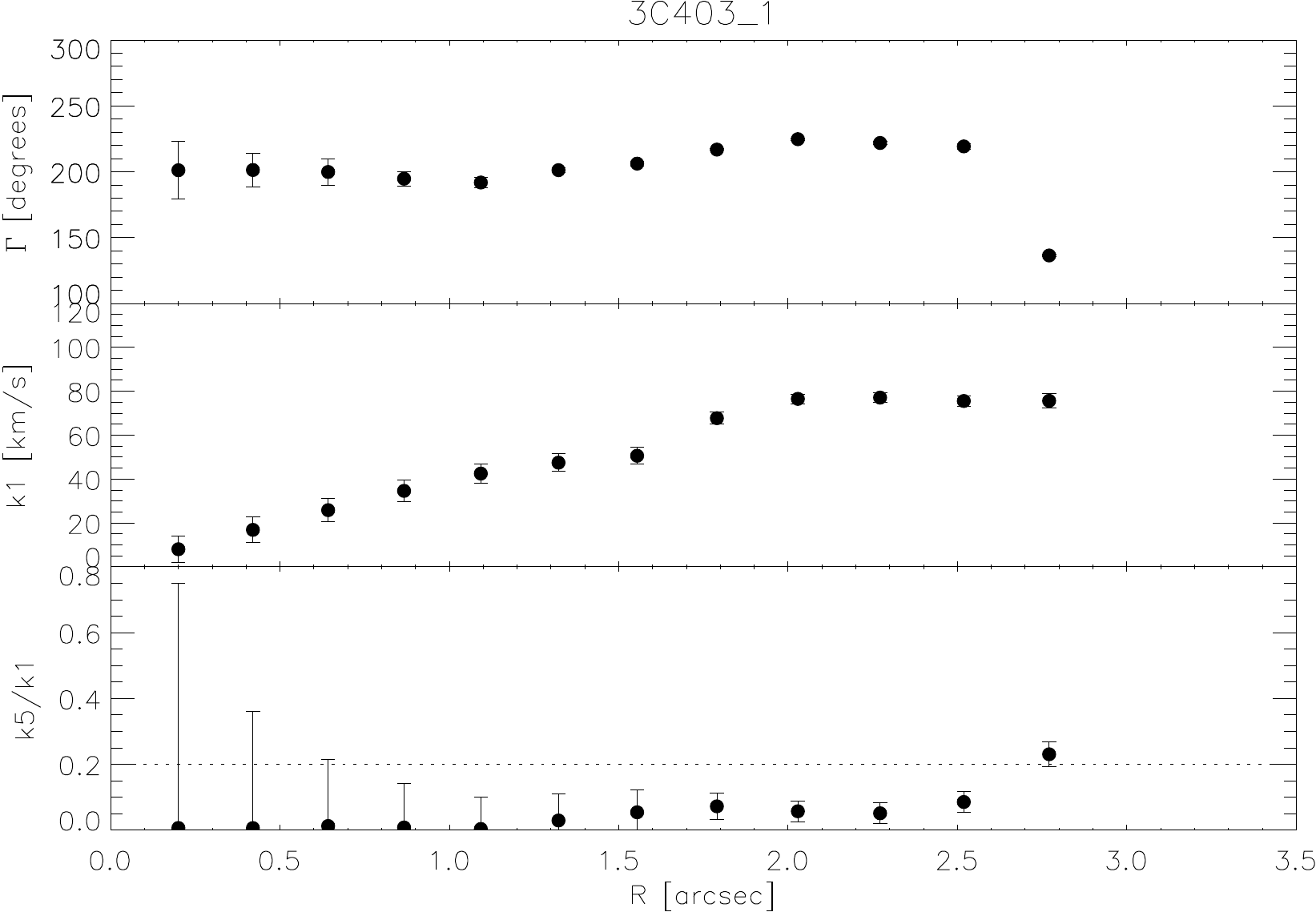}
\includegraphics[width=8.5cm]{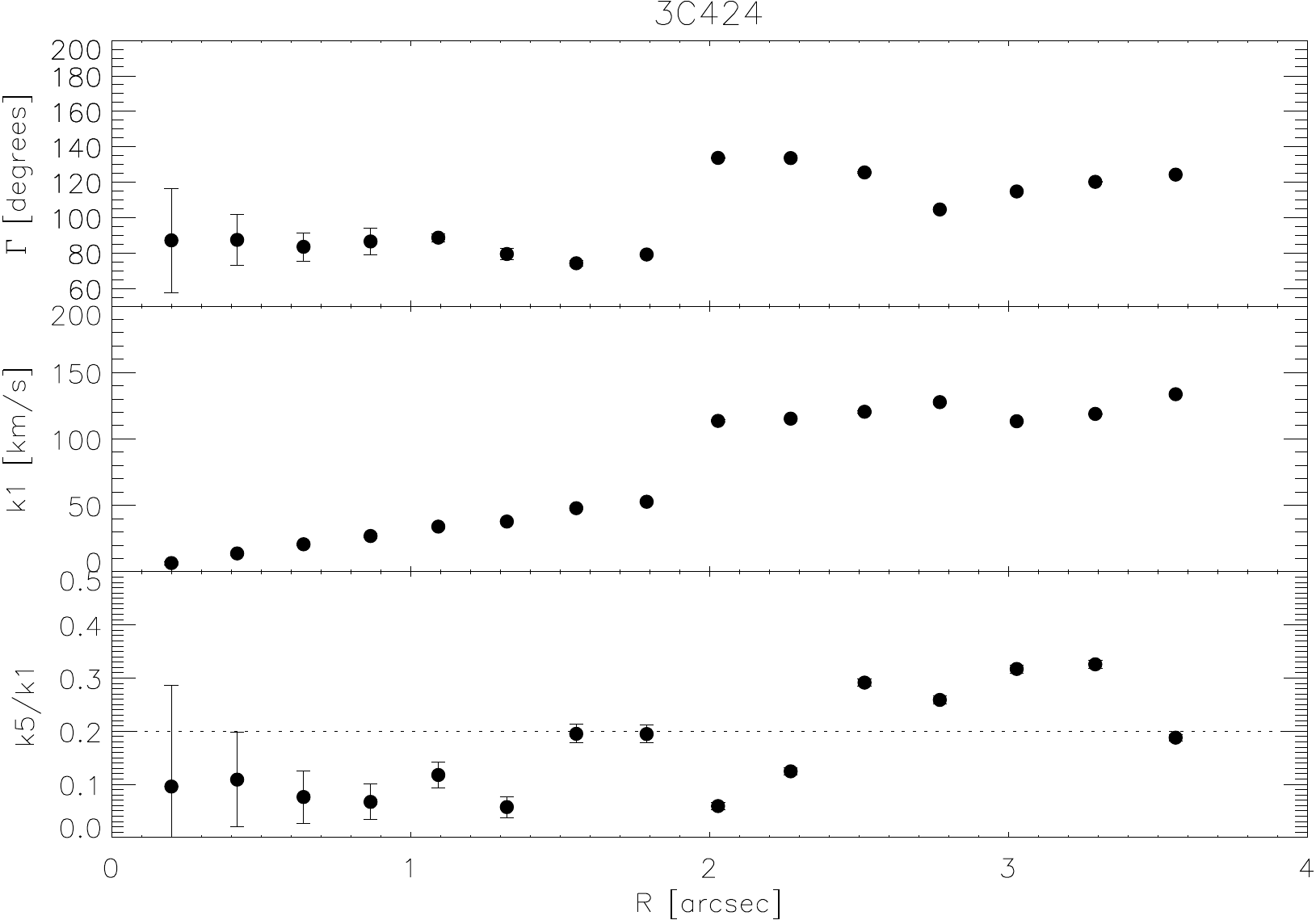}
\includegraphics[width=8.5cm]{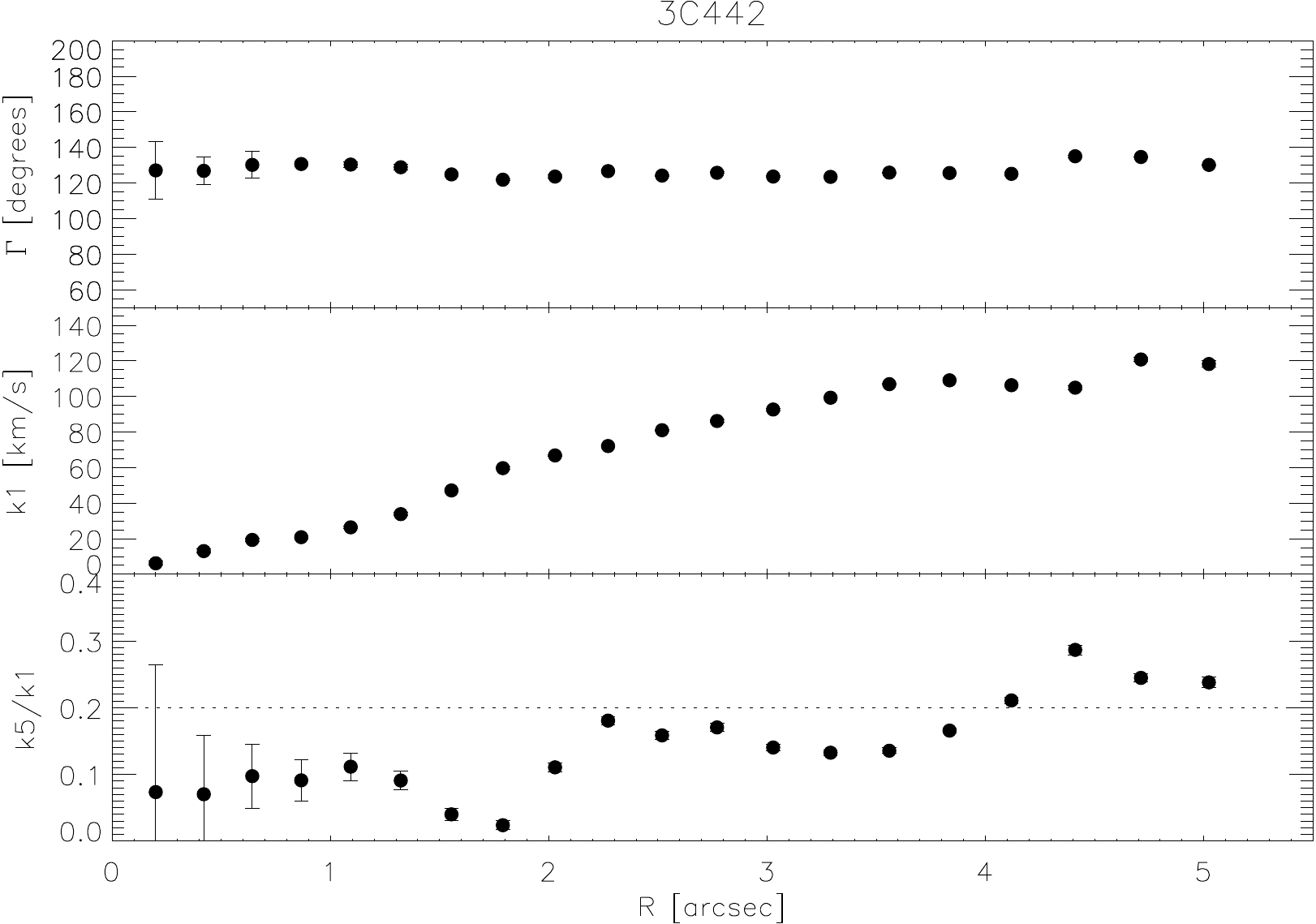}
}
\caption{- continued.}
\end{figure*}   

\addtocounter{figure}{-1}
\begin{figure*}  
\centering{ 
\includegraphics[width=8.5cm]{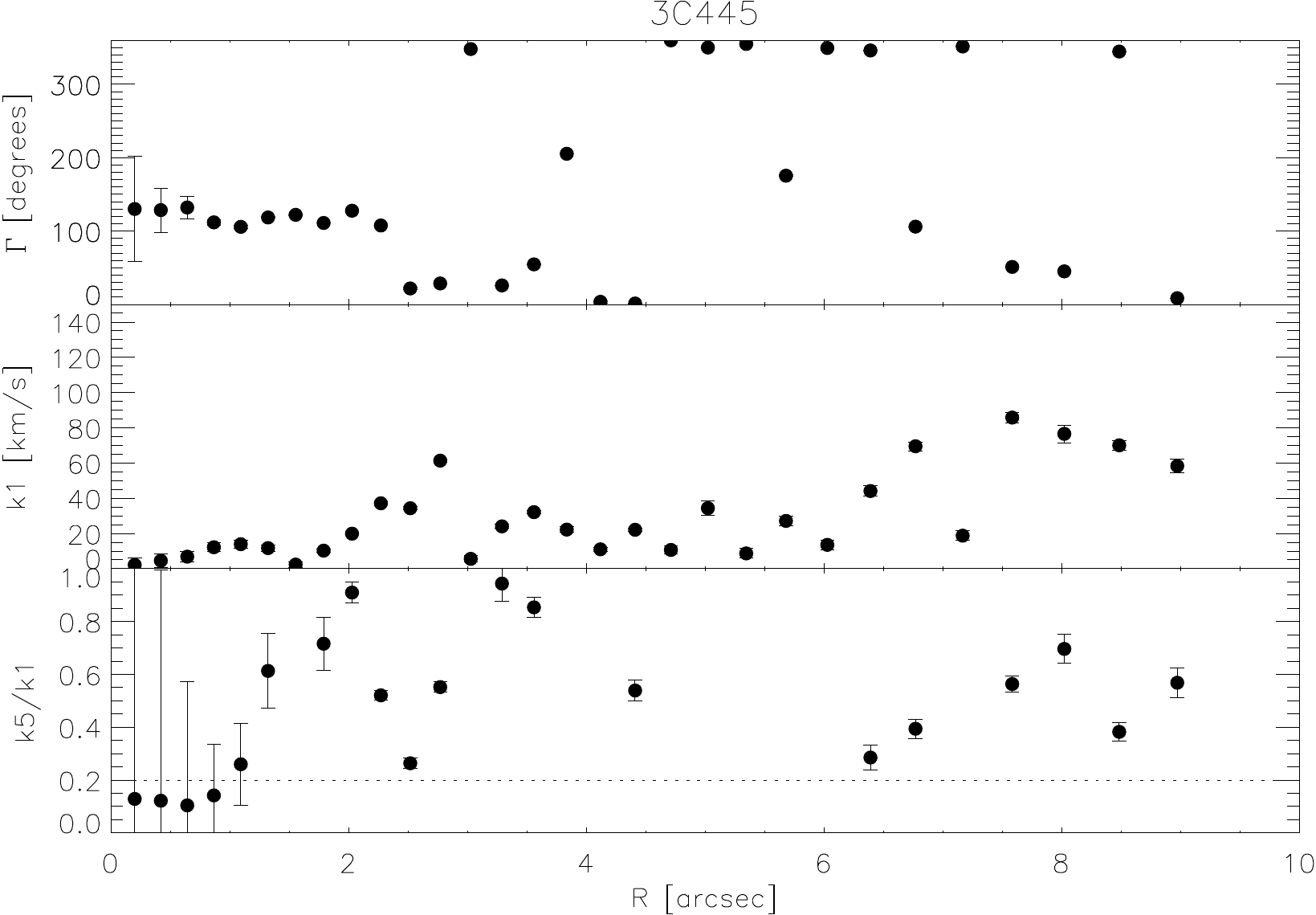}
\includegraphics[width=8.5cm]{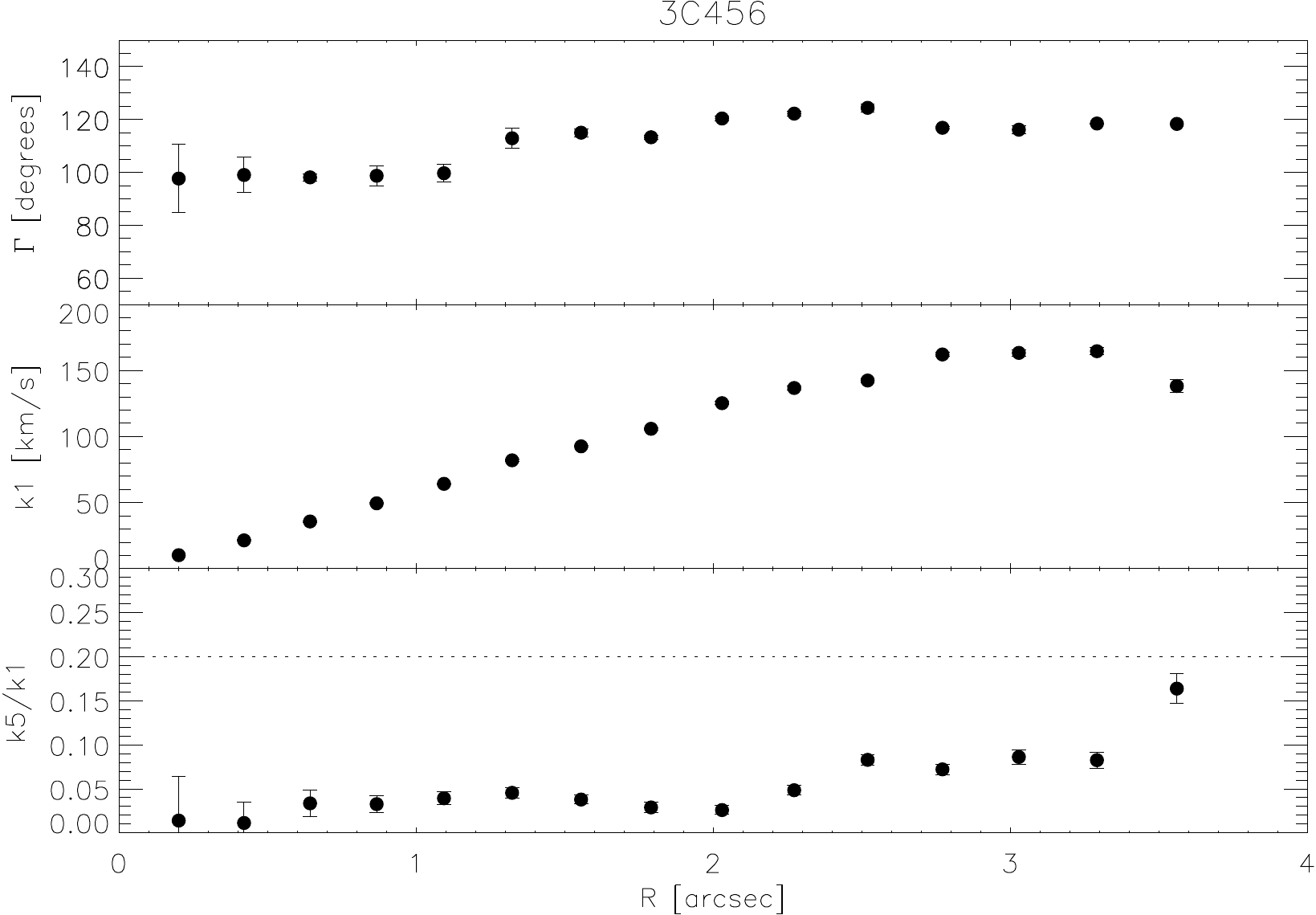}
\includegraphics[width=8.5cm]{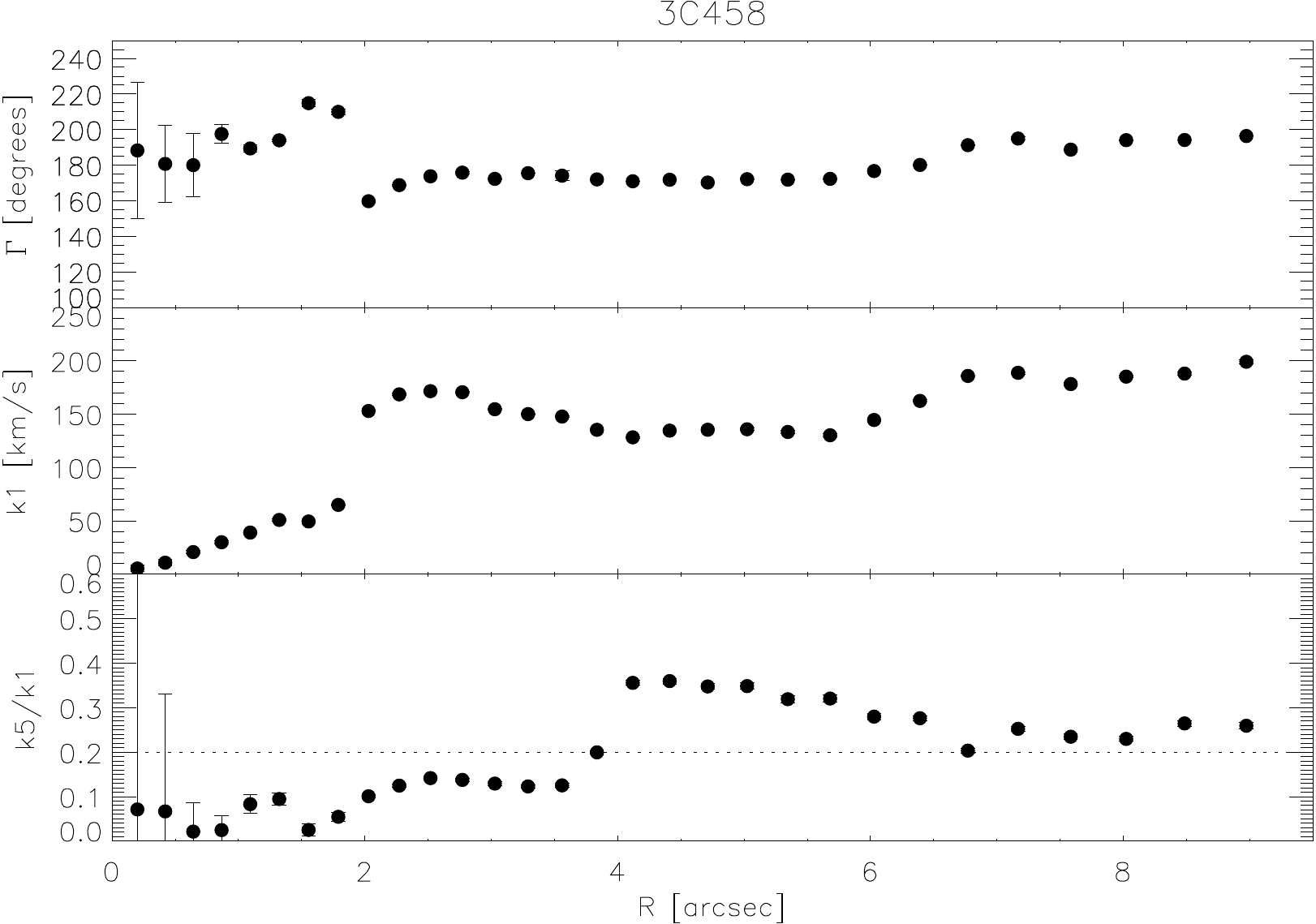}
\includegraphics[width=8.5cm]{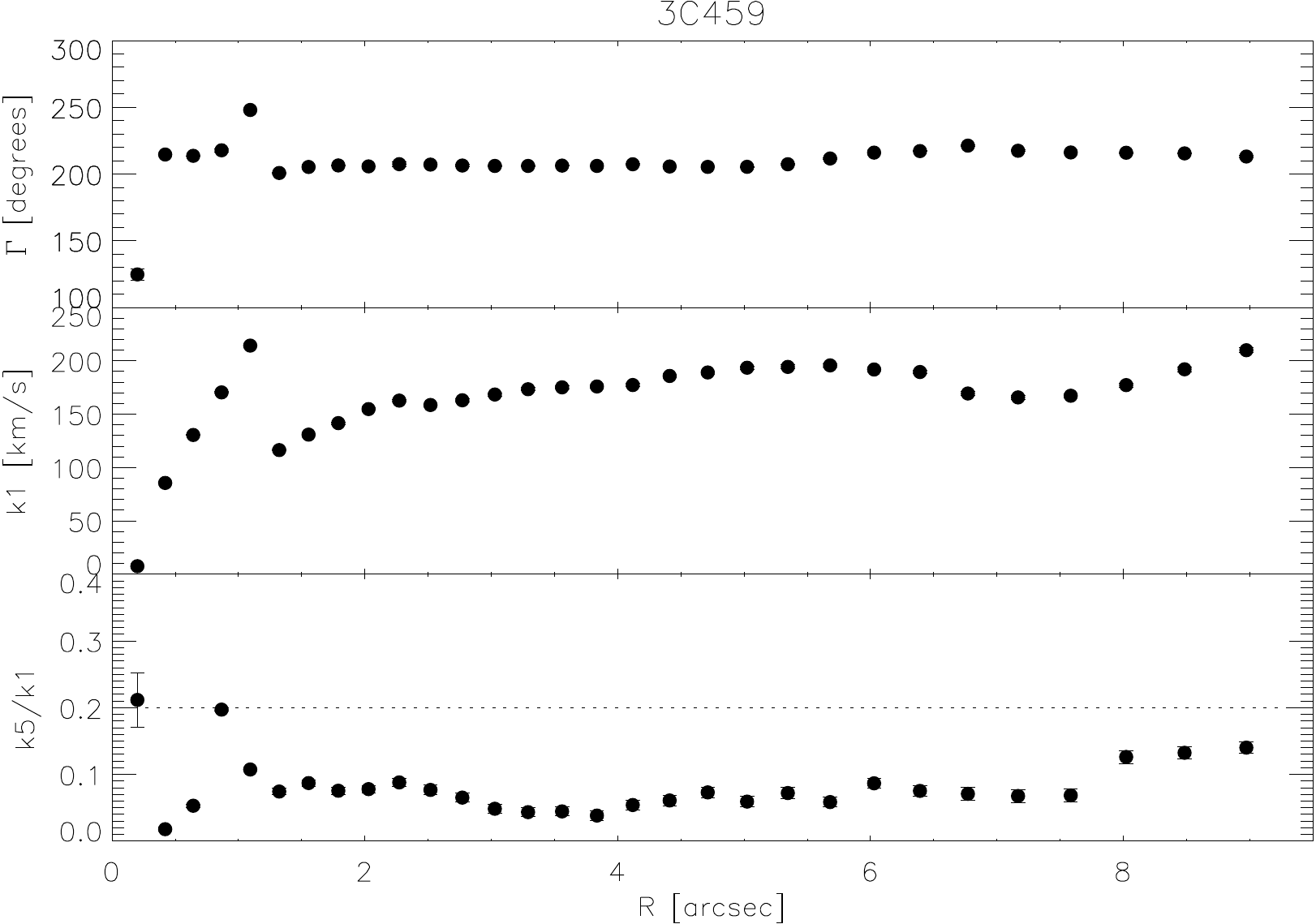}
}
\caption{- continued.}
\end{figure*}   

\end{appendix}


This research has made use of the NASA/IPAC Extragalactic Database
(NED), which is operated by the Jet Propulsion Laboratory, California
Institute of Technology, under contract with the National Aeronautics
and Space Administration. The radio images were produced as part of
the NRAO VLA Archive Survey, (c) AUI/NRAO. The NVAS can be browsed
through http://archive.nrao.edu/nvas/. S. Baum and C. O'Dea are grateful to the Natural Sciences and Engineering Research Council (NSERC) of Canada.
We thank the anonymous referee for her/his useful comments and suggestions.

\end{document}